\documentclass[manuscript,nonacm]{acmart}

\acmJournal{CSUR}
\acmVolume{1}
\acmNumber{1}
\acmArticle{1}
\acmMonth{1}
\acmYear{2026}
\setcopyright{acmlicensed}

\usepackage{pifont}
\usepackage{wasysym}
\usepackage{makecell}
\usepackage{multirow}
\usepackage[figuresleft]{rotating}
\usepackage{longtable,supertabular,booktabs}
\usepackage{threeparttablex,caption}
\usepackage{subcaption}
\usepackage{ltablex}
\usepackage[nolist]{acronym}
\usepackage{cleveref}
\usepackage{enumitem}

\usepackage{pgfplots}
\usepackage{pgfplotstable}
\pgfplotsset{compat=1.17}
\usepackage{color}
\usepackage{tikz}
\usepackage{adjustbox}
\usetikzlibrary{arrows.meta, positioning, shapes}
\usepackage[edges]{forest}
\definecolor{hiddendraw}{RGB}{205, 44, 36}
\definecolor{hidden-blue}{RGB}{194,232,247}
\definecolor{hidden-orange}{RGB}{243,202,120}
\definecolor{hidden-yellow}{RGB}{242,244,193}
\tikzstyle{mybox}=[
    rectangle,
    draw=hiddendraw,
    rounded corners,
    text opacity=1,
    minimum height=1.5em,
    minimum width=5em,
    inner sep=2pt,
    align=center,
    fill opacity=.5,
    ]

\usepackage[]{todonotes}
\let\oldtodo\todo
\renewcommand{\todo}[1]{\oldtodo[inline]{#1}}

\AtBeginDocument{%
  }

\usepackage{xspace}

\newcommand{\etc}{etc.\xspace}
\newcommand{\ie}{i.e.,\xspace}
\newcommand{\eg}{e.g.,\xspace}
\newcommand{\cf}{cf.\xspace}

\newcommand{\aka}{a.k.a.\xspace}
\let\oldnl\nl
\newcommand{\nonl}{\renewcommand{\nl}{\let\nl\oldnl}}

\begin{document}

\begin{acronym}[ASCII]
 \setlength{\itemsep}{0.2em}
 \acro{3DES}{Triple DES}
 \acro{AES}{Advanced Encryption Standard}
 \acro{AI}{Artificial Intelligence}
 \acro{ANF}{Algebraic Normal Form}
 \acro{ALU}{Arithmetic Logic Unit}
 \acro{ANN}{Artificial Neural Network}
  
 \acro{API}{Application Programming Interface}
 \acro{ABI}{Application Binary Interface}
 \acro{APIC}{Advanced Programmable Interrupt Controller}
 \acro{ARX}{Addition, Rotate, XOR}
 \acro{ASLR}{Address Space Layout Randomization}
 \acro{ASK}{Amplitude-Shift Keying}
 \acro{ASIC}{Application Specific Integrated Circuit}
 \acro{ASID}{Address Space Identifier}
 \acro{BL}{Bootloader Enable}
 \acro{BOR}{Brown-Out Reset}
 \acro{BPSK}{Binary Phase Shift Keying}
 \acro{BTB}{Branch Target Buffer}
 \acro{CBC}{Cipher Block Chaining}
 \acro{CBS}{Critical Bootloader Section}
 \acro{CGM}{Continuous Glucose Monitoring System}
 \acro{CMOS}{Complementary Metal Oxide Semiconductor}
 \acro{CNN}{Convolutional Neural Network}
 \acro{COPACOBANA}{Cost-Optimized Parallel Code Breaker and Analyzer}
 \acro{CPA}{Correlation Power Analysis}
 \acroplural{CPA}[CPAs]{Correlation Power Analyzes}
 \acro{CPU}{Central Processing Unit}
 \acro{CTR}{Counter \acroextra{(mode of operation)}}
 \acro{CRC}{Cyclic Redundancy Check}
 \acro{CRP}{Code Readout Protection}
 \acro{CVE}{Common Vulnerabilities and Exposures}
 \acro{DES}{Data Encryption Standard}
 \acro{DC}{Direct Current}
 \acro{DDR}{Double Data Rate RAM}
 \acro{DDS}{Digital Direct Synthesis}
 \acro{DFT}{Discrete Fourier Transform}
 \acro{DMA}{Direct Memory Access}
 \acro{DNN}{Deep Neural Network}
 \acro{DoS}{Denial-of-Service}
 \acro{DFA}{Differential Fault Analysis}
 \acro{DPA}{Differential Power Analysis}
 \acro{DRAM}{Dynamic Random-Access Memory}
 \acro{DRM}{Digital Rights Management}
 \acro{DSO}{Digital Storage Oscilloscope}
 \acro{DSP}{Digital Signal Processing}
 \acro{DST}{Digital Signature Transponder}
 \acro{DUT}{Device Under Test}
 \acroplural{DUT}[DUTs]{Devices Under Test}
 \acro{DVFS}{Dynamic Voltage and Frequency Scaling}
 \acro{ECB}{Electronic Code Book}
 \acro{ECC}{Elliptic Curve Cryptography}
 \acro{ECU}{Electronic Control Unit}
 \acro{EDE}{Encrypt-Decrypt-Encrypt \acroextra{(mode of operation)}}
 \acro{EEPROM}{Electrically Erasable Programmable Read-Only Memory}
 \acro{EL}{Exception Level}
 \acro{EM}{Electro-Magnetic}
 \acro{EPID}{Enhanced Privacy ID}
 \acro{FFT}{Fast Fourier Transform}
 \acro{FCC}{Federal Communications Commission}
 \acro{FIR}{Finite Impulse Response} 
 \acro{FIVR}{Fully Integrated Voltage Regulator}
 \acro{FPGA}{Field Programmable Gate Array}
 \acro{FSK}{Frequency Shift Keying}
 \acro{GMSK}{Gaussian Minimum Shift Keying}
 \acro{GPC}{Granule Protection Check}
 \acro{GPT}{Granule Protection Table}
 \acro{GPIO}{General Purpose I/O}
 \acro{GPU}{Graphics Processing Unit}
 \acro{GPGPU}{General Purpose GPU}
 \acro{GVT}{Graphics Virtualization Technology}
 \acro{HD}{Hamming Distance}
 \acro{HDL}{Hardware Description Language}
 \acro{HE}{Homomorphic Encryption}
 \acro{HF}{High Frequency}
 \acro{HMAC}{Hash-based Message Authentication Code}
 \acro{HW}{Hamming Weight}
 \acro{MPS}{Multi-Process Service}
 \acro{IC}{Integrated Circuit}
 \acro{ID}{Identifier}
 \acro{ISM}{Industrial, Scientific, and Medical \acroextra{(frequencies)}}
 \acro{IIR}{Infinite Impulse Response}
 \acro{IP}{Intellectual Property}
 \acro{IoT}{Internet of Things}
 \acro{ISA}{Instruction Set Architecture}
 \acro{IV}{Initialization Vector}
 \acro{JTAG}{Joint Test Action Group}
 \acro{LLC}{Last-Level Cache}
 \acro{LF}{Low Frequency}
 \acro{LFSR}{Linear Feedback Shift Register}
 \acro{LQI}{Link Quality Indicator}
 \acro{LSB}{Least Significant Bit}
 \acro{LSByte}{Least Significant Byte}
 \acro{LUT}{Look-Up Table}
 \acro{MAC}{Message Authentication Code}
 \acro{MKTME}{Multi-Key Total Memory Encryption}
 \acro{MDS}{Microarchitectural Data Sampling}
 \acro{MEE}{Memory Encryption Engine}
 \acro{MF}{Medium Frequency}
 \acro{ML}{Machine Learning}
 \acro{MMIO}{Memory-Mapped I/O}
 \acro{MMU}{Memory Management Unit}
 \acro{MITM}{Man-In-The-Middle}
 \acro{MPE}{Memory Protection Engine}
 \acro{MIG}{Multi-Instance GPU}
 \acro{MPC}{Multi-Party Computation}
 \acro{MSB}{Most Significant Bit}
 \acro{MSByte}{Most Significant Byte}
 \acro{MSK}{Minimum Shift Keying}
 \acro{MSR}{Model Specific Register}
 \acro{muC}[$\mathrm{\upmu C}$]{Microcontroller}
 \acro{NLFSR}{Non-Linear Feedback Shift Register}
 \acro{NLF}{Non-Linear Function}
 \acro{NFC}{Near Field Communication}
 \acro{NN}{Neural Network}
 \acro{NPU}{Neural Processing Unit}
 \acro{NRZ}{Non-Return-to-Zero \acroextra{(encoding)}}
 \acro{NVM}{Non-Volatile Memory}
 \acro{NW}{Normal World}
 \acro{SW}{Secure World}
 \acro{OOK}{On-Off-Keying}
 \acro{OP}{Operational Amplifier}
 \acro{OTP}{One-Time Password}
 \acro{ORAM}{Oblivious RAM}
 \acro{PC}{Personal Computer}
 \acro{PCB}{Printed Circuit Board}
 \acro{PCIe}{Peripheral Component Interconnect Express}
 \acro{PHT}{Page History Table}
 \acro{PhD}{Patiently hoping for a Degree}
 \acro{PKE}{Passive Keyless Entry}
 \acro{PKES}{Passive Keyless Entry and Start}
 \acro{PKI}{Public Key Infrastructure}
 \acro{PLL}{Phase Locked Loop}
 \acro{PMBus}{Power Management Bus}
 \acro{PMIC}{Power Management Integrated Circuit}
 \acro{PoC}{Proof-of-Concept}
 \acro{POR}{Power-On Reset}
 \acro{PPC}{Pulse Pause Coding}
 \acro{PRNG}{Pseudo-Random Number Generator}
 \acro{PSK}{Phase Shift Keying}
 \acro{PWM}{Pulse Width Modulation}
 \acro{RAPL}{Running Average Power Limit}
 \acro{RDP}{Read-out Protection}
 \acro{REE}{Rich Execution Environment}
 \acro{RMM}{Realm Management Monitor}
 \acro{RMP}{Reverse Map Table} 
 \acro{RF}{Radio Frequency}
 \acro{RFID}{Radio Frequency IDentification}
 \acro{RKE}{Remote Keyless Entry}
 \acro{RNG}{Random Number Generator}
 \acro{ROM}{Read Only Memory}
 \acro{ROP}{Return-Oriented Programming}
 \acro{RSA}{Rivest Shamir and Adleman}
 \acro{RSB}{Return Stack Buffer}
 \acro{RTL}{Register Transfer Level}
 \acro{SCA}{Side-Channel Analysis}
 \acro{SDK}{Software Development Kit}
 \acro{SDR}{Software-Defined Radio}
 \acro{SGX}{Software Guard Extensions}
 \acro{SEV}{Secure Encrypted Virtualization}
 \acro{SNR}{Signal to Noise Ratio}
 \acro{SNP}{Secure Nested Paging}
 \acro{SHA}{Secure Hash Algorithm}
 \acro{SHA-1}{Secure Hash Algorithm 1}
 \acro{SHA-256}{Secure Hash Algorithm 2 (256-bit version)}
 \acro{SMA}{SubMiniature version A \acroextra{(connector)}}
 \acro{SMBus}{System Management Bus}
 \acro{I2C}{Inter-Integrated Circuit}
 \acro{SPA}{Simple Power Analysis}
 \acro{SPI}{Serial Peripheral Interface}
 \acro{SPOF}{Single Point of Failure}
 \acro{SoC}{System on Chip}
 \acro{SVID}{Serial Voltage Identification}
 \acro{SWD}{Serial Wire Debug}
 \acro{TCB}{Trusted Computing Base} 
 \acro{TEE}{Trusted Execution Environment}
 \acro{TLB}{Translation Lookaside Buffer}
 \acro{TMTO}{Time-Memory Tradeoff}
 \acro{TMDTO}{Time-Memory-Data Tradeoff}
 \acro{TA}{Trusted Application}
 \acro{TZ}[TrustZone]{TrustZone}
 \acro{RSSI}{Received Signal Strength Indicator}
 \acro{SHF}{Superhigh Frequency}
 \acro{SMC}{Secure Monitor Call}
 \acro{UART}{Universal Asynchronous Receiver Transmitter}
 \acro{UCODE}[$\mathrm{\upmu Code}$]{Microcode}
 \acro{UHF}{Ultra High Frequency}
 \acro{UID}{Unique Identifier} 
 \acro{USRP}{Universal Software Radio Peripheral}
 \acro{USRP2}{Universal Software Radio Peripheral (version 2)}
 \acro{USB}{Universal Serial Bus} 
 \acro{VHF}{Very High Frequency}
 \acro{VLF}{Very Low Frequency}
 \acro{VHDL}{VHSIC (Very High Speed Integrated Circuit) Hardware Description Language}
 \acro{VR}{Voltage Regulator}
 \acro{WLAN}{Wireless Local Area Network}
 \acro{XEX}{Xor-Encrypt-Xor}
 \acro{XTS}{XEX-based Tweaked-Codebook Mode with Ciphertext Stealing}
 \acro{XOR}{Exclusive OR}
 \acro{IR}{Intermediate Representation}
 \acro{OCD}{On-Chip Debug}
 \acro{OS}{Operating System}
 \acro{TSX}{Transactional Synchronization Extensions}
 \acro{TZASC}{TrustZone Address Space Controller}
 \acro{VM}{Virtual Machine}
 \acro{vGPU}{Virtual Graphics Processing Unit}
 \acro{CC}{Confidential Computing}
 \acro{CCA}{Confidential Compute Architecture}
 \acro{TDX}{Trust Domain Extensions}
 \acro{LLM}{Large Language Model}
\end{acronym}

\acused{AES}
\acused{CRC}
\acused{DES}
\acused{EEPROM}
\acused{RSA}
\acused{USB}
\acused{SHA-1}
\acused{IC}
\acused{CPU}
\acused{USB}
\acused{DRAM}

\title{Confidential Computing on Heterogeneous CPU-GPU Systems: Survey and Future Directions}

\author{Qifan Wang}
\affiliation{%
  \institution{Durham University}
  \city{Durham}
  \country{United Kingdom}
}
\affiliation{%
  \institution{University of Birmingham}
  \city{Birmingham}
  \country{United Kingdom}
}
\email{qifan.wang2@durham.ac.uk}

\author{David Oswald}
\affiliation{%
  \institution{Durham University}
  \city{Durham}
  \country{United Kingdom}
}
\affiliation{%
  \institution{University of Birmingham}
  \city{Birmingham}
  \country{United Kingdom}
}
\email{david.f.oswald@durham.ac.uk}


\begin{abstract}
In recent years, the widespread informatization and rapid data explosion have increased the demand for high-performance heterogeneous systems that integrate multiple computing cores such as \acp{CPU}, \acp{GPU}, \acp{ASIC}, and \acp{FPGA}.
The combination of \ac{CPU} and \ac{GPU} is particularly popular due to its versatility. 
However, these heterogeneous systems face significant security and privacy risks. 
Advances in privacy-preserving techniques, especially hardware-based \acp{TEE}, offer effective protection for \ac{GPU} applications. 
Nonetheless, the potential security risks involved in extending \acp{TEE} to \acp{GPU} in heterogeneous systems remain uncertain and need further investigation. 
To investigate these risks in depth, we study the existing popular \ac{GPU} TEE designs and summarize and compare their key implications.
Additionally, we review existing powerful attacks on \acp{GPU} and traditional \acp{TEE} deployed on CPUs, along with the efforts to mitigate these threats. 
We identify potential attack surfaces introduced by \ac{GPU} \acp{TEE} and provide insights into key considerations for designing secure \ac{GPU} \acp{TEE}. 
This survey is timely as new \acp{TEE} for heterogeneous systems, particularly \acp{GPU}, are being developed, highlighting the need to understand potential security threats and build both efficient and secure systems.
\end{abstract}



\maketitle

\section{Introduction}
\label{sec:intro}

Heterogeneous computing systems have become a cornerstone of modern technology, driven by the need for enhanced energy efficiency and performance across a wide range of applications. 
From the immense computational power of supercomputers and high-performance clusters to the everyday use of smartphones and laptops, diverse computational architectures are now integrated to achieve the best balance between performance and power consumption~\cite{khokhar1993heterogeneous,mittal2015survey}.
Heterogeneous computing refers to systems that leverage multiple types of computing cores, including \acp{CPU}, \acp{GPU}, \acp{ASIC}, \acp{FPGA}, and \acp{NPU}. 
Although the concept of heterogeneous computing has been explored for over two decades, recent advancements in accelerators like \acp{GPU}, coupled with the explosive growth of \ac{ML}, have significantly transformed these systems.
In particular, since the mid-2000s, research has demonstrated that \acp{GPU} can accelerate the training of \acp{CNN} and other popular \ac{ML} models by 10--30${\times}$ compared to traditional \acp{CPU}~\cite{shi2016benchmarking,tan2021cryptgpu}. 
This acceleration makes heterogeneous computing systems using \acp{GPU} indispensable in today’s \ac{ML} infrastructure, providing critical support for the rapid development and deployment of \ac{AI} technologies.
While heterogeneous computing systems deliver exceptional performance and power efficiency, they also introduce new challenges. 
The complex architectures of these systems can create potential vulnerabilities and security risks, especially for the tasks they execute, such as \ac{ML} applications. Ensuring the security and reliability of these systems is essential as they continue to evolve and expand their role in the computing landscape.

To ensure the security and confidentiality of heterogeneous computing systems, researchers have focused on adapting privacy-preserving techniques traditionally used on \acp{CPU} to accelerators like \acp{GPU}. 
These techniques include cryptographic methods such as \ac{HE}~\cite{mishra2020delphi,ng2021gforce,fan2023tensorfhe} and Secure \ac{MPC}~\cite{mishra2020delphi,ng2021gforce,tan2021cryptgpu,watson2022piranha,jawalkar2024orca}, as well as \acp{TEE} tailored for accelerators like \acp{GPU}~\cite{volos2018graviton,jang2019heterogeneous,zhu2020enabling,NVIDIAh100,jiang2022cronus,lee2022tnpu,yudha2022lite,deng2022strongbox,ivanov2023sage,mai2023honeycomb,wang2024cage}.
\acp{TEE} offer certain advantages over cryptographic methods, particularly in performance and ease of deployment, providing a robust hardware-based security mechanism that is relatively straightforward to implement and manage. However, in contrast to cryptographic techniques, the security of \acp{TEE} relies on trust in the manufacturer and is difficult to analyse and prove rigorously.

Recent research has explored extending \acp{TEE} to accelerators like \acp{GPU}, both for x86~\cite{volos2018graviton,jang2019heterogeneous,zhu2020enabling,NVIDIAh100,lee2022tnpu,yudha2022lite,ivanov2023sage,mai2023honeycomb} and Arm~\cite{jiang2022cronus,deng2022strongbox,wang2024cage} architectures.
Most of these approaches involve modifying existing \ac{CPU} \acp{TEE} (especially process-based \acp{TEE} like Intel \ac{SGX} and Arm \ac{TZ}) to secure the \ac{GPU} software stack, such as \ac{GPU} drivers, and proposing strategies to ensure the isolation and confidentiality of \ac{GPU} kernel executions.
Apart from this, NVIDIA has introduced a pioneering commercial platform in its Hopper architecture~\cite{NVIDIAh100}, which supports confidential computing through features like memory encryption and isolated execution contexts.
However, the security threats associated with extending \acp{TEE} across both \acp{CPU} and accelerators in heterogeneous computing systems remain uncertain and require further investigation.

\begin{figure}[htbp]
\centering
\includegraphics[width=0.40\linewidth]{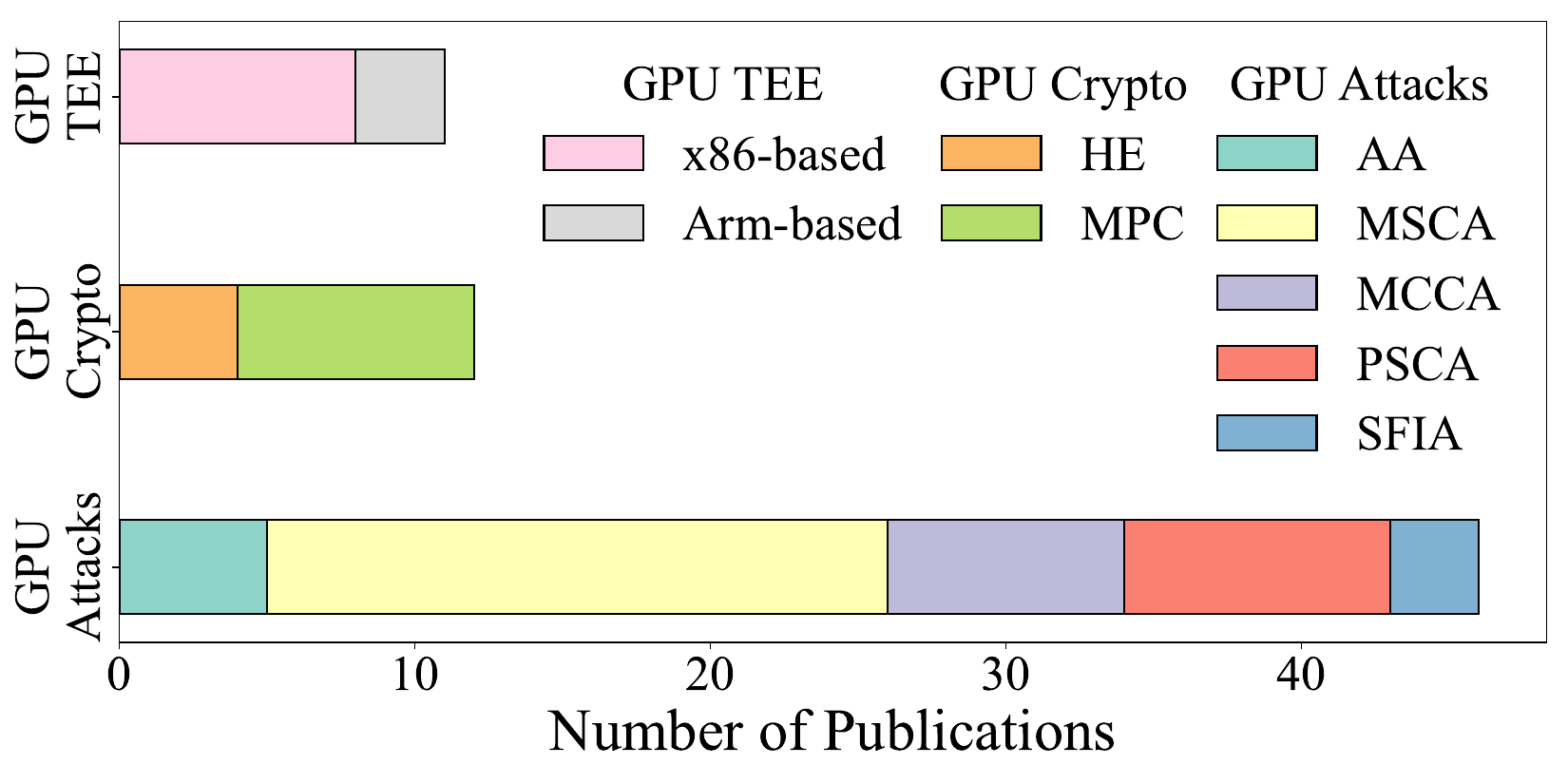}
\caption{Current state of security research on heterogeneous computing systems with \ac{GPU} (from the past 15 years). \emph{GPU Crypto} refers to papers that protect the privacy of \ac{GPU}-based applications using cryptographic techniques such as \ac{HE} and \ac{MPC}. \emph{GPU attacks} refer to architectural (AA), microarchitectural side-channel/covert-channel (MSCA/MCCA), physical side-channel (PSCA), and software-based fault injection attacks (SFIA).}
\label{fig:num_pubs}
\end{figure}

\begin{figure}[htbp]
\centering
\begin{adjustbox}{max width=\textwidth}
\begin{tikzpicture}[xscale=1.5, yscale=2.8]
    \draw[thick, -{Triangle[width=12pt,length=12pt]}, line width=1.5pt] (0,0) -- (16,0);
    
    \foreach \x/\year in {0/1999, 0.5/, 1.0/, 1.5/, 2.0/, 2.5/2004, 3.0/, 3.5/, 4.0/, 4.5/, 5/2009, 5.5/, 6.0/, 6.5/, 7.0/, 7.5/2014, 8.0/, 8.5/, 9.0/, 9.5/, 10/2019, 10.5/, 11.0/, 11.5/, 12.0/, 12.5/2024, 13.0/, 13.5/, 14.0/, 14.5/, 15.0/2029} {
        \draw (\x,0) -- (\x,-0.08);
        \node[anchor=north] at (\x,-0.2) {\year};
    }

    \node[draw, fill=gray!30, anchor=south west, xshift=0cm, align=center] (1) at (0,0.07) {\shortstack{First GPU,\\ NVIDIA GeForce256}};
    \draw[thick] (1.west) -- (0,0);

    \node[draw, fill=red!30, anchor=south west, xshift=3.75cm, align=center] (2) at (0,0.07) {\shortstack{ARM \ac{TZ} Launched in ARM11}};
    \draw[thick] (2.west) -- (2.5,0);

    \node[draw, fill=blue!30, anchor=south west, xshift= 4.1cm, align=center] (3) at (0,0.3) {\shortstack{First Virtualized GPU (NVIDIA VGX)}};
    \draw[thick] (3.east) -- (6.5,0);

    \node[draw, fill=green!30, anchor=south east, xshift= 11.25cm, align=center] (4) at (0,0.55) {\shortstack{First Attack Exploiting GPU Vuln.~\cite{lee2014stealing,maurice2014confidentiality}}};
    \draw[thick] (4.east) -- (7.5,0);

    \node[draw, fill=purple!30, anchor=south east, xshift=15.6cm, align=center] (5) at (0,0.81) {\shortstack{First MSCA on GPU~\cite{luo2015side}}};
    \draw[thick] (5.west) -- (8,0);
    \node[draw, fill=purple!30, anchor=south east, xshift=11.8cm, align=center] (5) at (0,0.81) {\shortstack{Intel SGX Launched in Skylake}};
    \draw[thick] (5.east) -- (8,0);

    \node[draw, fill=orange!30, anchor=south west, xshift=9.2cm, align=center] (6) at (0,-0.56) {\shortstack{First MCCA on GPU~\cite{naghibijouybari2017constructing}}};
    \draw[thick] (6.east) -- (9,0);
    \node[draw, fill=orange!30, anchor=south west, xshift=8.9cm, align=center] (6) at (0,-0.83) {\shortstack{AMD SEV Launched in EPYC}};
    \draw[thick] (6.east) -- (9,0);

    \node[draw, fill=yellow!30, anchor=south east, xshift=17.5cm, align=center] (7) at (0,-0.56) {\shortstack{First FIA on GPU~\cite{frigo2018grand}}};
    \draw[thick] (7.west) -- (9.5,0);
    \node[draw, fill=yellow!30, anchor=south west, xshift=14.2cm, align=center] (7) at (0,-0.83) {\shortstack{First x86-based GPU TEE~\cite{volos2018graviton}}};
    \draw[thick] (7.west) -- (9.5,0);

    \node[draw, fill=cyan!30, anchor=south east, xshift=15.7cm, align=center] (8) at (0,0.40) {\shortstack{First MIG-enabled\\ GPU, NVIDIA A100}};
    \draw[thick] (8.east) -- (10.5,0);
    \node[draw, fill=cyan!30, anchor=south east, xshift=15.6cm, align=center] (8) at (0,0.15) {\shortstack{Intel TDX Announced}};
    \draw[thick] (8.east) -- (10.5,0);

    \node[draw, fill=green!30, anchor=south east, xshift= 23.0cm, align=center] (9) at (0,-0.56) {\shortstack{AMD SEV-SNP Available in Milan}};
    \draw[thick] (9.west) -- (11,0);

    \node[draw, fill=pink!30, anchor=south east, xshift=21.8cm, align=center] (10) at (0,0.56) {\shortstack{First Arm-based GPU TEE~\cite{deng2022strongbox}}};
    \draw[thick] (10.west) -- (11.5,0);
    \node[draw, fill=pink!30, anchor=south east, xshift=21.9cm, align=center] (10) at (0,0.81) {\shortstack{First Commercial GPU TEE~\cite{NVIDIAh100}}};
    \draw[thick] (10.west) -- (11.5,0);

    \node[draw, fill=green!30, anchor=south east, xshift= 23.6cm, align=center] (11) at (0,0.32) {\shortstack{First MCCA on MIG-enabled GPU~\cite{zhang2023t}}};
    \draw[thick] (11.west) -- (12,0);

    \node[draw, fill=green!30, anchor=south east, xshift= 23.9cm, align=center] (12) at (0,0.09) {\shortstack{First Arm CCA-based GPU TEE~\cite{wang2024cage}}};
    \draw[thick] (12.west) -- (12.5,0);

\end{tikzpicture}
\end{adjustbox}
\caption{Timeline of significant events in secure heterogeneous computing systems.}
\label{fig:timeline}
\end{figure}
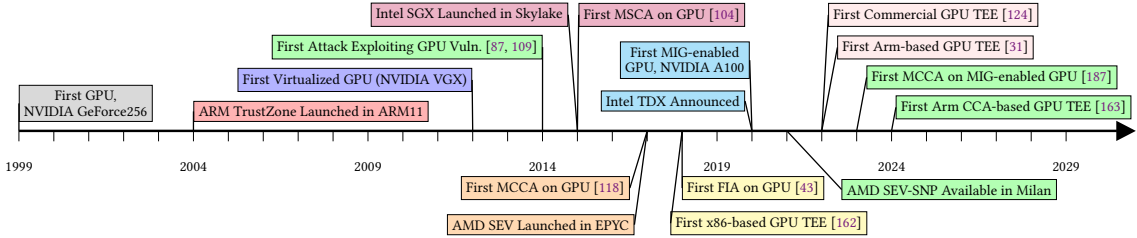

In this survey, we delve into the potential vulnerabilities in current \ac{TEE} designs on accelerators, with a specific emphasis on \acp{GPU} (referred to as \emph{\ac{GPU} \acp{TEE}}).
To this end, we begin by reviewing and classifying research efforts dedicated to creating \ac{GPU} \ac{TEE} prototypes. These solutions target various aspects: some aim to secure the \ac{GPU} software stack, others focus on safeguarding the confidentiality of data processed on the \ac{GPU}, and some provide robust isolation mechanisms. Consequently, each design presents different attack surfaces.
To identify these potential attack surfaces, we categorize and analyze existing attacks on \acp{GPU}, including architectural/logical attacks, microarchitectural side-channel and covert-channel attacks, physical side-channel attacks, and fault injection. 
The unique architecture and accessibility of \acp{GPU} within heterogeneous computing systems pose significant differences compared to traditional \acp{CPU}, which have historically been the primary targets of these attacks.
Additionally, we also review existing attacks against \ac{CPU} \acp{TEE} such as Intel \ac{SGX}, Arm \ac{TZ}, and AMD \ac{SEV}, and investigate their implications for code and data running on \ac{GPU} \acp{TEE}. 
Finally, by synthesizing these insights, we discuss the potential attack surfaces in \ac{GPU} \acp{TEE} exposed by these powerful attack vectors.

As shown in \Cref{fig:num_pubs} and \Cref{fig:timeline}, the evolution of \acp{GPU} has led to the development of increasingly high-performance architectures. 
Consequently, research has increasingly focused on investigating potential attack surfaces in \acp{GPU} by exploring new architectural vulnerabilities.
Likewise, we expect the potential security threats of recently emerging \ac{GPU} \acp{TEE} to receive growing attention.
We believe that this survey provides critical insights into key considerations for designing \ac{GPU} \acp{TEE}, the primary workflows of existing attacks targeting heterogeneous computing systems, and the potential attack surface introduced for both CPU \acp{TEE} and \acp{GPU}. 
This understanding can guide future research into developing defenses for secure high-performance computing on heterogeneous systems and enhance confidential computing on \ac{GPU} \acp{TEE} by refining the \ac{TCB}. 
Moreover, we believe our work can inspire the creation of more \ac{GPU} \ac{TEE} prototypes that are well-suited for various architectures and real-world applications.

\section{Background}
\label{sec:bg}
In this section, we introduce the basics of \acp{GPU}, common attack types, and the most existing popular \ac{TEE} designs.

\subsection{General Purpose Graphics Processing Units}
\label{sec:GPU basics}

\begin{table}[htbp]
\footnotesize
\centering
\caption{Summary of existing popular \acp{GPU}}
\label{tab:gpu}

\resizebox{\linewidth}{!}{
\begin{tabular}{lp{2cm}lp{3cm}p{1.5cm}cccllll}
\toprule
    \multicolumn{2}{c}{\multirow{2}{*}{\textbf{Accelerator}}} & \multicolumn{1}{c}{\multirow{2}{*}{\textbf{Year}}} & \multicolumn{1}{c}{\multirow{2}{*}{\textbf{Features}}} & \multirow{2}{*}{\textbf{Virtualization}} & \multicolumn{3}{c}{\textbf{Shared Memory Hierarchy}} & \multicolumn{3}{c}{\textbf{Memory Size}} \\ 

    \cmidrule{6-11}
    
    \multicolumn{2}{c}{} & \multicolumn{1}{c}{} & \multicolumn{1}{c}{} &  & \multicolumn{1}{c}{\textbf{None}} & \multicolumn{1}{c}{\textbf{L2/L3 Cache}} & \multicolumn{1}{c}{\textbf{DRAM}} & \multicolumn{1}{c}{\textbf{L1}} & \multicolumn{1}{c}{\textbf{L2}} & \multicolumn{1}{c}{\textbf{DRAM}} \\ 

    \midrule

    \multirow{1}{*}{Intel} & HD/Iris/UHD/Xe & 2010-2020 & Integrated graphics & GVT &  & $\checkmark$ & $\checkmark$ & Shared & Shared & Shared \\

    \midrule
    
    \multirow{4}{*}{NVIDIA} & P100 & 2016 & HBM2 & vGPU/MPS & $\checkmark$ &  &  & 64 KB & 4MB & 16 GB \\
     & V100 & 2017 & Optimized SM for \ac{ML} & vGPU/MPS & $\checkmark$ &  &  & 128 KB & 6MB & 16 GB \\
     & A100 & 2020 & 3G Tensor Cores and NVLink & vGPU/MPS/MIG & $\checkmark$ &  &  & 192 KB & 40 MB & 40 GB \\
     & H100 & 2022 & Confidential computing & vGPU/MPS/MIG & $\checkmark$ &  &  & 256 KB & 50 MB & 80 GB \\

    \midrule

    \multirow{2}{*}{AMD} & Radeon RX 580 & 2017 & FreeSync, CrossFire & N/A & $\checkmark$ &  &  & 16 KB & 2 MB & 4/8 GB \\
    & Radeon RX 6900 & 2020 & Infinity Cache, SAM & N/A & $\checkmark$ &  &  & 80 KB & 4 MB & 16 GB \\

    \midrule

     \multirow{1}{*}{Arm} & Mali-T624 \ac{GPU} & 2013 & Unified Shader Architecture & N/A &  &  & $\checkmark$ & 16 KB & 256 KB & Shared \\

     \midrule

     \multirow{1}{*}{Qualcomm} & Adreno \ac{GPU} 800 & 2020 & \ac{AI} Acceleration & N/A &  &  & $\checkmark$ & 32-64 KB & 256 KB-1 MB & Shared \\
    
    \bottomrule
\end{tabular}
}
\begin{flushleft}
In \textbf{Virtualization}, vGPU, GVT, MPS, and MIG denote Virtual Graphics Processing Unit, Graphics Virtualization Technology, Multi-Process Service, and Multi-Instance \ac{GPU}.
\textbf{Memory Sharing} denotes whether a level of the memory hierarchy is shared between the CPU  and the accelerator. 
\end{flushleft}
\end{table}

\acp{GPGPU} are specialized accelerators designed to handle diverse compute-intensive workloads beyond traditional graphics rendering. With their robust parallel computing capabilities, these \acp{GPU} are ideal for applications such as data analysis, \ac{ML}, cryptography, \etc
Typically, a \ac{GPGPU} consists of multiple processing units (\ie CUDA cores in NVIDIA \acp{GPU} and shader cores and execution units in Arm \acp{GPU}) capable of executing thousands of threads simultaneously.
Aside from \acp{GPU}, there are several prominent accelerators designed to boost performance for specific computational tasks. These include \acp{FPGA}, \acp{ASIC}, and \acp{NPU}. 
In this survey, our primary focus is on heterogeneous systems utilizing \acp{GPU}.

With the evolution of \acp{GPU}, vendors like Intel, NVIDIA, AMD, Arm, Qualcomm, and others have introduced \acp{GPU} with distinct architectural features and capabilities.
Different architectural features, such as the memory hierarchy, may introduce different attack surfaces on \acp{GPU}.
Existing widely used \acp{GPU}, detailed in \Cref{tab:gpu}, vary in their architectural designs, particularly concerning their memory hierarchy, which can affect their vulnerability to certain types of attacks (\eg microarchitectural attacks on integrated and discrete \acp{GPU} in \Cref{sec:GPU-sec}). 
To provide a comprehensive overview of these \acp{GPU}, we consider the aspects in the following sections.

\noindent\textbf{Integrated/Discrete \ac{GPU}.} 
Integrated \acp{GPU}, \eg Intel HD/Iris/UHD/Xe Graphics, Arm, and Qualcomm \acp{GPU} in \Cref{tab:gpu}, are integrated on the same die or package as the CPU. 
These \acp{GPU} share the system's memory with the CPU. 
Discrete \acp{GPU} are standalone and dedicated graphics processing units separate from the CPU.
They have their dedicated memory, which is distinct from the system memory (\eg NVIDIA and AMD \acp{GPU} in \Cref{tab:gpu}). 
NVIDIA GPUs have long supported host memory access via PCIe and Unified Memory, but recent architectures like the H100---especially in Grace Hopper configurations---enable tighter CPU-GPU integration with coherent memory sharing over NVLink-C2C~\cite{nvidia2022grace}.

\noindent\textbf{\ac{GPU} virtualization.} 
\emph{\ac{GVT}}, \aka Intel GVT, enables a single physical \ac{GPU} to be shared among multiple \acp{VM} with dedicated \ac{GPU} resources.
\emph{\ac{MPS}} from NVIDIA manages multiple CUDA applications concurrently on a single \ac{GPU} but lacks the address space isolation of true GPU virtualization, making it unsuitable for multi-tenant systems~\cite{zhang2023t}.
NVIDIA's \emph{\ac{vGPU}}, unlike \ac{MPS}, ensures separate \ac{GPU} address spaces and execution contexts.
\emph{\ac{MIG}}, introduced in NVIDIA's A100 \ac{GPU} architecture, partitions a single physical \ac{GPU} into multiple instances. Each \ac{MIG} instance operates as a separate \ac{GPU} with dedicated resources, including compute cores, memory, and cache.

\noindent\textbf{Open-source status of \ac{GPU} source code.} 
The availability of \ac{GPU} source code, including drivers and low-level architectural details, varies depending on the vendor's policies and the specific \ac{GPU} architecture. 
For instance, vendors like AMD (\eg AMD's ROCm~\cite{hunt2020telekine}) and Intel (Mesa 3D Graphics~\cite{intelMesa}) offer open-source \ac{GPU} drivers for specific platforms.
However, NVIDIA, ARM, and Qualcomm typically provide proprietary drivers, limiting developers' access to design details.
As a result, many research efforts on \ac{GPU} attacks often rely on reverse engineering to acquire details like \ac{TLB} levels, warp schedulers, and more (see \Cref{sec:GPU-sec}).

\subsection{Types of \ac{GPU} Attacks}
\label{sec:attack-types}

In recent decades, a range of attacks has emerged as significant threats to modern \ac{CPU}-based computing systems. As the use of \acp{GPU} becomes more prevalent, there is growing interest in exploring potential security vulnerabilities that extend from the \ac{CPU} to the \ac{GPU}. 
In this survey, we detail various attack vectors, including architectural/logical attacks, microarchitectural side-channel and covert-channel attacks, physical side-channel attacks, and fault injection.

\subsubsection{Architectural Attacks}

These attacks exploit vulnerabilities in the design and configuration of hardware components. This includes misuse of GPU or processor features, privilege escalation, and weak memory isolation. Attackers may target processing units (\eg processors, hardware accelerators, and co-processors), the memory subsystem, and communication components responsible for interfacing with other system elements to bypass protections or leak sensitive data~\cite{ghasempouri2023survey}. \emph{Logical attacks}, a subset of architectural attacks, exploit flaws in the intended logic or functional behavior of the hardware. These may involve undocumented instructions, incorrect permission checks, or logic bugs that allow attackers to alter control flow or data access patterns.

\subsubsection{Microarchitectural Side-Channel and Covert-Channel Attacks}
Microarchitectural side-channel attacks exploit unintended behaviours of a processor's microarchitecture to extract sensitive information. 
They target components like cache behaviour, branch prediction, or speculative execution. Initially demonstrated on \acp{CPU}, these attacks exploit various optimization structures such as caches~\cite{gruss2016flush+,liu2015last,gruss2015cache,osvik2006cache} and branch predictors~\cite{evtyushkin2016jump,evtyushkin2016understanding}. 
Recently, a new category of microarchitectural attacks, known as transient-execution attacks~\cite{xiong2021survey}, has emerged on \acp{CPU}. 
Examples include Meltdown~\cite{lipp2020meltdown}, Spectre~\cite{kocher2020spectre}, and \ac{MDS}~\cite{van2019ridl,ragab2021crosstalk}, which rely on out-of-order and transient (speculative) execution in modern \acp{CPU}.
Unlike physical side-channel attacks, microarchitectural attacks necessitate an understanding of the specific microarchitecture and its inherent vulnerabilities.

A microarchitectural covert-channel attack is a subtype of a side-channel attack. 
Covert channels involve two processes establishing an unauthorized communication pathway through an unintended channel to leak information, making them relevant in cloud computing environments~\cite{muchene2013reporting,zhenyu2012whispers}. 
Due to imperfections in isolation and shared resource usage, malicious processes can exploit such resources to establish a communication channel with the victim process. 
The covert channel typically involves a sender, known as the ``trojan'', and a receiver, often referred to as the ``spy''. 
Through covert channels, the trojan can transmit sensitive information or leak data to the spy~\cite{ahn2021network}.

\subsubsection{Physical Side-Channel Attacks}
These passive attacks exploit the physical characteristics of a computing device, such as power consumption, \ac{EM}, or timing information, to gather insights into the device's operations. Examples include attacks such as \ac{DPA}, which exploits fluctuations in power consumption to extract cryptographic keys or other sensitive data. 
Unlike microarchitectural attacks, physical side-channel attacks do not always demand knowledge of the device's internal architecture or vulnerabilities. 
Instead, they use the unintentional leakage of information through physical signals and behaviour.

\subsubsection{Fault Injection Attack}
Fault injection attacks involve actively introducing faults or errors into a system to alter its behavior or extract sensitive information. 
These attacks can be generally categorized into non-invasive, semi-invasive, and invasive~\cite{shuvo2023comprehensive}. 
Non-invasive attacks allow adversaries to use methods such as clock and voltage glitching, overheating, \ac{EM} faults, software faults, and remote hardware faults, with either physical or non-physical access. 
Semi-invasive attacks require the (partial) decapsulation of the target chip for techniques like laser and optical fault injection, while invasive attacks involve more direct methods, such as micro-probing and \ac{IC} modification, both of which necessitate deep physical access. 
In this work, we primarily focus on non-invasive fault injection attacks.

\subsubsection{Comparison of Attacks on \acp{CPU} and \acp{GPU}}
While attacks on \ac{CPU} and \ac{GPU} share some fundamental principles---such as exploiting timing variations, memory leakage, or shared resources---their feasibility, targeted attack surfaces, and required attacker capabilities diverge significantly due to underlying architectural and operational differences.
CPUs generally enforce stricter privilege separation and memory isolation, making attacks like residual memory leakage less effective. 
In contrast, GPUs have historically lacked such safeguards, making them more vulnerable to memory remanence and weak access control. For microarchitectural attacks, GPUs introduce new leakage vectors due to their massive parallelism and SIMT execution model, allowing attackers to exploit fine-grained resource contention (\eg, warps, shared caches) at scale. 
GPUs also expose performance counters and APIs more openly than CPUs, increasing their attack surface. 
In physical side-channel attacks, GPUs tend to emit more observable signals due to their higher power usage and bandwidth, and their close coupling with CPUs introduces additional leakage through shared interconnects or MMIO. 
For fault injection, while CPUs are often targeted through physical glitching, GPU attacks frequently exploit software-exposed features like DVFS or power-state manipulation to induce faults during kernel execution. 
ML workloads on GPUs are resilient to random faults but remain vulnerable to precise, targeted injections. 
Overall, while CPUs benefit from more mature defense mechanisms, GPU security is still evolving. Adapting proven CPU defenses to GPU-specific designs and threat models remains an important direction for future research.

\subsection{\ac{TEE}}
\label{sec:tee}
Most existing \ac{GPU} \acp{TEE} rely on a \ac{CPU} \ac{TEE} to protect the \ac{GPU} drivers on the \ac{CPU} host, which can be compromised by the adversary. This section introduces \ac{CPU} \acp{TEE}, including process-based \acp{TEE} such as Intel \ac{SGX} and Arm \ac{TZ}, which provide isolation at the process level, and virtualization-based \acp{TEE} like AMD \ac{SEV}, Arm \ac{CCA}, and Intel \ac{TDX}, which offer isolation at the \ac{VM} level.

\subsubsection{Intel \ac{SGX} Overview}
\ac{SGX} offers a secure and isolated environment, termed \emph{enclave}, for sensitive workloads without the need to rely on the security of the untrusted host \ac{OS}. \ac{SGX} ensures the authenticity of executed code, maintains the integrity of runtime states (\eg \ac{CPU} registers, memory, and sensitive I/O), and preserves the confidentiality of enclave code, data, and runtime states stored in persistent memory through a sealing process~\cite{costan2016intel}.

\noindent\textbf{Threat model of Intel \ac{SGX}.} 
An attacker is assumed to control the entire software stack of the underlying computing platform, including the hypervisor, BIOS, and \ac{OS}.
Specifically, within this context, the attacker can (1) execute arbitrary instructions at privileged levels; (2) launch, pause, and destroy enclave instances as desired; (3) control the software stack within the non-enclave environment, including tasks such as memory mapping, I/O device, and \ac{CPU} scheduling~\cite{li2024sok}.

\noindent\textbf{\ac{TCB} of Intel \ac{SGX}.}
The \ac{TCB} of \ac{SGX} refers to all of a system's hardware, firmware, and software components considered trusted and involved in providing a secure environment.
A vulnerability in any \ac{TCB} component would jeopardize the security of \ac{SGX}. 
\ac{TCB} size serves as a common metric to assess the security of a \ac{TEE} design, where a smaller \ac{TCB} size typically indicates fewer lines of code or hardware elements, implying a reduced vulnerability risk.
However, comparing the security of two \ac{TEE} designs based solely on \ac{TCB} size is difficult and unreasonable~\cite{li2024sok}.
Ongoing research efforts are focused on building smaller \ac{TEE} \ac{OS} with memory-safe programming languages (\eg Rust-SGX~\cite{wang2019towards}).

\subsubsection{Arm \ac{TZ} Overview}
\ac{TZ} has been integrated into Arm processors (Cortex-A) since 2004~\cite{armtrustzone}, primarily employed in mobile devices.
Unlike Intel \ac{SGX}, \ac{TZ} is not inherently tailored for cloud platforms and lacks support for remote attestation, which would enable remote parties to verify the initial states of the \ac{TEE} instances~\cite{li2024sok}.
\ac{TZ} operates around the notion of protection domains referred to as the \ac{SW} and the \ac{NW}. 
It also extends the concept of "privilege rings" by introducing \acp{EL} in the ARMv8 \ac{ISA}. 
The "Secure Monitor" operates in the highest privilege level EL3, overseeing both the \ac{SW} and \ac{NW}. 
The untrusted OS runs within the \ac{NW}, referred to as the \ac{REE}.
Within the \ac{SW}, the trusted OS kernel operates in EL1, while the secure users-pace operates in EL0. 
The separation between \ac{SW} and \ac{NW} ensures that specific RAM regions and bus peripherals are designated as secure, limiting access solely to the \ac{SW}. 

\noindent\textbf{Threat model of Arm \ac{TZ}.} 
Currently, there is a lack of a formal threat model for \ac{TZ}. 
Commonly assumed attacker capabilities are as follows:
The attacker (1) has full access to the \ac{REE} of the device, encompassing the OS. This could be an authenticated (root) user, malicious third-party software installed on the device, or a compromised OS; 
(2) exploits unauthorized access to resources within the SW and extract sensitive information~\cite{pinto2019demystifying}. 

\noindent\textbf{\ac{TCB} of Arm \ac{TZ}.}
Within the \ac{SW}, the trusted OS operates in EL1 and provides runtime support to sustain the lifecycle of \acp{TA}, ensuring their secure execution in user mode (EL0).
The core of the trusted OS is the trusted kernel, which furnishes key OS primitives for the scheduling and management of \acp{TA}. 
The secure monitor implements mechanisms for secure context switching between worlds and operates with the highest privilege in EL3. Meanwhile, the \ac{TEE} bootloader initiates the \ac{TEE} system into a secure state~\cite{cerdeira2020sok}. 
In a nutshell, the trusted OS, secure monitor, and \ac{TEE} bootloader constitute the \ac{TCB} of a \ac{TZ}-assisted \ac{TEE}.

\subsubsection{Virtualization-based \acp{TEE}}
\label{sec:vm tee}
Process-based \acp{TEE} such as Intel \ac{SGX} require partitioning an application into a trusted part, the enclave, and an untrusted part, which can complicate development efforts as developers must adhere to this security model and segment their applications accordingly.
In response, vendors now iterated on their previous trusted computing \ac{CPU} extensions to facilitate virtualization-based confidential computing across various server platforms, eliminating the need for extensive modifications to application code.
Here, we outline three such virtualization-based confidential computing solutions: Intel \ac{TDX}~\cite{inteltdx}, AMD \ac{SEV}~\cite{amdsev} and Arm \ac{CCA}~\cite{armcca}.

\noindent\textbf{AMD \ac{SEV}.} 
AMD introduced \ac{SEV} with the Zen architecture to enable confidential computing in cloud environments. 
Recent AMD \ac{SEV}-\ac{SNP}~\cite{sev2020strengthening} extends the capabilities of AMD \ac{SEV} by introducing features such as secure nested paging to enhance the integrity of encrypted memory.

\noindent\textbf{Arm \ac{CCA}.} 
Arm \ac{CCA} is an advanced isolation technology introduced in Armv9.
It preserves \ac{TZ}'s \ac{NW} and \ac{SW} while introducing a new \emph{realm} world designed for running multiple confidential ``realms''.
These realms, though isolated from other worlds, are managed by untrusted software components (\eg hypervisor). A lightweight \ac{RMM}~\cite{armRME}, operating within the hypervisor layer of the realm world, ensures memory isolation between realms. \ac{CCA} incorporates a \emph{root} world to accommodate the highest privilege monitor code stored in firmware~\cite{wang2024cage}.

\noindent\textbf{Intel \ac{TDX}.} 
Intel \ac{TDX} leverages Intel's \ac{MKTME}~\cite{intelMKTME} and a \ac{CPU}-attested software module to facilitate the isolated execution of secure \acp{VM}. The \ac{TDX} module, digitally signed and executed in a novel processor mode called SEAM, functions as a separation kernel. It provides an interface to the hypervisor for scheduling, creating, and managing secure \acp{VM}, which are called \emph{trusted domains}.
For an in-depth comparison of AMD \ac{SEV}, Arm \ac{CCA}, and Intel \ac{TDX}, we refer to the survey in~\cite{guanciale2022sok}.

\section{Overview and Taxonomy}
\label{sec:overview}

\tikzstyle{leaf}=[mybox,minimum height=1em,
fill=hidden-orange!40, text width=20em,  text=black,align=left,font=\footnotesize,
inner xsep=2pt,
inner ysep=1pt,
]

\begin{figure}[tp]
    \centering
\resizebox{\linewidth}{!}{
    \begin{forest}
          forked edges,
          for tree={
              grow=east,
              reversed=true,
              anchor=base west,
              parent anchor=east,
              child anchor=west,
              base=left,
              font=\small,
              rectangle,
              draw=hiddendraw,
              rounded corners,align=left,
              minimum width=2.5em,
            s sep=3pt,
            inner xsep=2pt,
            inner ysep=1pt,
            ver/.style={rotate=90, child anchor=north, parent anchor=south, anchor=center},
        },
      where level=1{text width=6.2em,font=\footnotesize,}{},
      where level=2{text width=5.6em,font=\footnotesize}{},
      where level=3{text width=9.0em,font=\footnotesize}{},
      [Confidential Computing on Heterogeneous CPU-GPU Systems, ver
        [GPU \acp{TEE}\\ (\Cref{sec:cc-gpu})
            [x86-based\\ \ac{GPU} \acp{TEE}
               [Graviton~\cite{volos2018graviton}{,}  HIX~\cite{jang2019heterogeneous}{,}  HETEE~\cite{zhu2020enabling}{,}  NVIDIA H100~\cite{NVIDIAh100}{,}  TNPU~\cite{lee2022tnpu}{,}  LITE~\cite{yudha2022lite}{,}  SAGE~\cite{ivanov2023sage}{,}  Honeycomb~\cite{mai2023honeycomb},leaf,text width=38em]
            ]
            [Arm-based\\ \ac{GPU} \acp{TEE}
               [StrongBox~\cite{deng2022strongbox}{,}  CRONUS~\cite{jiang2022cronus}{,}  CAGE~\cite{wang2024cage},leaf,text width=15em]
            ]
        ]
        [Security of \acp{GPU}\\ (\Cref{sec:GPU-sec})
            [Architectural\\ Attacks
                [\citeauthor{lee2014stealing}~\cite{lee2014stealing}{,}  \citeauthor{maurice2014confidentiality}~\cite{maurice2014confidentiality}{,} \citeauthor{pietro2016cuda}~\cite{pietro2016cuda}{,} \citeauthor{zhou2016vulnerable}~\cite{zhou2016vulnerable}{,} \citeauthor{wenjian2020igpu}~\cite{wenjian2020igpu},leaf,text width=30em]
            ]
            [Microarchitectural\\ Side/Covert-\\Channel Attacks
                [Attacks on Integrated \ac{GPU}
                    [\citeauthor{frigo2018grand}~\cite{frigo2018grand}{,} \citeauthor{wenjian2020igpu}~\cite{wenjian2020igpu}{,} \citeauthor{dutta2021leaky}~\cite{dutta2021leaky}{,} \citeauthor{taneja2023hot}~\cite{taneja2023hot}{,} \citeauthor{wang2024gpu}~\cite{wang2024gpu},leaf,text width=31em]
                ]
                [Attacks on Discrete \ac{GPU}
                    [\citeauthor{jiang2016complete}~\cite{jiang2016complete}{,}  \citeauthor{jiang2017novel}~\cite{jiang2017novel}{,} \citeauthor{ahn2021trident}~\cite{ahn2021trident}{,} \citeauthor{giner2024generic}~\cite{giner2024generic}{,} \citeauthor{luo2019side}~\cite{luo2019side}{,} \citeauthor{zhu2021hermes}~\cite{zhu2021hermes}{,}\\ \citeauthor{hu2020deepsniffer}~\cite{hu2020deepsniffer}{,} \citeauthor{naghibijouybari2017constructing}~\cite{naghibijouybari2017constructing}{,} \citeauthor{wei2020leaky}~\cite{wei2020leaky}{,} \citeauthor{ahn2021network}~\cite{ahn2021network}{,} \citeauthor{nayak2021mis}~\cite{nayak2021mis}{,}\\ \citeauthor{cronin2021exploration}~\cite{cronin2021exploration}{,} \citeauthor{yang2022eavesdropping}~\cite{yang2022eavesdropping}{,} \citeauthor{dutta2023spy}~\cite{dutta2023spy}{,} \citeauthor{zhang2023t}~\cite{zhang2023t},leaf,text width=32em]
                ]
                [Countermeasures
                    [\citeauthor{wang2007new}~\cite{wang2007new}{,} \citeauthor{kong2009hardware}~\cite{kong2009hardware}{,} \citeauthor{liu2016catalyst}~\cite{liu2016catalyst}{,} \citeauthor{zhang2023t}~\cite{zhang2023t}{,} \citeauthor{wang2014timing}~\cite{wang2014timing}{,}\\ \citeauthor{shafiee2015avoiding}~\cite{shafiee2015avoiding}{,} \citeauthor{di2021tlb}~\cite{di2021tlb}{,} \citeauthor{kadam2018rcoal}~\cite{kadam2018rcoal}{,} \citeauthor{kadam2020bcoal}~\cite{kadam2020bcoal}{,} \citeauthor{lin2019scatter}~\cite{lin2019scatter},leaf,text width=30em]
                ]
            ]
            [Physical Side-\\Channel Attacks
                [Physical Access-based
                    [\citeauthor{luo2015side}~\cite{luo2015side}{,} \citeauthor{gao2018electro}~\cite{gao2018electro}{,} \citeauthor{hu2020deepsniffer}~\cite{hu2020deepsniffer}{,} \citeauthor{chmielewski2021reverse}~\cite{chmielewski2021reverse}{,} \citeauthor{maia2022can}~\cite{maia2022can}{,}\\ \citeauthor{horvath2023barracuda}~\cite{horvath2023barracuda}{,} \citeauthor{taneja2023hot}~\cite{taneja2023hot},leaf,text width=32em]
                ]
                [Remote Access-based
                    [\citeauthor{zhan2022graphics}~\cite{zhan2022graphics}{,} \citeauthor{liang2022clairvoyance}~\cite{liang2022clairvoyance},leaf,text width=12em]
                ]
                [Countermeasures
                    [\citeauthor{mangard2008power}~\cite{mangard2008power}{,} \citeauthor{zhan2022graphics}~\cite{zhan2022graphics},leaf,text width=13em]
                ]
            ]
            [Fault Injection\\ Attacks
                [Software-based
                    [\citeauthor{frigo2018grand}~\cite{frigo2018grand}{,} \citeauthor{sabbagh2020novel}~\cite{sabbagh2020novel}{,} \citeauthor{sabbagh2021gpu}~\cite{sabbagh2021gpu}{,} \citeauthor{sun2023lightning}~\cite{sun2023lightning}{,} \citeauthor{dos2021revealing}~\cite{dos2021revealing},leaf,text width=32em]
                ]
            ]
        ]
        [Security of \ac{CPU}\\ \acp{TEE} (supplemental\\  material) 
            [Security of Intel\\ \ac{SGX}
                [Controlled Channel/Address\\ Translation-based Attacks
                    [\citeauthor{xu2015controlled}~\cite{xu2015controlled}{,}
                    \citeauthor{shinde2016preventing}~\cite{shinde2016preventing}{,}
                    \citeauthor{van2017sgx}~\cite{van2017sgx}{,}
                    \citeauthor{van2017telling}~\cite{van2017telling}{,} \citeauthor{wang2017leaky}~\cite{wang2017leaky}{,}\\ \citeauthor{gyselinck2018off}~\cite{gyselinck2018off}{,}
                    \citeauthor{kim2019sgx}~\cite{kim2019sgx}{,}
                    \citeauthor{fei2021security}~\cite{fei2021security},leaf,text width=33em] 
                ]
                [Cache-based Attacks        
                    [\citeauthor{gotzfried2017cache}~\cite{gotzfried2017cache}{,} \citeauthor{moghimi2017cachezoom}~\cite{moghimi2017cachezoom}{,} \citeauthor{schwarz2017malware}~\cite{schwarz2017malware}{,} \citeauthor{brasser2017software}~\cite{brasser2017software}{,} \citeauthor{dall2018cachequote}~\cite{dall2018cachequote}{,}\\ \citeauthor{moghimi2019memjam}~\cite{moghimi2019memjam},leaf,text width=32em]
                ]
                [Branch Prediction Attacks
                    [\citeauthor{fei2021security}~\cite{fei2021security}{,} \citeauthor{lee2017inferring}~\cite{lee2017inferring}{,} \citeauthor{chen2019sgxpectre}~\cite{chen2019sgxpectre}{,} \citeauthor{evtyushkin2018branchscope}~\cite{evtyushkin2018branchscope}{,} \citeauthor{huo2020bluethunder}~\cite{huo2020bluethunder}{,} \citeauthor{koruyeh2018spectre}~\cite{koruyeh2018spectre},leaf,text width=34em]
                ]
                [Transient Execution Attacks
                    [\citeauthor{chen2019sgxpectre}~\cite{chen2019sgxpectre}{,} \citeauthor{van2018foreshadow}~\cite{van2018foreshadow}{,} \citeauthor{van2020sgaxe}~\cite{van2020sgaxe}{,} \citeauthor{ragab2021crosstalk}~\cite{ragab2021crosstalk},leaf,text width=27em]
                ]
                [Software-based Fault\\ Injection Attacks
                    [\citeauthor{murdock2020plundervolt}~\cite{murdock2020plundervolt}{,} \citeauthor{kenjar2020v0ltpwn}~\cite{kenjar2020v0ltpwn}{,} \citeauthor{qiu2020voltjockey}~\cite{qiu2020voltjockey}{,} \citeauthor{jang2017sgx}~\cite{jang2017sgx},leaf,text width=24em]
                ] 
                [Hardware-based Attacks
                    [\citeauthor{chen2021voltpillager}~\cite{chen2021voltpillager}{,} \citeauthor{Chen23pmfault}~\cite{Chen23pmfault}{,} \citeauthor{lee20membuster}~\cite{lee20membuster},leaf,text width=18em]
                ]
                [Interface-based Attacks
                    [\citeauthor{van2019tale}~\cite{van2019tale}{,} \citeauthor{alder2020faulty}~\cite{alder2020faulty}{,} \citeauthor{cloosters2020teerex}~\cite{cloosters2020teerex}{,} \citeauthor{khandaker2020coin}~\cite{khandaker2020coin}{,} \citeauthor{alder2024pandora}~\cite{alder2024pandora},leaf,text width=32em]
                ]
            ]
            [Security of\\ Arm \ac{TZ}
                [Implementation Bugs
                    [\citeauthor{beniamini2016trustzone}~\cite{beniamini2016trustzone}{,} \citeauthor{shen2015exploiting}~\cite{shen2015exploiting}{,} \citeauthor{Danielunbox}~\cite{Danielunbox},leaf,text width=15em] 
                ]
                [Cache-based Attacks        
                    [\citeauthor{zhang2016cachekit}~\cite{zhang2016cachekit}{,} \citeauthor{guanciale2016cache}~\cite{guanciale2016cache}{,} \citeauthor{lipp2016armageddon}~\cite{lipp2016armageddon}{,} \citeauthor{zhang2016truspy}~\cite{zhang2016truspy}{,} \citeauthor{cho2018prime+}~\cite{cho2018prime+},leaf,text width=30em]
                ]
                [Branch Prediction Attacks
                    [\citeauthor{ryan2019hardware}~\cite{ryan2019hardware},leaf,text width=4em]
                ]
                [Software-based Fault\\ Injection Attacks
                [\citeauthor{tang2017clkscrew}~\cite{tang2017clkscrew}{,} \citeauthor{qiu2020voltjockey}~\cite{qiu2020voltjockey}{,} \citeauthor{Pierrerowhammer}~\cite{Pierrerowhammer},leaf,text width=16em]
                ] 
            ]
            [Security of\\ AMD \ac{SEV}
                [Memory Encryption Issues
                    [\citeauthor{hetzelt2017security}~\cite{hetzelt2017security}{,} \citeauthor{buhren2017fault}~\cite{buhren2017fault}{,} \citeauthor{du2017secure}~\cite{du2017secure}{,} \citeauthor{morbitzer2018severed}~\cite{morbitzer2018severed}{,} \citeauthor{li2021cipherleaks}~\cite{li2021cipherleaks}{,}\\ \citeauthor{li2022understanding}~\cite{li2022understanding}{,} \citeauthor{li2021crossline}~\cite{li2021crossline}{,} \citeauthor{li2021tlb}~\cite{li2021tlb},leaf,text width=32em] 
                ]
                [Software-based Fault\\ Injection Attacks
                    [\citeauthor{morbitzer2021severity}~\cite{morbitzer2021severity}{,} \citeauthor{buhren2021one}~\cite{buhren2021one}{,} \citeauthor{zhang2024cachewarp}~\cite{zhang2024cachewarp}{,} \citeauthor{schluter2024wesee}~\cite{schluter2024wesee}{,} \citeauthor{schluter2024heckler}~\cite{schluter2024heckler},leaf,text width=33em]
                ]
                [Interrupt-driven Attacks
                    [\citeauthor{wilke2024sev}~\cite{wilke2024sev},leaf,text width=6em]
                ]
            ]
        ]
        [Potential Attacks\\ against \ac{GPU} \acp{TEE}\\ (\Cref{sec:GPU tee attacks})
            [Attacks against\\ Intel \ac{SGX}
                [x86-based \ac{GPU} \ac{TEE},leaf,text width=7em]
            ]
            [Attacks against\\ \ac{VM}-based \acp{TEE}
                [NVIDIA Hopper H100-based \ac{GPU} \ac{TEE},leaf,text width=13em]
            ]
            [Attacks against\\ Arm \ac{TZ}
                [Arm-based \ac{GPU} \ac{TEE},leaf,text width=7em]
            ]
            [Architectural\\ Attacks
                [x86-based \ac{GPU} \ac{TEE},leaf,text width=7em]
            ]
            [Microarchitectural\\ Side/Covert-\\Channel Attacks
                [x86-based \ac{GPU} \ac{TEE},leaf,text width=7em]
                [NVIDIA Hopper H100-based \ac{GPU} \ac{TEE},leaf,text width=13em]
                [Arm-based \ac{GPU} \ac{TEE},leaf,text width=7em]
            ]
            [Physical Side-\\Channel Attacks
                [x86-based \ac{GPU} \ac{TEE},leaf,text width=7em]
                [Arm-based \ac{GPU} \ac{TEE},leaf,text width=7em]
            ]
            [Fault Injection\\ Attacks
                [x86-based \ac{GPU} \ac{TEE},leaf,text width=7em]
                [NVIDIA Hopper H100-based \ac{GPU} \ac{TEE},leaf,text width=13em]
                [Arm-based \ac{GPU} \ac{TEE},leaf,text width=7em]
            ]
        ]
      ]
    \end{forest}
}
    \caption{Overview of confidential computing on heterogeneous CPU-GPU systems}
    \label{fig:taxonomy}
\end{figure}
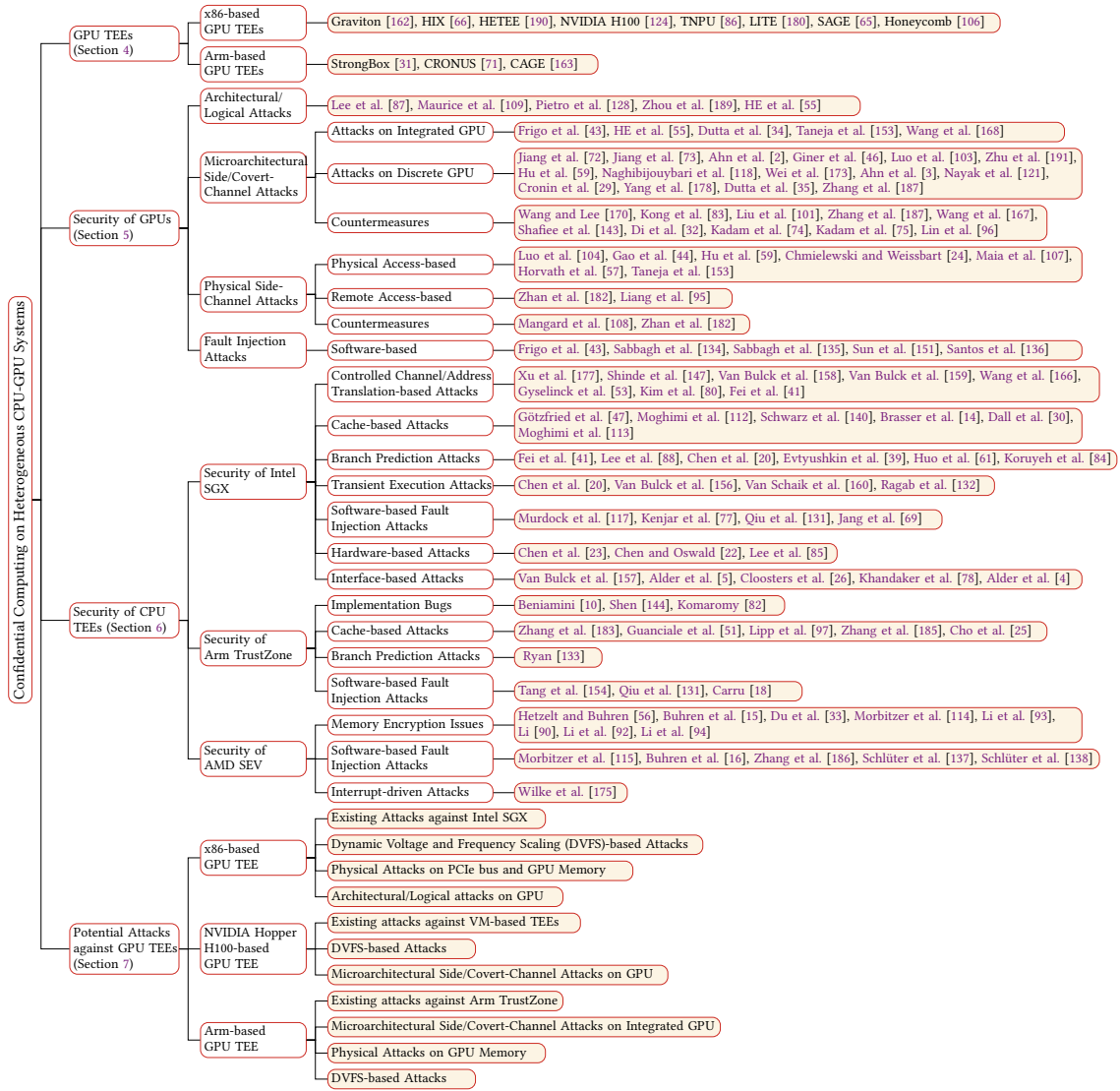

Confidential computing on \acp{GPU} is an emerging field that ensures sensitive data processed on \acp{GPU} remains protected from unauthorized access and tampering even for attackers with privileged access. 
Other privacy-preserving techniques, such as secure multi-party computation, homomorphic encryption, and \acp{TEE}, offer different security guarantees. 
However, in this work, our primary focus is on the extension of \acp{TEE} to \acp{GPU} (\ie \ac{GPU} \acp{TEE}).
Moving from homogeneous to such heterogeneous systems inevitably creates new threats and security vulnerabilities. 
It is crucial to understand these new attack vectors to inform the design of systems that can effectively mitigate them.
In comparison to existing \ac{CPU} \acp{TEE}, \ac{GPU} \acp{TEE} present unique opportunities for attackers from several perspectives:

    \emph{\textbf{\ac{GPU} software stack on \ac{CPU}:}} \ac{GPU} drivers handle critical aspects such as resource allocation, task scheduling, and dynamic power management. Specifically, in resource allocation, \ac{GPU} drivers manage and allocate resources, including memory, processing units, and other hardware components. For task scheduling, they efficiently schedule and manage tasks to ensure optimal performance, balancing workloads across \ac{GPU} cores. In dynamic power management, drivers adjust parameters such as voltage based on workload, optimizing performance and efficiency to extend battery life in mobile devices and reduce energy costs in data centers.
    In current \ac{GPU} \acp{TEE}, the protection of \ac{GPU} drivers, particularly their positioning in the \ac{CPU} host, may introduce new attack surface. Furthermore, some state-of-the-art attacks against \ac{CPU} \acp{TEE} may also affect \ac{GPU} drivers.
    
    \emph{\textbf{Hardware components on \ac{GPU}:}} \ac{GPU} \acp{TEE} are still in the early stages of development. Most designs are experimental research and often lack extensive hardware modifications, resulting in insufficient hardware protection for components such as \ac{GPU} memory, the \ac{PCIe} bus, and power management components. Even products designed by \ac{GPU} manufacturers (\eg NVIDIA H100 \ac{GPU}) might lack protection for certain hardware components, such as power management components. Thus, existing attacks against these components may also translate to weaknesses in \ac{GPU} \acp{TEE}.
    
    \emph{\textbf{Architectural complexity:}} Although the security of \acp{GPU} is just starting to be explored, several vulnerabilities and attacks have already been documented. Due to the architectural complexity of \acp{GPU}, such as memory sharing among different concurrent kernels and the variety of \ac{GPU} types (\eg integrated and discrete), some existing \ac{GPU} \acp{TEE} may not provide adequate protection. Additionally, many designs overlook cache-level access protection, making them susceptible to cache-based attacks on \acp{GPU}.

In \Cref{fig:taxonomy}, we summarise our taxonomy underlying this survey based on relevant related work.
In the remainder of this paper, we first review and discuss existing \ac{GPU} \acp{TEE} in \Cref{sec:cc-gpu}, covering their threat models, hardware modifications, \ac{TCB} size, and performance overhead. 
In \Cref{sec:GPU-sec}, we give an overview of existing attacks targeting \acp{GPU} based on the attack types listed in \Cref{sec:bg}. 
We also review state-of-the-art attacks against existing \ac{CPU} \acp{TEE}, such as Intel \ac{SGX}, Arm \ac{TZ}, and AMD \ac{SEV} in the supplemental material due to the limited space of this article. 
Finally, based on these \ac{GPU} \ac{TEE} designs and the summarized attacks, we discuss potential attacks against existing (and future) \ac{GPU} \acp{TEE} in \Cref{sec:GPU tee attacks}.

\section{Confidential Computing on \acp{GPU}}
\label{sec:cc-gpu}

We summarize the state-of-the-art \ac{GPU} \ac{TEE} designs based on Intel, AMD, and ARM in \Cref{tab:gpu_tee}.
Next, we first review all existing schemes and then discuss them based on aspects such as threat models, hardware modifications, \ac{TCB} size and performance overhead resulting from implementing trusted execution on accelerators.

\subsection{Trusted Execution Environments on \acp{GPU}}

Traditional CPUs and cloud systems have slowly adopted hardware-based \acp{TEE} to securely isolate computations from a malicious \ac{OS} or certain hardware attacks, such as snooping on the memory bus. 
However, \acp{GPU} and their cloud deployments have largely yet to incorporate such support for hardware-based trusted computing. 
With large amounts of sensitive data being offloaded to \acp{GPU} for acceleration in cloud environments, especially for sensitive \ac{ML} workloads, there is an urgent need to address the trusted computing requirements for ubiquitous \acp{GPU}. 
Researchers and commercial vendors have recently proposed various designs to support confidential computing on \acp{GPU}.

\begin{table}[!ht]
\footnotesize
\centering
\caption{Summary of proposed heterogeneous \ac{TEE} for accelerators}
\label{tab:gpu_tee}

\resizebox{\linewidth}{!}{
\begin{tabular}{llllllcccccc}
\toprule

    \multirow{2}{*}{\textbf{Scheme}} & \multirow{2}{*}{\textbf{ISA}} & \multirow{2}{*}{\textbf{Accelerator}} & \multirow{2}{*}{\textbf{Memory}} & \multirow{2}{*}{\textbf{GPU Driver Location}} & \multirow{2}{*}{\textbf{Methods}} & \multirow{2}{*}{\textbf{\makecell[l]{HW\\ Changes}}} & \multicolumn{5}{l}{\textbf{Key Components of \ac{GPU} \ac{TEE}}} \\ 
    \cline{8-12} 
    &  &  &  &  &  &  & \textbf{Attestation} & \textbf{Secure RM} & \textbf{Secure I/O} & \textbf{Integrity} & \textbf{GPU ME} \\
\midrule
     Graviton~\cite{volos2018graviton} & Intel x86 & NVIDIA \ac{GPU} & Dedicated & Untrusted \ac{OS} & \makecell[l]{Modify \ac{GPU} command\\ processor} & \CIRCLE & $\checkmark$ & $\checkmark$ & $\checkmark$ & $\checkmark$ & \ding{53} \\
     HIX~\cite{jang2019heterogeneous} & Intel x86 & NVIDIA \ac{GPU} & Dedicated & Part inside \ac{CPU} enclave & \ac{SGX} Enclave & \CIRCLE & $\checkmark$ & $\checkmark$ & $\checkmark$ & $\checkmark$ & \ding{53} \\
     HETEE~\cite{zhu2020enabling} & Intel x86 & NVIDIA \ac{GPU} & Dedicated & Inside \ac{CPU} enclave & \makecell[l]{Dynamic allocation of\\ accelerators within \ac{CPU}\\ enclave and extra FPGA} & \CIRCLE & $\checkmark$ & $\checkmark$ & $\checkmark$ & $\checkmark$ & \ding{53} \\
     NVIDIA H100~\cite{NVIDIAh100} & Intel x86 & NVIDIA \ac{GPU} & Dedicated & Inside confidential \ac{VM} & Hardware-level protection & \CIRCLE & $\checkmark$ & $\checkmark$ & $\checkmark$ & $\checkmark$ & $\checkmark$ \\
     %
     TNPU~\cite{lee2022tnpu} & Intel x86 & NPU & Unified & Inside \ac{CPU} enclave & \ac{SGX} Enclave & \CIRCLE & ${\checkmark}$ & ${\checkmark}$ & $\checkmark$ & $\checkmark$ & $\checkmark$ \\
     LITE~\cite{yudha2022lite} & Intel/AMD x86 & NVIDIA \ac{GPU} & Dedicated & N/A & \makecell[l]{Software-based unified \\ encryption domain} & \CIRCLE & \ding{53}$^{\dag}$ & \ding{53}$^{\dag}$ & $\checkmark$ & $\checkmark$ & $\checkmark$ \\
     StrongBox~\cite{deng2022strongbox} & Arm & Arm Mali \ac{GPU} & Unified & Inside normal world & \ac{TZ} & \Circle & ${\checkmark}$ & ${\checkmark}$ & $\checkmark$ & $\checkmark$ & \ding{53} \\
     CRONUS~\cite{jiang2022cronus} & Arm & General accelerators & Dedicated & Inside secure world & \ac{TZ} virtualization & \CIRCLE & ${\checkmark}$ & ${\checkmark}$ & $\checkmark$ & $\checkmark$ & \ding{53}$^{\dag}$ \\
     SAGE~\cite{ivanov2023sage} & Intel x86 & NVIDIA \ac{GPU} & Dedicated & N/A & \makecell[l]{Software-based\\ verification function} & \Circle & \ding{53}$^{\dag}$ & \ding{53}$^{\dag}$ & $\checkmark$ & $\checkmark$ & \ding{53} \\
     Honeycomb~\cite{mai2023honeycomb} & AMD x86 & AMD \ac{GPU} & Dedicated & Inside confidential \ac{VM} & AMD \ac{SEV}-SNP & \Circle & ${\checkmark}$ & ${\checkmark}$ & $\checkmark$ & $\checkmark$ & \ding{53} \\
     CAGE~\cite{wang2024cage} & Arm & Arm Mali \ac{GPU} & Unified & Inside normal world & RME, GPC & \Circle & \ding{53}$^{\dag}$ & ${\checkmark}$ & $\checkmark$ & $\checkmark$ & \ding{53} \\
\bottomrule
\end{tabular}
}
\begin{flushleft}
        \textbf{Memory} indicates whether the accelerators share the memory with the \ac{CPU} or not. \textbf{HW Modification} explains if need modification on hardware (\CIRCLE~= yes, \Circle~= no). \textbf{Secure RM} and \textbf{GPU ME} are \emph{Secure Resource Management} and \emph{GPU Memory Encryption}.
        ${\dag}$ indicates that scheme needs additional support (\eg hardware security features provided in NVIDIA H100 \ac{GPU} or next-generation Arm devices), as they are not currently considering them.
\end{flushleft}
\end{table}

\noindent\textbf{x86-based \ac{GPU} \acp{TEE}.} 
Graviton~\cite{volos2018graviton} initiated the work of trusted execution on hardware accelerators like \acp{GPU}.
Graviton directly establishes a \ac{TEE} within the \ac{GPU}, necessitating hardware modifications but avoiding the need for \ac{CPU}-side \ac{TEE} like \ac{SGX}. 
Graviton establishes a secure channel between the untrusted host and \ac{GPU}, monitoring command submission and data transfer by directing all resource allocation requests through the \ac{GPU}'s command processor instead of \ac{GPU} driver.
Nevertheless, for data protection, Graviton assumes the \ac{GPU}'s on-package memory can be trusted.
The assumption is that even with physical access, it is extremely difficult for an attacker to breach the integrity of the \ac{GPU} package and snoop on the silicon interconnect between \ac{GPU} and stacked memory.

HIX~\cite{jang2019heterogeneous} removes the key functionalities for controlling the \ac{GPU} from the OS-resident driver and places those within a trusted \emph{GPU enclave}, extending an \ac{SGX} enclave and incorporating new \ac{SGX}-style instructions like $\mathtt{ECREATE}$ and $\mathtt{EADD}$.
Furthermore, HIX establishes a secure hardware I/O path, preventing the OS from altering the virtual-to-physical address mapping for the \ac{GPU} \ac{MMIO} region and from deducing commands and data transmitted via \ac{MMIO} and \ac{DMA}. Like Graviton, HIX lacks a trusted \ac{GPU} memory region, leaving data in \ac{GPU} memory accessible in plaintext.

To eliminate the need for modifications to \ac{CPU} or \ac{GPU} chips, HETEE~\cite{zhu2020enabling} introduces a data-center level \ac{GPU} \ac{TEE} designed to support extensive confidential computing, without any chip-level change. 
It facilitates dynamic allocation of computing resources for both secure and non-sensitive computing tasks across all servers within a rack. 
They incorporate a small \ac{TCB}, such as a single security controller with customized code supporting task isolation, reducing design complexity and minimizing the attack surface associated with resource sharing with the untrusted OS.

Similar to HIX, Honeycomb~\cite{mai2023honeycomb} ensures secure communication between the user and the \ac{GPU} by using a \ac{CPU}-side \ac{TEE} (\eg AMD \ac{SEV}-SNP) to transfer sensitive data to the \ac{GPU} and manage access to \ac{GPU} \ac{MMIO}.
Unlike previous designs, Honeycomb employs static analysis to confirm that mutually distrusting \ac{GPU} applications are contained within their enclaves. This approach enables more efficient implementations by moving runtime checks to load time. 
Moreover, Honeycomb reduces the \ac{TCB} by excluding both the user-space and kernel-space \ac{GPU} drivers from it, instead using two security monitors to intercept and control all traffic between applications and the \ac{GPU}.
Honeycomb makes the same assumption as Graviton in trusting the \ac{GPU}'s device memory based on the fact that modern \acp{GPU} typically integrate device memory using 2.5D/3D silicon interposers within the same package.

SAGE~\cite{ivanov2023sage} offers a software-based alternative for trusted \ac{GPU} execution.
It employs an \ac{SGX} enclave running on the host as a local verifier and uses this enclave to bootstrap a software primitive that establishes a dynamic root of trust on the \ac{GPU}. It ensures that the user kernel on the untrusted device remains unmodified, is invoked for execution on the untrusted \ac{GPU}, and runs without tampering.
It complements existing hardware-based \ac{GPU} \acp{TEE}.

Given that previous designs ignore memory security (\ie they do not encrypt \ac{GPU} memory), 
LITE~\cite{yudha2022lite} introduces a co-designed \ac{GPU} \ac{TEE} with \ac{CPU} to establish a unified encryption domain, where data is stored in ciphertext form in \ac{GPU} caches and memory.
Specifically, LITE ensures that data enters the \ac{GPU} chip encrypted. The software then decrypts the data using the \ac{ALU}, operating on the data in registers. Additionally, data may be temporarily stored in plaintext in the on-chip shared memory.

TNPU~\cite{lee2022tnpu} proposes tree-less off-chip memory protection for \acp{DNN} accelerators.
The core idea is to group contiguous memory blocks into tiles and provide freshness guarantees at the tile level. 
TNPU employs a dedicated software module running on the host \ac{CPU} to manage version numbers, which are stored in a ``Tensor Table'' within the host \ac{CPU}'s secure memory. The tensor table is protected by an integrity tree.

NVIDIA's H100 Tensor Core \ac{GPU}~\cite{NVIDIAh100}, based on the Hopper architecture, introduces advanced features for confidential computing. 
To enable confidential computing with the H100, it requires specific \ac{CPU} \acp{TEE}, such as Intel \ac{TDX}, AMD \ac{SEV}-\ac{SNP}, and Arm \ac{CCA}. 
In addition to ensuring the confidentiality and integrity of data and code, H100 can defend against basic physical attacks targeting the \ac{PCIe} bus and \ac{DDR}. 
H100 supports three different operation modes, namely \emph{CC-Off}, \emph{CC-On}, and \emph{CC-DevTools}.
In CC-On mode, the H100, along with the drivers on the \ac{CPU}, fully activates all available confidential computing features. 
Some hardware resources, such as performance counters, are disabled in CC-On mode to prevent potential side-channel attacks, as performance counters could be used to infer the behavior of device usage~\cite{naghibijouybari2018rendered}.
However, details about specific hardware changes, such as memory encryption in the H100, remain unclear at present.

\noindent\textbf{Arm-based \ac{GPU} \acp{TEE}.} 
Previous designs for Intel/AMD-based platforms cannot be readily extended to Arm \acp{GPU} due to numerous architectural differences. 
StrongBox~\cite{deng2022strongbox} thus introduces the first \ac{GPU} \ac{TEE} for secure general computation on Arm endpoints without necessitating hardware modifications or architectural changes. 
It deploys the StrongBox runtime within the secure monitor to ensure that the \ac{GPU} executes in isolation.
Through a combination of Stage-2 translation and the \ac{TZASC}, StrongBox specifies six distinct access permission types to effectively manage secure access from \ac{DMA}, \ac{GPU}, and other peripherals.

CRONUS~\cite{jiang2022cronus} introduces a design that supports various heterogeneous accelerators, enables spatial sharing within a single accelerator, and ensures robust isolation across multiple accelerators. 
Leveraging Arm secure virtualization technology, CRONUS partitions heterogeneous computation into isolated \ac{TEE} enclaves. 
Each enclave encapsulates a specific type of computation like \ac{GPU} computation and allows multiple enclaves to spatially share a single accelerator.

To support confidential \ac{GPU} computing on next-generation Arm devices, CAGE~\cite{wang2024cage} presents a new scheme for Arm \ac{CCA}'s realm-style architecture.
It offloads resource-intensive but data-independent tasks to \ac{NW}, while protecting sensitive data within realms through a shadow task mechanism. By configuring \ac{GPC}, CAGE restricts access from untrusted components. Additionally, it isolates realms in \ac{GPU} computing by providing each realm with a distinct \ac{GPU} memory view in the \ac{GPU} \ac{GPC}. 
CAGE retains high hardware compatibility, as it requires no additional hardware or modifications to existing \ac{CPU} and \ac{GPU} configurations.

\subsection{Discussion}
In this section, we discuss several key metrics in designing an effective and secure \ac{GPU} \acp{TEE}.

\noindent\textbf{Comparison to \ac{CPU} \acp{TEE}.}
While both \ac{CPU} and \ac{GPU} \acp{TEE} aim to provide secure and isolated execution environments, their designs differ significantly due to fundamental differences in microarchitecture and usage patterns. 
\ac{CPU} \acp{TEE}, such as Intel SGX and AMD SEV-SNP, operate in control-centric environments with centralized execution and predictable memory behavior, making them well-suited for fine-grained access control, secure paging, and privilege enforcement. 
In contrast, \acp{GPU} are optimized for massive data-parallel workloads, with thousands of lightweight threads, wide memory bandwidth, and deep memory hierarchies. 
These characteristics make it difficult to directly apply traditional \ac{TEE} techniques like enclave management or synchronized memory encryption. 
Moreover, \acp{GPU} often lack built-in privilege separation and rely on the CPU for task scheduling and memory control, expanding the attack surface. 
As a result, \ac{GPU} \ac{TEE} designs typically require either hardware modifications to decouple from the host (\eg Graviton, HIX) or leverage CPU-side \acp{TEE} for protection and communication (\eg Honeycomb, SAGE). This architectural gap explains the greater diversity and complexity in \ac{GPU} \ac{TEE} designs compared to their CPU counterparts.

\noindent\textbf{Threat models.} 
Essentially all \ac{GPU} \ac{TEE} designs adopt a threat model similar to that of \ac{CPU} \acp{TEE}, aiming to create secure execution environments for \acp{GPU}. 
Specifically, they consider an adversary capable of controlling the entire software stack, including device drivers, the guest \ac{OS}, and the hypervisor. 
Additionally, the adversary is assumed to have limited physical access to the hardware, allowing for passive physical attacks, such as snooping on \ac{PCIe} traffic. 
This means that such an adversary could potentially leak or tamper with sensitive data and execution results of \ac{GPU} applications. 
They could also access or tamper with user data in \ac{DMA} buffers or with commands submitted by the victim application to the \ac{GPU}.

Regarding \ac{GPU} memory protection, many designs assume that the adversary cannot access secrets in the \ac{GPU} memory or explicitly state that physical attacks targeting memory are out of their scope. 
For instance, Graviton~\cite{volos2018graviton} and Honeycomb~\cite{mai2023honeycomb} trust the device memory of the \ac{GPU}, as modern \acp{GPU} typically integrate device memory using 2.5D/3D silicon interposers within the same package. 
These designs mainly focus on preventing adversaries from accessing sensitive data stored in \ac{GPU} memory with finer-grained access control.
To fill this gap, LITE~\cite{yudha2022lite} designs a software-based unified encryption domain to protect sensitive data stored in \ac{GPU} memory, including L1/L2 cache and device memory.
Additionally, NVIDIA H100~\cite{NVIDIAh100} and TNPU~\cite{lee2022tnpu} support hardware-based memory encryption, while CAGE~\cite{wang2024cage} requires hardware-assisted memory encryption support built into future CCA devices.

\noindent\textbf{Hardware changes.} 
Earlier works like Graviton~\cite{volos2018graviton} and HIX~\cite{jang2019heterogeneous} require non-trivial modifications to peripheral components, such as the \ac{GPU}'s command processor, the introduction of new \ac{SGX} instructions, specific \ac{MMU} page table walkers, and other hardware changes.
The extent to which a \ac{GPU} \ac{TEE} design modifies hardware components determines its practical compatibility, \ie its suitability for various \acp{GPU}.
Many designs necessitate hardware changes that have long lead times before they can be deployed in production environments. 
Consequently, works like LITE~\cite{yudha2022lite}, SAGE~\cite{ivanov2023sage}, and Honeycomb~\cite{mai2023honeycomb}, seek purely software-based approaches for trusted \ac{GPU} execution.
CAGE~\cite{wang2024cage} primarily leverages generic hardware features in Arm \ac{CCA} and \acp{GPU} without requiring additional hardware changes or customization of existing \acp{CPU} and \acp{GPU}.

\noindent\textbf{\ac{TCB} size.} 
One of the key challenges in \ac{GPU} \ac{TEE} design is reducing the \ac{TCB} size as much as possible.
A large software stack increases the \ac{TCB}, which can introduce more vulnerabilities and hence threaten security.
An effective and efficient way to reduce the \ac{TCB} is to place most of the \ac{GPU} software stack in the untrusted \ac{OS}, while keeping only the key functions for controlling the \ac{GPU} in a trusted area of the \ac{CPU}, such as an \ac{SGX} enclave.
For example, CAGE~\cite{wang2024cage} achieves a `thin' \ac{TCB} for realms by using its shadow task mechanism instead of loading the heavyweight \ac{GPU} software into the \ac{TEE}.
It introduces 2–-26K lines of code for a \ac{CPU}-side isolation and 1,301 lines of code for monitor, resulting in the lowest \ac{TCB} size among current Arm-based \ac{GPU} \acp{TEE}.

\noindent\textbf{Performance overhead.} 
The ultimate goal of designing a \ac{GPU} \ac{TEE} is to ensure strong isolation of \ac{GPU} execution while minimizing the performance overhead associated with providing security guarantees. 
Comparing the performance overhead of different schemes is challenging due to the costly re-implementation efforts required, as each scheme targets different platforms and is implemented under varying conditions.
For example, in HIX, the software components are implemented on top of an emulated system (like KVM and QEMU), which supports hardware modifications.
Several factors, such as the number of cryptographic operations, the frequency of switching between secure and normal worlds, and the integrity checks, play significant roles in optimizing performance compared to insecure baselines on \acp{GPU}.
For example, CAGE~\cite{wang2024cage} outperforms StrongBox~\cite{deng2022strongbox} by 1.96\%--13.16\% on six Rodinia benchmarks.
This performance gain is attributed to CAGE's ability to securely transfer plaintext data to buffers and store the results without introducing additional cryptographic operations.
It is reasonable to speculate that the NVIDIA H100 \ac{GPU} introduces a smaller performance burden due to its hardware-based encryption support. 
It would be insightful to conduct a comprehensive performance comparison across different platforms once the entire workflow for confidential computing on the NVIDIA H100 \ac{GPU} is fully available in the future.

\noindent\textbf{Trade-off between cryptographic methods and GPU TEEs.} 
In general, while cryptographic approaches like \ac{HE} and \ac{MPC} offer strong confidentiality guarantees, they remain impractical for many GPU workloads due to significant computation/communication overhead, memory usage, and developer complexity. 
Recent GPU-based MPC systems such as CryptGPU~\cite{tan2021cryptgpu}, Piranha~\cite{watson2022piranha}, and Orca~\cite{jawalkar2024orca} show that performance is highly sensitive to the number of computing participants and choice of operators, making them difficult to develop.
Piranha further highlights that communication overhead becomes a major bottleneck in wide-area network settings---an important consideration since real-world deployments often involve geographically distributed GPU providers.
For HE, even with dedicated HE accelerator support, as in TensorFHE~\cite{fan2023tensorfhe}, significant modifications are needed to underlying schemes (\eg CKKS) and operations (\eg NTT). Applying HE to \ac{ML} workloads also requires careful tuning of HE parameters and may necessitate changes to model architecture, complicating deployment in applications.
Therefore, GPU TEEs offer a more practical path forward, enabling protection of existing workloads with less developer effort and minimal modification to \ac{ML} models.
While GPU TEEs are still in their early stages and only applicable on certain machines, we believe they hold strong potential for future GPU-based applications, especially in \ac{ML}.

\section{Security of \acp{GPU}}
\label{sec:GPU-sec}

Due to their widespread adoption as accelerators, \acp{GPU} have come under  scrutiny from both offensive and defensive perspectives in recent years. 
Researchers have explored \ac{GPU} vulnerabilities at hardware architecture and software levels, showcasing practical attacks and suggesting mitigations. 
We believe that summarizing and analyzing attacks on \acp{GPU} will facilitate the exploration of potential attacks on \ac{GPU} \acp{TEE}.
While \citeauthor{naghibijouybari2022microarchitectural}~\cite{naghibijouybari2022microarchitectural} offer a comprehensive survey categorizing \ac{GPU} security vulnerabilities and potential countermeasures, their analysis does not cover significant advancements that have appeared since the latter half of 2022.
Moreover, their emphasis primarily is on microarchitectural attacks across diverse accelerators without considering attacks that leverage physical access or measurements. 
Nonetheless, the consideration of physical channels is essential for potential attacks on \ac{GPU} \ac{TEE}. 

In this section, we review existing \ac{GPU} attacks, including microarchitectural side-channel/covert-channel attacks, physical side-channel attacks, and fault injection attacks.
\Cref{tab:GPU attacks} provides a summary of attacks, target \acp{GPU}, threat model, attack goals and methods.
Based on \Cref{tab:GPU attacks}, we highlight several interesting take-away points: 
\ding{182} Most attacks target single discrete GPU systems, with memory and timing behavior serving as the primary leakage vectors. However, recent research has begun to explore attacks on multi-GPU and multi-instance GPU environments, revealing new vulnerabilities in GPU-to-GPU communication and resource partitioning.
\ding{183} A majority of attacks can be carried out with only user-level permissions, indicating that isolation mechanisms in modern GPU systems remain inadequate, particularly in multi-tenant or cloud-based deployments.
\ding{184} Attacks span across multiple GPU vendors (\eg NVIDIA, AMD, Qualcomm) and consistently aim to recover sensitive information, such as AES keys, \ac{NN} model parameters, and user input keystrokes. Notably, attacks targeting more complex GPU workloads, such as \acp{LLM}, remain largely unexplored.
\ding{185} While impactful, fault injection and covert channels appear less frequently in the literature. These often require stronger attacker capabilities or intricate coordination between software and hardware layers, which may explain their limited exploration.
\ding{186} Emerging attacks have begun to explore cross-GPU timing channels and remote leakage via browsers, suggesting that cloud-hosted GPU services—especially those accessible by multiple users—may become an increasingly important and vulnerable attack surface.


\begin{ThreePartTable}
    \tiny
    \begin{TableNotes}
    \setlength{\leftskip}{-7pt}
        \item \textbf{CNN/DNN} denotes Convolutional and Deep Neural Network.
        In \textbf{Attacks}, \textbf{SCA}, \textbf{CovA}, and \textbf{FIA} denote \acl{SCA}, Covert-Channel, and Fault Injection Attacks, respectively. \textbf{AA} represents an attack that exploits an architectural or logical vulnerability on \ac{GPU}.
        \textbf{Physical} denotes whether this attack leverages physical properties (\eg power consumption, \ac{EM}, acoustics, \etc).
        In \textbf{Model}, \textbf{sGPU}, \textbf{mGPU}, and \textbf{iGPU} denote the attack is deployed on a single \ac{GPU}, multiple \acp{GPU} or integrated \ac{GPU}, respectively. Note that multiple \acp{GPU} represent multiple discrete \ac{GPU} cards or multiple \ac{GPU} instances (\eg NVIDIA's MIG feature).
        In \textbf{Main results}, we report the central outcomes, \eg bandwidth and classification accuracy for CovA and SCA, respectively.
    \end{TableNotes}
    
    \begin{longtable}{p{1.4cm}p{0.8cm}p{0.3cm}p{2.5cm}p{0.2cm}p{2.5cm}p{2.8cm}p{1.5cm}}
    \caption{Comparison of different attacks on heterogeneous systems (oldest results first)} \label{tab:GPU attacks} \\
        
        \toprule
            \multirow{2}{*}{\textbf{Work\&Year}} & \multirow{2}{*}{\textbf{Accelerator}}& \multicolumn{2}{l}{\textbf{Threat Model}} & \multirow{2}{*}{\textbf{\makecell[l]{Phy-\\sical}}} & \multirow{2}{*}{\textbf{Attack Scenarios \& Goals}} & \multirow{2}{*}{\textbf{Method}} & \multirow{2}{*}{\textbf{Main results}} \\ 
            
            \cline{3-4}
            
            &  & \textbf{Model} & \textbf{Permissions} &  &  &  &  \\
            \midrule

            \multicolumn{7}{l}{\textbf{Architectural Attacks}} \\
            \midrule
             \citeauthor{lee2014stealing}'14~\cite{lee2014stealing} & NVIDIA/AMD \ac{GPU} & sGPU & User-level permissions & \ding{53} & Infer which web pages a victim user has visited & Exploit that \acp{GPU} do not initialize newly allocated memory pages & $95.4\%$ accuracy \\
             \citeauthor{maurice2014confidentiality}'14~\cite{maurice2014confidentiality} & NVIDIA \ac{GPU} & sGPU & Require root privileges based on accessing \ac{GPU} memory via \ac{GPU} runtime or PCI configuration & \ding{53} & Recover data of victim's previously execution on \ac{GPU} in virtualized environment & Exploit the vulnerability of memory clearing on \ac{GPU} & Access data stored in \ac{GPU} global memory after soft reboot \\
             \citeauthor{pietro2016cuda}'16~\cite{pietro2016cuda} & NVIDIA \ac{GPU} & sGPU & User-level permissions  & \ding{53} & Read information stored in memory and infer the \ac{AES} keys & Leverage the lack of adequate data allocation and data cleaning & $70\%$, $28\%$  leakages \\
             \citeauthor{zhou2016vulnerable}'16~\cite{zhou2016vulnerable} & NVIDIA/AMD \ac{GPU} & sGPU & User-level permissions & \ding{53} & Recover users' profile images and email contents & Reconstruct images from identified image-like tiles in \ac{GPU} memory & $79.3\%$ successful recovery \\
             
            \midrule
            \multicolumn{7}{l}{\textbf{Microarchitectural Side-Channel/Covert-Channel Attacks}} \\
            \midrule
             \citeauthor{luo2015side}'15~\cite{luo2015side} & NVIDIA \ac{GPU} & sGPU & Unprivileged attacker has physical access to the victim \ac{GPU} & $\checkmark$ & Recover all of the last round \ac{AES} key byte & Apply correlation power analysis to analyze the power consumption of the \ac{GPU} & Recover the key with 160,000 power traces \\
             \citeauthor{jiang2016complete}'16~\cite{jiang2016complete} & NVIDIA \ac{GPU} & sGPU & Send plaintext to victim \ac{GPU} for encryption with user-level permission & \ding{53} & Recover the 16-byte last-round key of the \ac{AES} & Leverage the relation between encryption kernel's runtime and the key & 30 min with 1M timing samples \\
             \citeauthor{jiang2017novel}'17~\cite{jiang2017novel} & NVIDIA \ac{GPU} & sGPU & Send plaintext to victim \ac{GPU} for encryption with user-level permission & \ding{53} & Recover the 16-byte last-round key of the \ac{AES} & Exploit timing variability due to shared memory bank conflicts. & Use 10M timing samples \\
             \citeauthor{naghibijouybari2017constructing}'17~\cite{naghibijouybari2017constructing} & NVIDIA \ac{GPU} & sGPU & Trojan and spy co-exist on a \ac{GPU}, both with user-level permission & \ding{53} & Infer transmitted data & Reverse engineer warp schedulers; learn contention on caches and memory & $4$\,MB/s (error-free) \\
             \citeauthor{gao2018electro}'18~\cite{gao2018electro} & NVIDIA \ac{GPU} & sGPU & Unprivileged attacker has physical access to the victim \ac{GPU} & $\checkmark$ & Recover \ac{AES} private key & Sample \ac{MMIO} traces via trigger; exploit parallel \ac{MMIO} leakages & Recover the key with 11k \ac{MMIO} traces \\
             \citeauthor{naghibijouybari2018rendered}'18~\cite{naghibijouybari2018rendered} & NVIDIA \ac{GPU} & sGPU & Spy co-locates with the victim program with user-level permission & \ding{53} & Infer victim's accessed website data and the number of neurons & Learn contention via memory allocation, performance counter, and timing & $90\%$, $94.8\%$, $88.6\%$ accuracy \\ 
             \citeauthor{luo2019side}'19~\cite{luo2019side} & NVIDIA \ac{GPU} & sGPU & User-level permissions; available to timing information & \ding{53} & Recover RSA private key & Capture the parallel characteristics of an RSA on \ac{GPU} via timing model & Recover 512/2048-bit key in 15\,min/8\,h  \\
             \citeauthor{wei2020leaky}'20~\cite{wei2020leaky} & NVIDIA \ac{GPU} & sGPU & Spy co-locates with the victim program with user-level permission & \ding{53} & Recover the layer composition and hyper-parameters of \ac{DNN} & Leverage context-switching penalties when switching different kernels & $76.9\%$-$100\%$ accuracy \\
             \citeauthor{hu2020deepsniffer}'20~\cite{hu2020deepsniffer} & NVIDIA \ac{GPU} & sGPU & Physically access the victim \ac{GPU} with user-level permissions & $\checkmark$ & Obtain the complete model architecture information & Learn the inter-layer temporal association information from kernel execution & $75.9\%$ accuracy \\
             \citeauthor{wenjian2020igpu}'20~\cite{wenjian2020igpu} & Intel HD 530 & iGPU & Unprivileged adversary and the victim co-locate on the same GPU, but  are isolated & \ding{53} & (1) Recover \ac{AES} key; (2) monitor the browser activity; (3) conduct a covert channel to exchange data & Exploit context management with uncleared registers and shared memory during context switches & Recover \ac{AES}-128/192 in 0.15\,s/2\,min; $90.7\%$ accuracy; 1.3--8\,Gbps \\
             \citeauthor{ahn2021network}'21~\cite{ahn2021network} & NVIDIA \ac{GPU} & sGPU & Trojan and spy co-exist on a \ac{GPU}, both with user-level permissions & \ding{53} & Infer transmitted data & Exploit interconnect channel contention & $24$\,MB/s \\
             \citeauthor{cronin2021exploration}'21~\cite{cronin2021exploration} & Apple \ac{GPU} & sGPU & Exploit JavaScript website and trick user into installing a malicious app & \ding{53} & Track the user's web activity & Exploit the shared cache between the \ac{CPU}-cores and peripherals & $90\%$, $95\%$ accuracy \\
             \citeauthor{zhu2021hermes}'21~\cite{zhu2021hermes} & NVIDIA \ac{GPU} & sGPU & Install a \ac{PCIe} bus snooping device to monitor and log the \ac{PCIe} traffic & \ding{53} & Recover the \ac{DNN} model (semantically identical architecture) & Reverse engineer \ac{GPU} command headers; collect \ac{PCIe} packets via snooping  & Similar accuracy as original ones \\
             \citeauthor{ahn2021trident}'21~\cite{ahn2021trident} & NVIDIA \ac{GPU} & sGPU & Send plaintext to victim \ac{GPU} for encryption with user-level permission & \ding{53} & Recover the 16-byte last-round key of the \ac{AES} & Recover early/later key bytes via timing correlation and cache-collision & Use 100k timing samples \\
             \citeauthor{dutta2021leaky}'21~\cite{dutta2021leaky} & Intel HD Graphics & iGPU & Spy and trojan run on \ac{CPU} and \ac{GPU} with user-level permissions & \ding{53} & Infer transmitted data and track user's cache activity & CovA1: prime+probe attack on {LLC}; CovA2: observe delays on the ring bus & CovA1: 120\,KB/s, CovA2: 400\,KB/s \\
             %
             %
             \citeauthor{nayak2021mis}'21~\cite{nayak2021mis} & NVIDIA \ac{GPU} & sGPU & User-level permissions & \ding{53} & Infer the accessed data from Virginian database & Leverage the shared last-level \ac{TLB} (reverse engineering) and prime+probe & 81\,KB/s (MPS-enabled) \\
             %
             %
             \citeauthor{yang2022eavesdropping}'22~\cite{yang2022eavesdropping} & Qualcomm Adreno & sGPU & Install and launch an eavesdropping application on the victim device & \ding{53} & Eavesdrop the user's input credentials (login usernames and passwords) through onscreen keyboard & Read raw values from the \ac{GPU} device file and identify \ac{GPU} events from key presses via \ac{ML} models & Correctly infer more than $80\%$ of user's credential inputs \\
             \citeauthor{zhan2022graphics}'22~\cite{zhan2022graphics} & NVIDIA/AMD \ac{GPU} & sGPU & Unprivileged attacker eavesdrops on a victim & $\checkmark$ & Website fingerprinting and keystroke timing inference & Exploit EM caused by \ac{DVFS} & $95.6\%$ accuracy and $90\%$ keystrokes \\
             \citeauthor{liang2022clairvoyance}'22~\cite{liang2022clairvoyance} & NVIDIA \ac{GPU} & sGPU & Attacker and victim \acp{GPU} may be separated by meters & $\checkmark$ & Infer the number of layers/kernels, sizes of layers and strides & Exploit far-field EM signals emanated from a \ac{GPU} & Recover VGG's and ResNet's parameters \\
             \citeauthor{maia2022can}'22~\cite{maia2022can} & NVIDIA \ac{GPU} & sGPU & Unprivileged attacker with physical proximity & $\checkmark$ & Infer NN architectures and layer parameters & Exploit the magnetic flux emanating from a \ac{GPU}'s power cable & Up to $96.8\%$ accuracy \\
             \citeauthor{horvath2023barracuda}'23~\cite{horvath2023barracuda} & NVIDIA \ac{GPU} & sGPU & User-level permission with physical access and know the \ac{CNN} architecture & $\checkmark$ & Recover the weights and biases of \ac{CNN} & Correlation electromagnetic analysis & N/A \\
             \citeauthor{dutta2023spy}'23~\cite{dutta2023spy} & NVIDIA \ac{GPU} & mGPU & Spy and victim connect via NVLink with user-level permissions & \ding{53} & CovA: infer message; SCA: infer the number of neurons of NN & Mount Prime+Probe attack on the L2 cache of victim \ac{GPU} & CovA: 4\,MB/s; SCA: $99.91\%$ accuracy \\
             \citeauthor{zhang2023t}'23~\cite{zhang2023t} & NVIDIA \ac{GPU} & mGPU & Unprivileged spy and trojan on an MIG-enabled GPU & \ding{53} & Infer transmitted data and classify the running \ac{ML} frameworks & Mount Prime+Probe attack on the shared L3-\ac{TLB} & $31.47$\,KB/s \\ 
             %
             %
             \citeauthor{taneja2023hot}'23~\cite{taneja2023hot} & Various \acp{GPU} & iGPU, sGPU & Read frequency, power, and temperature from internal sensors (may require privileged access) & $\checkmark$ & (1) Observe \ac{CPU}/\ac{GPU} behaviour; (2) recover images and user's browsing history; (3) fingerprint websites & (1) Use data from internal sensors; (2) observe throttling via timing; (3) leverage frequency and power bursts on iGPUs & Pixel stealing with accuracy of 60\%--90\% \\
             \citeauthor{wang2024gpu}'24~\cite{wang2024gpu} & Various \acp{GPU} & iGPU, sGPU & Observe coarse-grained pattern redundancy using timer & \ding{53} & Perform cross-origin pixel stealing attacks in the browser & Infer data-dependent DRAM traffic and cache usage via compression channel & Pixel stealing with accuracy 68.6\%--99.6\% \\
             \citeauthor{giner2024generic}'24~\cite{giner2024generic} & NVIDIA/AMD \ac{GPU} & sGPU & SCA: Victim passively visits attacker's webpage; CovA: Spy accesses the website visited by victim & \ding{53} & SCA: Detect typed characters and recover the victim's \ac{AES} key; CovA: Infer transmitted data & Exploit cache contention on the L2 cache via Prime+Probe attack, parallelize eviction set construction & SCA: $98\%$ accuracy, recover key in 6\, min; CovA: $10.9$\,KB/s \\
            \midrule
            \multicolumn{7}{l}{\textbf{Fault Injection Attacks}} \\
            \midrule
             \citeauthor{frigo2018grand}'18~\cite{frigo2018grand} & Intel HD, Qualcomm Adreno & iGPU & Access iGPU through a malicious application or when victim visits malicious JavaScript website & \ding{53} & Escape the Firefox sandbox on Android platforms and break ASLR & Detect contiguous memory allocation via SCA and then trigger Rowhammer bit flips & Hijack the browser and break {ASLR} in 116\,s and 27\,s \\ 
             \citeauthor{sabbagh2020novel}'20~\cite{sabbagh2020novel} & AMD \ac{GPU} & sGPU & \ac{DVFS} interfaces access & \ding{53} & Recover \ac{AES} key & Exploit \ac{DVFS} to introduce random faults during kernel execution & Recover key in 15\,min \\
             \citeauthor{dos2021revealing}'21~\cite{dos2021revealing} & NVIDIA \ac{GPU} & sGPU & Inject faults into pipeline registers, ALU, and memory & \ding{53} & Improve the probability of incorrect output by injecting faults & Combine register transfer level FIA with software FIA & Underestimate the error rate by up to $48\%$ \\
        
        \bottomrule
        \insertTableNotes \\
    \end{longtable}
\end{ThreePartTable}

\subsection{Architectural Attacks}
\label{sec:al attack}
There has been an increase in research examining attacks that target \acp{GPU} or their software stack on the \ac{CPU}.
Schemes \cite{lee2014stealing,maurice2014confidentiality,pietro2016cuda,zhou2016vulnerable} focus on information leakage due to leftover memory from processes that recently terminated.
\citeauthor{lee2014stealing}~\cite{lee2014stealing} exploit the vulnerability arising from the \acp{GPU} not initializing newly allocated memory pages, thereby revealing the victim's data stored in \ac{GPU} memory both during its execution and immediately after termination. In practice, they infer the webpages visited by a victim user when analyzing remaining textures from Chromium and Firefox.
\citeauthor{maurice2014confidentiality}~\cite{maurice2014confidentiality} recover data from previously executed \ac{GPU} applications by observing that \ac{GPU} global memory is not always zeroed.
\citeauthor{pietro2016cuda}~\cite{pietro2016cuda} leverage interleaved access and inadequate data handling to read information stored in shared or global memory by another process.
\citeauthor{zhou2016vulnerable}~\cite{zhou2016vulnerable} demonstrate how malicious programs can exploit \ac{GPU} memory management strategies to breach the memory isolation boundary and recovers images from other processes. 
\citeauthor{wenjian2020igpu}~\cite{wenjian2020igpu} observed a vulnerability in Intel integrated \ac{GPU} caused by flawed context management, where residual register values and shared memory are not cleared during a context switch.
This class of vulnerability can usually be fixed by clearing memory (\ie memory zeroing) before reallocating it to another application.

\noindent\textbf{Takeaway points.} Architectural attacks often exploit leftover data in GPU memory due to missing or incomplete memory clearing. These attacks require only user-level access and affect both integrated and discrete GPUs, showing that such flaws are common across vendors. While many of these vulnerabilities have been patched in recent years, they highlight past weaknesses in GPU memory isolation.

\subsection{Microarchitectural Side-Channel/Covert-Channel Attacks}
\label{sec:micro_sca_cca-attack}
In this subsection, we review existing attacks on \acp{GPU} w.r.t.\ microarchitectural side-channel and covert-channel attacks.

\subsubsection{Attacks on Integrated \acp{GPU}.}
Attacks on integrated \acp{GPU} are gaining attention due to their security risks. 
The integration opens up new attack opportunities by exploiting shared resources, leading to cross-component microarchitectural attacks. 
A key requirement for these attacks is co-location, where the attacker runs code close enough to interact with the victim~\cite{naghibijouybari2022microarchitectural}. This interaction could involve inducing faults or contention to measure side-channel leakage. 
Therefore, integrated \acp{GPU}, which share the same chip as the \ac{CPU}, offer more attack opportunities.

\citeauthor{dutta2021leaky}~\cite{dutta2021leaky} develop two covert channels, enabling two malicious applications located on separate components (\ac{CPU} and iGPU) to exchange confidential data through shared hardware resources. One channel exploits the shared \ac{LLC} within Intel's integrated \ac{GPU}, while the other relies on the contention from the ring bus.
\citeauthor{wang2024gpu}~\cite{wang2024gpu} develop cross-origin pixel stealing attacks in web browsers by exploiting the iGPU compression channel to induce data-dependent DRAM traffic and cache utilization. Given that the \ac{LLC} is shared between the \ac{CPU} and \ac{GPU}, they demonstrate how to use \ac{LLC} walk time to infer the \ac{GPU}-induced \ac{CPU} cache state changes. 
\citeauthor{wang2021towards}~\cite{wang2021towards} extend the previous timing side-channel attack on \ac{AES} implementation~\cite{jiang2016complete} to the iGPU.

Some recent studies have also delved into \acp{GPU} such as the Qualcomm Adreno \ac{GPU} and Apple \ac{GPU}. 
Building upon Arm's system-level cache, \citeauthor{cronin2021exploration}~\cite{cronin2021exploration} exploit the \ac{GPU} and shared cache architecture of Arm DynamIQ to create a website fingerprinting side channel.
\citeauthor{yang2022eavesdropping}~\cite{yang2022eavesdropping} target the Qualcomm Adreno \acp{GPU}, investigating the \ac{GPU} overdraw induced by rendering popups of user key presses. 
Each key press leads to unique variations in selected \ac{GPU} performance counters, allowing accurate inference of these key presses.
To access \ac{GPU} performance counters on Android, they read the raw values of \ac{GPU} PCs from the \ac{GPU} device file (\eg Qualcomm Adreno \ac{GPU} drivers' open-source header file, \texttt{msm\_kgsl.h}).

\noindent\textbf{Takeaway points.} 
Integrated GPUs introduce new security challenges due to their tight coupling with CPUs and shared microarchitectural resources, such as \ac{LLC} and interconnects. This shared environment enables cross-component attacks, exploiting shared caches, compression channels, and GPU performance counters. These attacks are particularly concerning in web and mobile environments, where integrated GPUs are widely deployed and users have minimal control over low-level hardware access.

\subsubsection{Attacks on Discrete \acp{GPU}.}
Attacks on discrete \acp{GPU} can generally be categorized in two ways: (i) where the spy and victim (or trojan) applications are co-located on a single \ac{GPU}, denoted as \emph{sGPU}; and (ii) where the spy and victim (or trojan) applications are co-located on different \acp{GPU}, denoted as \emph{mGPU}. For instance, this scenario could involve a system equipped with multiple \acp{GPU}, such as NVIDIA's Pascal-based DGX-1 system comprising eight Tesla P100 \acp{GPU}. Alternatively, it could involve a system equipped with an \ac{MIG}-enabled \ac{GPU}, where a single physical \ac{GPU} is partitioned into multiple instances, each acting as an independent \ac{GPU} with dedicated resources, such as NVIDIA A100 and H100.

\noindent\textbf{Attacks on a single \ac{GPU}.} 
This category of attacks typically assumes that an attacker with only user-level permissions either launches a \ac{GPU} kernel on a single \ac{GPU} to gain insights from measurements such as execution time, performance counters, and other leakage vectors; or co-locates with the victim process on a single \ac{GPU} to establish a covert channel. 
In the following, we review the existing research and categorize them into the following aspects:

\emph{Recovering private encryption keys in \ac{GPU}-based cryptography implementations: }
Recent works demonstrate the capability of recovering encryption keys (\eg \ac{AES} and \ac{RSA} private keys) from encryption applications running on \acp{GPU}.
\citeauthor{jiang2016complete}~\cite{jiang2016complete} exploit timing differences between addresses generated by different threads accessing memory, where execution time is affected by the number of unique memory requests after coalescing. Consequently, the key affects the address pattern accessed by threads, thereby impacting the observed runtime of the \ac{AES} encryption.
In their another study~\cite{jiang2017novel}, they establish a correlation between the execution time of a warp's table lookup and the number of shared memory bank conflicts generated by threads within the warp. These key-dependent timing differences are then used to correlate the measured execution time with the key at the last round of the \ac{AES} encryption as it executes on the \ac{GPU}.
\citeauthor{ahn2021trident}~\cite{ahn2021trident} exploits negative timing correlation to recover earlier key bytes of \ac{AES}, while employing cache-collision attacks for the latter \ac{AES} key bytes.
More recently, \citeauthor{giner2024generic}~\cite{giner2024generic} recover keys from an \ac{AES} T-table \ac{GPU} implementation. They allocate a sizable buffer to occupy a significant portion of the cache, execute an \ac{AES} encryption, and then use Prime+Probe to identify evicted buffer offsets that correlate with the encryption's inputs and outputs.
They demonstrate superior performance by recovering the entire \ac{AES} key within 6\,min, whereas previous studies took between 15 and 30\,min.
Furthermore, aside from \ac{GPU}-based \ac{AES} implementation, \citeauthor{luo2019side}~\cite{luo2019side} present a timing side-channel attack on an \ac{RSA} implementation.
They use the correlation between decryption time and the number of reductions per decryption window to extract the \ac{RSA} private key.

\begin{figure}[htbp]
  \centering
    \begin{subfigure}{0.45\linewidth}
      \centering   
      \includegraphics[width=0.75\linewidth]{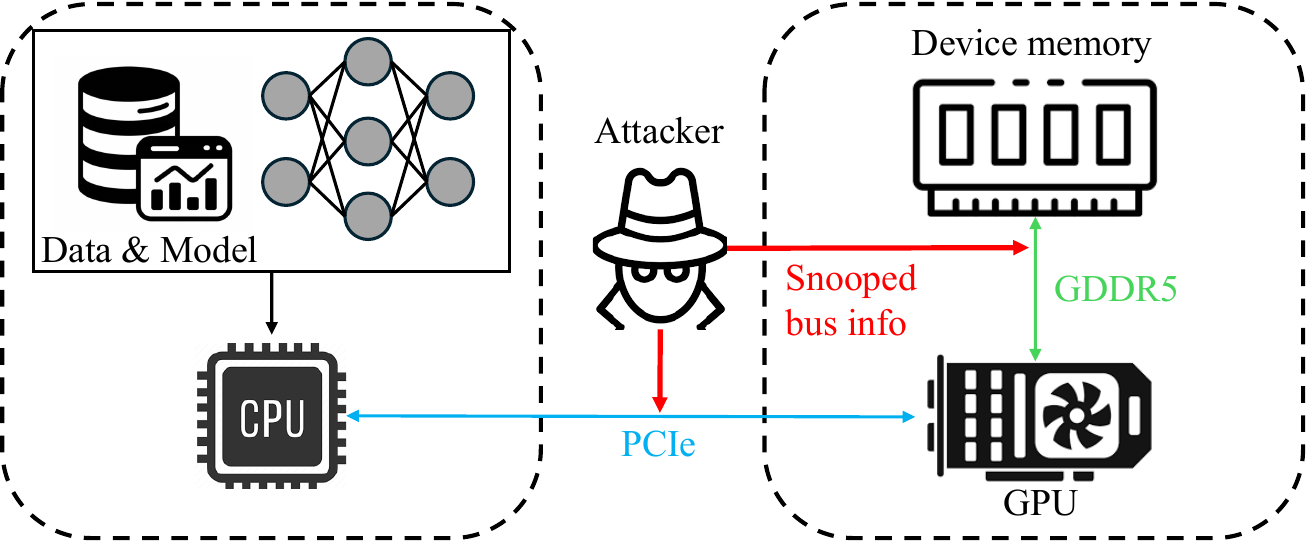}
        \caption{The attacker can snoop the \ac{PCIe} traffic or GDDR5 \\ using a bus snooping device}
        \label{fig:attacker model:sub1}
    \end{subfigure}   
    \begin{subfigure}{0.45\linewidth}
      \centering   
      \includegraphics[width=0.95\linewidth]{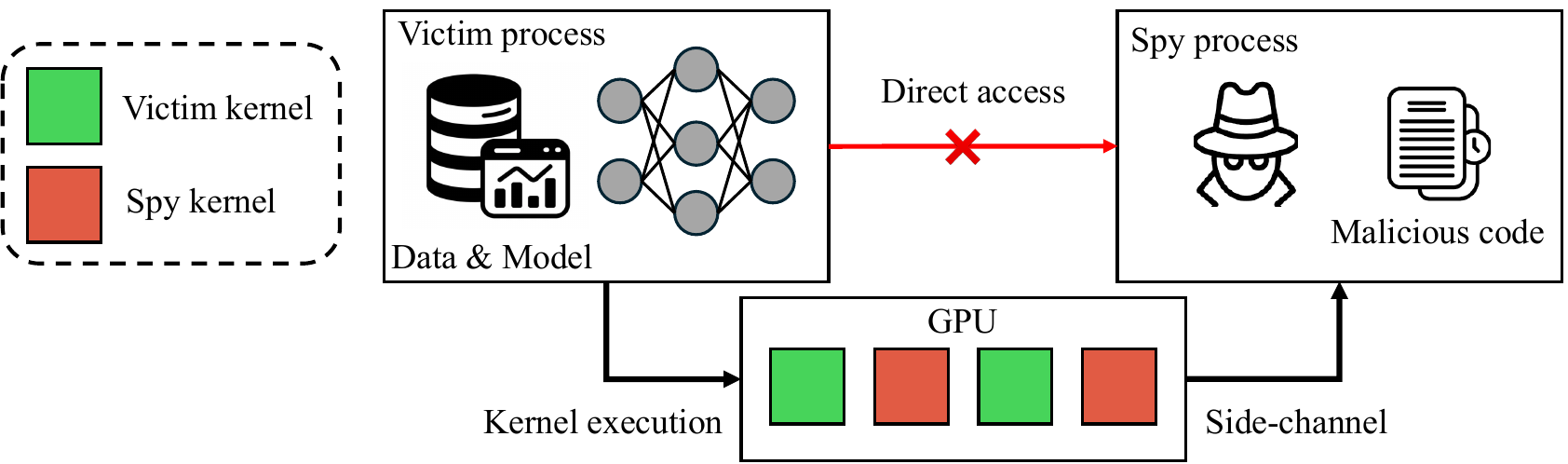}
        \caption{Spy and victim are on two different applications (or processes and \acp{VM}) sharing the same \ac{GPU}}
        \label{fig:attacker model:sub2}
    \end{subfigure}
\caption{
\label{fig:attacker model}
Attacker models for inferring \acs{NN} models on a single \ac{GPU}.}  
\end{figure}

\emph{Inferring sensitive information from \ac{GPU}-based \ac{NN}: }
Other attacks exploit side-channel leakage to infer information from \ac{ML} models like \acp{NN}.
\citeauthor{zhu2021hermes}~\cite{zhu2021hermes} and \citeauthor{hu2020deepsniffer}~\cite{hu2020deepsniffer} focus on the attack surface of the bus, particularly \ac{PCIe} (as depicted in \Cref{fig:attacker model:sub1}).
\citeauthor{zhu2021hermes}~\cite{zhu2021hermes} reverse-engineer the critical data structures (\eg \ac{GPU} command headers) offline and collect victim \ac{PCIe} packets via a snooping device to reconstruct models in the online phase.
The stolen \ac{DNN} models have the same parameters and semantically identical architecture as the original ones.
One potential mitigation for this attack is encrypting \ac{PCIe} traffic. 
In contrast, \citeauthor{hu2020deepsniffer}~\cite{hu2020deepsniffer} assume that the adversary cannot access data passing through buses, only the addresses, allowing attacks even with encrypted data.
They conduct bus snooping attacks on discrete \ac{GPU}'s memory and \ac{PCIe} bus, obtaining kernel read/write access volume and memory address traces. Correlating extracted architectural events with model internal architectures.
Another class of attacks occurs between two concurrent \ac{GPU} applications, as shown in \Cref{fig:attacker model:sub2}.
In~\cite{naghibijouybari2017constructing}, they (i) implement website fingerprinting attack via the \ac{GPU} memory utilization \ac{API} or \ac{GPU} performance counters; (ii) monitor user's web activities during interactions or while typing characters on a keyboard through keystroke timing analysis; (iii) utilize a CUDA spy within the computational stack of \ac{GPU} to deduce the \ac{NN} structure by leveraging shared resources.
In contrast to~\cite{naghibijouybari2017constructing}, \citeauthor{wei2020leaky}~\cite{wei2020leaky} consider fine-grained time-sliced sharing of applications on different \acp{VM} and exploit context-switching penalties on performance counters as leakage vector to recover the layer composition and hyperparameters of a \ac{DNN}.

Most of these attacks require extensive reverse-engineering of NVIDIA \ac{GPU} details. This is because\ac{GPU} driver and internal \ac{GPU} workings are closed-source, and crucial data structures are left undocumented (\cf \Cref{sec:GPU basics}).
In other words, these factors make the attack more difficult to some extent, but do not fundamentally prevent them.

\begin{figure}[htbp]
\centering
\includegraphics[width=0.35\linewidth]{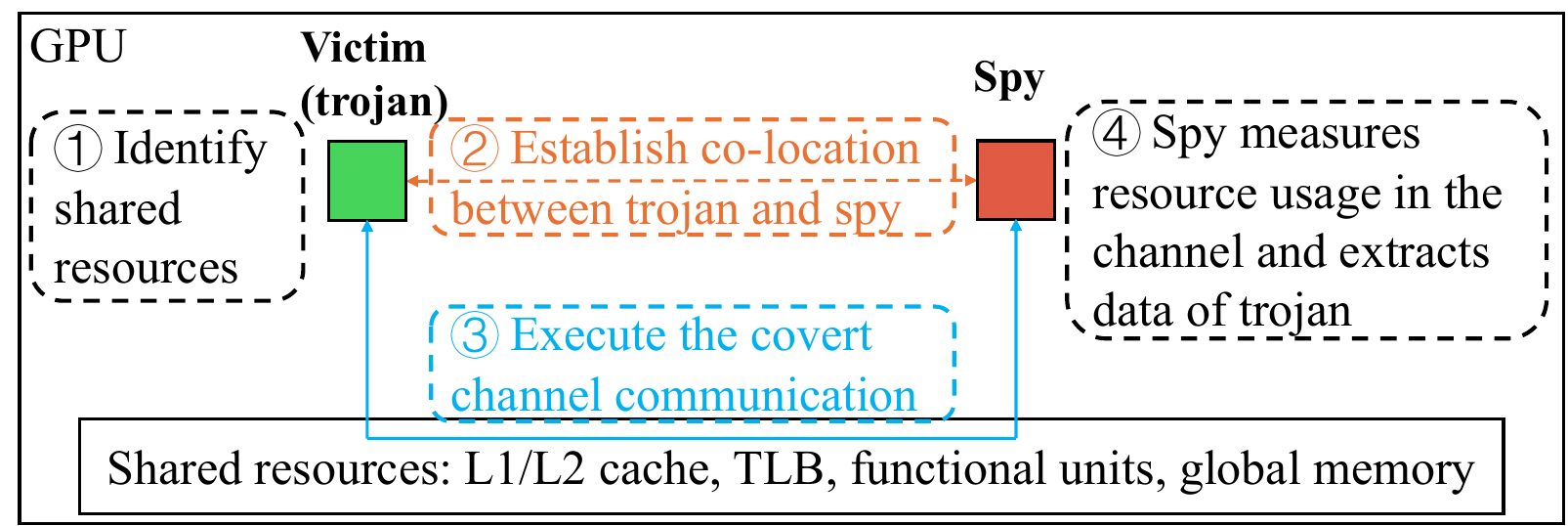}
\caption{Key steps of covert-channel attacks on a single \ac{GPU}.}
\label{fig:cca_sgpu}
\end{figure}

\emph{Covert-channel attacks where both spy and victim concurrently run on a single \ac{GPU}: }
As summarized in \Cref{fig:cca_sgpu}, the key point of a successful covert-channel attack is identifying the shared resources utilized by both the trojan and spy, often requiring extensive reverse engineering of \ac{GPU} architectures. 
This process is necessary to uncover details about \ac{GPU} components, such as the \ac{TLB} and other shared resources, which can be exploited for covert communication.
\citeauthor{naghibijouybari2017constructing}~\cite{naghibijouybari2017constructing} first reverse-engineer the hardware block and warp-to-warp schedulers to establish the co-location, then learn contention on caches, functional units and memory, and finally construct covert channels based on these resources with an error-free bandwidth exceeding 4\,Mbps. With this, they recover the number of neurons in an \ac{NN}.
\citeauthor{ahn2021network}~\cite{ahn2021network} initially reverse-engineer the structure of on-chip networks and find the hierarchical organization of the \ac{GPU} leads to the sharing of interconnect bandwidth among adjacent cores.
They then exploit the contention of the interconnect channel, leverage clock registers for sender-receiver synchronization and ultimately establish microarchitectural covert channels with a high bandwidth of 24\,Mbps.
\citeauthor{nayak2021mis}~\cite{nayak2021mis} reverse-engineer the details of the \ac{TLB} hierarchy and the details of all \ac{TLB} levels, such as the number of entries and associativity, of the NVIDIA 1080Ti.
They create a covert timing channel via the shared L3 \ac{TLB} using Prime+Probe.
With MPS, they increase the bandwidth of the \ac{GPU}'s L3 \ac{TLB}-based covert channel to 81\,Kbps.
Unlike previous work relies on access to the native \ac{GPU} \acp{API} through CUDA or OpenGL for high-precision performance counters monitoring, \citeauthor{giner2024generic}~\cite{giner2024generic} do not rely on WebGPU providing hardware timers. 
Instead, they use a shared memory buffer and a dedicated thread to increment a shared variable constantly, functioning as a timer. 
With this timer and their \ac{GPU}-accelerated cache eviction set construction, they establish a Prime+Probe cache covert channel for the L2 cache with a bandwidth of 10.9\,KB/s.

\noindent\textbf{Attacks on multiple \acp{GPU}.} 
In NVIDIA's DGX machines, where discrete \acp{GPU} are connected via NVLink and \ac{PCIe}, \citeauthor{dutta2023spy}~\cite{dutta2023spy} reverse-engineer the cache hierarchy and show that an attacker on one \ac{GPU} can create contention in the L2 cache of another \ac{GPU} through NVLink.
They establish a Prime+Probe covert-channel across two \acp{GPU}, achieving a bandwidth of 4\,MB/s.
Additionally, their side-channel attack allows an attacker to monitor another application running on different \acp{GPU}, enabling the recovery of the number of neurons of the \ac{NN}.
By reverse-engineering the \ac{TLB} structure in recent NVIDIA \acp{GPU} such as A100 and H100, \citeauthor{zhang2023t}~\cite{zhang2023t} discover a design flaw in the NVIDIA \ac{MIG} feature: \ac{MIG} does not partition the last-level \ac{TLB}, which is shared by all \ac{GPU} instances. 
They dump \ac{GPU} memory to locate corresponding table entries and access the target entry through the 1\,MB \ac{GPU} memory window (\ac{MMIO}) for runtime modifications. 
They finally establish a cross-\ac{GPU}-instance covert-channel based on the shared \ac{TLB}.
In summary, both studies begin by identifying shared resources present in multi-\ac{GPU} systems, such as the shared remote L2 cache~\cite{dutta2023spy} and the shared L3 \ac{TLB}~\cite{zhang2023t}, and then create visible contention on these shared resources.

\noindent\textbf{Takeaway points.} 
\ding{182} Reverse engineering remains key to GPU attacks, highlighting both the limitations of security through obscurity and the lack of transparency from vendors that impedes systematic evaluation.
\ding{183} Attack bandwidths have significantly improved and are now comparable to early CPU-based covert channels, reflecting the growing sophistication of GPU exploitation.
\ding{184} Isolation mechanisms such as \ac{MIG} may have flaws. Logical partitioning alone does not guarantee security and requires careful validation.
\ding{185} Attacks leveraging NVLink and PCIe show that GPU interconnects are an emerging and under-protected attack surface, particularly in AI datacenter environments.
\ding{186} \ac{ML} workloads, especially \acp{DNN}, remain prime attack targets. Yet, the security of newer models such as \acp{LLM} and diffusion models across multi-GPU systems remains largely unexplored.

\subsubsection{Countermeasures}
The countermeasures can be summarized as follows:

\emph{(i) Partitioning resources:} Co-location between the attacker and the victim is essential for those attacks that exploit shared microarchitectural resources.  
One solution to mitigate this threat is to partition resources (\eg \ac{LLC}) either statically or dynamically~\cite{wang2007new,kong2009hardware,liu2016catalyst}. By assigning the spy and trojan to different cache partitions, they cannot replace each other's cache lines.
The \ac{MIG} introduced in \Cref{sec:GPU basics} potentially mitigates these attacks by isolating resources between independent \ac{GPU} instances.
However, \citeauthor{zhang2023t}~\cite{zhang2023t} point out that the last-level \ac{TLB} between \ac{MIG}-enabled \ac{GPU} instances is not properly partitioned, allowing for the creation of a covert channel using this vulnerability.

\emph{(ii) Eliminating contention:} Contentions among processes can be eliminated by controlling traffic in memory controllers. This involves grouping memory requests from each processor into the same queue, potentially accessing the same memory bank/port~\cite{wang2014timing,shafiee2015avoiding}. 
For example, TLB-pilot~\cite{di2021tlb} binds thread blocks from different kernels to different groups of streaming multiprocessors by considering hardware isolation of last-level \ac{TLB} and the application's resource needs. It gathers kernel information before kernel launches and reduces load imbalance by coordinating software and hardware-based scheduling and employing a kernel splitting strategy.
There are also defences against coalescing-based correlation timing attacks. 
For example, \citeauthor{kadam2018rcoal}~\cite{kadam2018rcoal,kadam2020bcoal} add extra memory accesses to make execution time less predictable.
To protect against timing and cache-based attacks in \ac{GPU}-based \ac{AES}, \citeauthor{lin2019scatter}\cite{lin2019scatter} use scatter and gather techniques to reorganize \ac{AES} tables, avoiding key-dependent lookups from timing or address leakage.


\subsection{Physical Side-Channel Attacks}
We classify attacks allowing an attacker to have direct proximity or contact with the \ac{GPU} or its surrounding hardware components as \emph{physical access-based} attacks.
Such an attacker is physically close to the device and can directly interact with it. 
This may involve placing sensors or monitoring equipment directly on or near the device to capture physical signals or emissions.
Note that we term attacks that perform monitoring or manipulation of the physical properties of the target device from a distance, without direct physical access to the device itself, as \emph{remote access-based} attacks (called hybrid attacks in~\cite{taneja2023hot}).
Such attacks use \eg wireless sensors or network-connected devices to capture signals or emissions emitted by the target device from a distance.

\subsubsection{Physical Access-based.}
\citeauthor{luo2015side}~\cite{luo2015side} and \citeauthor{gao2018electro}~\cite{gao2018electro} focus on recovering the \ac{AES} private key by exploiting physical properties.
\citeauthor{luo2015side} utilize correlation power analysis to analyze the \ac{GPU}'s power consumption (\ie power traces), necessitating the attacker to instrument the power supply.
\citeauthor{gao2018electro} sample \ac{EM} traces employing a precise trigger mechanism and establish a heuristic leakage model to exploit the concurrent \ac{MMIO} leakages in parallel scenarios. This method requires removing the \ac{GPU}'s heat sink and positioning a probe near the \ac{GPU} chip surface.

\citeauthor{hu2020deepsniffer}~\cite{hu2020deepsniffer} employ \ac{MMIO} side-channel attacks to capture memory access volume and kernel execution time by targeting \ac{CPU}-\ac{GPU} interconnections or co-locating a CUDA spy. 
This allows them to deduce the model architecture without any prior knowledge of the \ac{NN} model.
Similarly, \citeauthor{chmielewski2021reverse}~\cite{chmielewski2021reverse} use simple \ac{MMIO} analysis to recover partial information of an \ac{NN} model.
\citeauthor{maia2022can}~\cite{maia2022can} show how to recover the full model architectures of 64 \acp{NN} by exploiting the magnetic flux emanating from a \ac{GPU}'s power cable via a physical sensor.
\citeauthor{horvath2023barracuda}~\cite{horvath2023barracuda} use correlation \ac{EM} analysis to recover the weights and biases in the convolutional layers of a \ac{CNN} model.

\citeauthor{taneja2023hot}~\cite{taneja2023hot} capture side-channel leakage of frequency, power, and temperature (monitored via internal sensors) in response to the workload executed in both integrated and discrete \acp{GPU}. 
Their attacks, which observe instruction and data-dependent behaviour, require different privilege levels depending on vendors.
For instance, Apple permits unprivileged access to both, whereas Google restricts temperature readings but allows unprivileged access to power.

\subsubsection{Remote Access-based.}
\citeauthor{zhan2022graphics}\cite{zhan2022graphics} identify the \ac{DVFS} feature used in \ac{GPU} as the underlying cause of an \ac{MMIO} side-channel vulnerability.
In scenarios where a victim employs an NVIDIA or AMD \ac{GPU}, an attacker can monitor the victim's activities and discern the visited webpages (\ie a website fingerprinting attack). 
The key idea is capturing the fluctuating workload of a \ac{GPU} during webpage rendering, which influences the \ac{MMIO} signal it emits.
These workload variations, coupled with the need to manage heat generation, fan noise, and minimize power consumption, lead to changes in the \ac{GPU}'s performance levels (\eg P-states in NVIDIA \acp{GPU}).
Consequently, when a performance level switches on or off, distinct \ac{MMIO} signals corresponding to the \ac{GPU}'s memory clock frequencies appear or vanish in the spectrum. 
Remarkably, this attack can intercept \ac{MMIO} signal leakage at distances of up to 6 meters, even through a wall.
\citeauthor{liang2022clairvoyance}~\cite{liang2022clairvoyance} conduct a similar far-field \ac{MMIO} side-channel attack to infer the number of layers and their types, the number of kernels, sizes of layers and strides in an \ac{NN} model.

\noindent\textbf{Takeaway points.} 
\ding{182} \ac{MMIO} channels are a consistent and critical leakage source, acting as a bridge between CPU and GPU for both local and remote side-channel attacks.
\ding{183} Access to internal sensors (e.g., temperature, power) varies by vendor. On platforms like Apple, where such access is unprivileged, this creates opportunities for exploitation.
\ding{184} Remote physical attacks are becoming increasingly feasible, with some able to succeed from several meters away, even through walls, without privileged access. This broadens the threat model to include fully remote, passive attackers.

\subsubsection{Countermeasures}
\ac{MMIO} side channels from \acp{GPU} can be exploited when these \ac{EM} signals are dependent on computations, revealing information about ongoing activities. Such side-channel signals are easily measurable when they are strong and can propagate over several meters. 
Two countermeasures can mitigate such vulnerabilities.

\emph{(i) Hiding and masking~\cite{mangard2008power}:} Hiding aims to reduce the \ac{SNR} available to attackers. Generating \ac{MMIO} noise is possible, but shielding the computer to decrease the intensity of emitted \ac{MMIO} signals is more effective. 
Hardware manufacturers could also focus on minimizing \ac{MMIO} emissions in future product designs. 
Masking involves combining data with random mask values to ensure the observable leakage is independent of the secret. 
For example, creating random \ac{GPU} workloads during sensitive tasks can disrupt usage patterns and reduce predictability~\cite{mahmoud2022electrical,zhan2022graphics}. 
However, masking can be complicated due to the need for random masks and the added complexity of the algorithms.

\emph{(ii) Coarsening the granularity:} Another approach is to make \ac{GPU} \ac{DVFS} less sensitive to workload changes. This would keep performance more stable, leading to less detailed leaked information and reducing the chance of sensitive data being inferred~\cite{zhan2022graphics}. However, such changes might negatively affect the efficiency benefits of \ac{DVFS}.

\subsection{Fault Injection Attacks}


\subsubsection{Software-based.}
\citeauthor{frigo2018grand}~\cite{frigo2018grand} use \ac{GPU}-based timers to execute a JavaScript-based side-channel attack, enabling attackers to identify contiguous regions of physical memory. 
This method allows for remote triggering of Rowhammer bit flips in the identified memory areas. 
Consequently, an attacker can breach \ac{ASLR} and evade the Firefox sandbox on Android platforms featuring integrated Qualcomm Adreno \acp{GPU}.
\citeauthor{sabbagh2020novel}~\cite{sabbagh2020novel} introduced the first non-invasive fault attack named ``overdrive'' on a \ac{GPU}.
By exploiting the \ac{DVFS} interface, an adversary can configure out-of-specification voltage and frequency combinations on the host \ac{CPU} and send malicious overdrive commands to the \ac{GPU} control registers, inducing random faults during \ac{GPU} kernel execution. 
They also perform a misclassification attack on victim \ac{CNN} inference~\cite{sabbagh2021gpu}, improving the overdrive attack by addressing its limitations in timing precision and the lack of analysis on how fault injection affects silent data corruption.

However, unlike encryption algorithms, where hardware fault injection attacks are highly effective, \ac{ML} models, especially well-trained commercial \ac{DNN} models, are inherently more fault-tolerant. This is because computation errors from random hardware faults are usually absorbed by later layers and do not greatly impact accuracy. 
Therefore, inducing hardware faults at specific timing points, optimizing the glitch voltage-frequency pair, and determining the appropriate fault duration are critical for generating controllable and impactful faults in \ac{ML} models.
To this end, \citeauthor{sun2023lightning}~\cite{sun2023lightning} focus on searching sensitive targets of a \ac{CNN} model, identifying proper parameters for timing and position, refining fault injection parameters, and ultimately reducing inference accuracy by an average of $69.1\%$.
\citeauthor{dos2021revealing}~\cite{dos2021revealing} propose a two-level fault injection framework combining the accuracy of \ac{RTL} fault injection with the efficiency of software fault injection.

Overall, researchers have primarily leveraged software interfaces to inject faults, often utilizing \ac{DVFS} or Rowhammer attacks. 
Rowhammer attacks exploit the DRAM vulnerability where repeated accesses to a memory row (\,  i.e. hammering) can cause bit flips in adjacent rows, enabling dangerous exploits like privilege escalation from a normal user to a system administrator. 
The high parallelism and memory bandwidth inherent in \acp{GPU} potentially make them more vulnerable to Rowhammer attacks, as they can generate memory accesses faster than \acp{CPU}.

\noindent\textbf{Takeaway points.} 
\ding{182} Fault injection `can be triggered entirely through software by abusing out-of-spec voltage-frequency settings, exposing a lack of safeguards in GPU power management and its tight coupling with CPU controls.
\ding{183} While \acp{DNN} are generally resilient to random faults, targeted injection at specific layers or operations can significantly reduce accuracy. Achieving this, however, requires precise control over timing, location, and fault parameters.
\ding{184} The high memory bandwidth and parallelism of \acp{GPU} make them especially prone to Rowhammer attacks, as they can generate rapid memory access patterns that increase the risk of bit flips.

\subsubsection{Countermeasures}
Protections against Rowhammer exploits are most effective when implemented in the main memory or the memory controller. 
\citeauthor{frigo2018grand}\cite{frigo2018grand} investigate preventing attackers from targeting valuable data by enforcing stricter memory reuse policies. 
One solution involves enhancing the physical compartmentalization suggested by CATT~\cite{brasser2017can} to user-space applications. 
However, this approach brings trade-offs in complexity, performance, and capacity, since dynamically allocating (isolated) pages increases complexity and can have performance implications.

Regarding \ac{DVFS}-based attacks, it is feasible to make it more difficult for adversaries to control and exploit the interfaces. Voltage and clock signal monitors can be used to detect unconventional changes\footnote{\cf \eg Intel's approach: \url{https://www.intel.com/content/www/us/en/newsroom/news/the-story-behind-new-intel-security-feature.html}}. Additionally, some vendors have disabled all software-accessible interfaces for \ac{DVFS}. For example, Intel defends against \ac{DVFS}-based attacks on Intel \ac{SGX}~\cite{murdock2020plundervolt} with a BIOS setting to disable the overclocking mailbox interface and with a microcode update that includes the current state of this setting in the \ac{SGX} \ac{TCB} attestation. A similar strategy could also be applied to \acp{GPU}.


\section{Discussions on future directions in \ac{GPU} \ac{TEE}}
\label{sec:GPU tee attacks}
In Section~\ref{sec:cc-gpu}, we provided a taxonomy of existing \ac{GPU} \ac{TEE} designs, highlighting their architectural differences, security assumptions, and threat models.
Building on this foundation, Sections~\ref{sec:GPU-sec} and the supplemental material (due to the limited space of this article) reviewed and compared a broad range of attacks targeting standalone \acp{GPU} and established \ac{CPU} \acp{TEE} such as Intel SGX, Arm TrustZone, and AMD SEV.
In this section, we extrapolate from these findings to identify potential attack vectors that may arise in the context of \ac{CPU} \acp{TEE}, mapping them to the specific components and trust boundaries involved in these systems.
We note that as \ac{GPU} \acp{TEE} are not widely commercially available, it is not yet possible to practically evaluate these attack vectors, but we believe that according vulnerabilities might be discovered in the future.

\subsection{Potential Attacks against x86-based \ac{GPU} \ac{TEE} Prototype}

\begin{figure}
\centering
\includegraphics[width=0.6\linewidth]{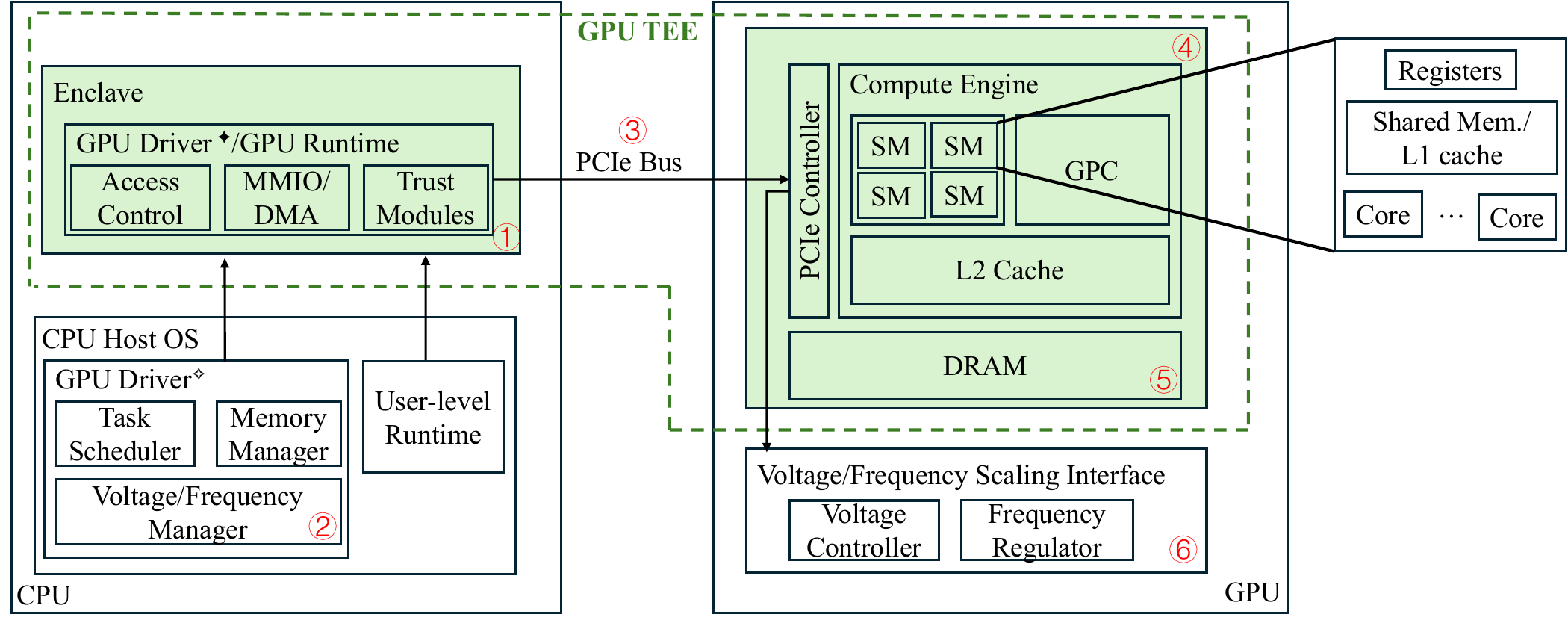}
\caption{Potential attacks against x86-based \ac{GPU} \ac{TEE} prototype.}
\label{fig:intel_gputee_attacks}
\end{figure}

As summarized in \Cref{sec:cc-gpu}, the state-of-the-art designs targeting x86 architecture each have unique characteristics aimed at reducing the \ac{TCB} size and minimizing hardware modifications.
The primary goal of these designs is to protect \ac{GPU} resources from a compromised \ac{OS}, including the \ac{GPU} driver. We outline the key design ideas and provide a general overview in \Cref{fig:intel_gputee_attacks}.
Specifically, in this threat model, the adversary controls the entire software stack, encompassing \ac{OS}, hypervisor, and device drivers. 
The adversary also aims to compromise the hardware and software I/O data path between a user application and the \ac{GPU}.
The functionalities of \ac{GPU} drivers are often partitioned into secure and non-secure parts instead of porting the entire driver into the \ac{TEE} enclave, which would increase the TCB size.
Given that the adversary can compromise the \ac{OS}, the \ac{GPU} driver located in the \ac{OS} is also within this scope.
The device memory of the \ac{GPU} is trusted and the adversary cannot observe or corrupt the data stored in it. 
Moreover, we recognize that the power and frequency management modules residing in both the \ac{CPU} and \ac{GPU} sides are outside the scope of protection.
Next, based on this security model, we will analyze potential attacks that could pose threats and leak information when such designs are deployed in real-life applications.

\subsubsection{Existing Attacks against Intel \ac{SGX}}
Given that the essential \ac{GPU} resource management functions within the \ac{GPU} driver reside within the \ac{TEE} enclave, the majority of attacks outlined in the supplemental material could potentially target this enclave to glean secrets (\cf attack surface \ding{192} in \Cref{fig:intel_gputee_attacks}).
It is worth noting that while some research may assert that most side-channel attacks against \ac{SGX} are out of scope, it is critical to provide the corresponding protection when considering the deployment of \ac{GPU} \ac{TEE} designs in real-world cloud platforms (where even some forms of physical access by malicious insiders might be relevant). 
These potential leakages can be categorized as follows:
(i) Observing memory access patterns allows attackers to deduce the control flow of programs within the \ac{SGX} enclave, including decision-making processes. Armed with this insight, attackers could monitor \ac{GPU} commands and data transfers, thereby inferring the tasks executed on the \ac{GPU}. 
For instance, it might become feasible to infer the quantity (and size) of detected objects in each frame of a \ac{CNN}-based video analysis task~\cite{poddar2020visor}.
(ii) Monitoring page fault patterns enables attackers to extract cryptographic keys from implementations of certain cryptographic open-source software libraries (\eg OpenSSL and Libgcrypt)~\cite{shinde2016preventing}. 
With this knowledge, attackers could subsequently access all sensitive information transmitted to the \ac{GPU}.
(iii) Exploiting timing channels allows attackers to infer sensitive information about kernels running on the \ac{GPU}. 
For instance, Telekine~\cite{hunt2020telekine} demonstrates how attackers can accurately classify images from ImageNet by solely observing the timing of \ac{GPU} kernel execution, without needing access to the images themselves.

\subsubsection{\ac{DVFS}-based Attacks}
To enhance energy efficiency, many current \ac{CPU} processors enable the \ac{DVFS} extension, which adjusts the frequency and voltage of processor cores based on real-time computational loads.
This capability now extends to accelerators like NVIDIA \acp{GPU}. 
As demonstrated by Lightning~\cite{sun2023lightning}, NVIDIA \ac{GPU} working voltages can be altered through software instructions executed on the \ac{CPU} via NVIDIA drivers, potentially enabling attackers to induce transient voltage glitches for fault injection attacks. 
Furthermore, current NVIDIA \acp{GPU} offer unrestricted control for \ac{PLL} and clock divider, allowing easy overclocking by adjusting the frequency offset using NVIDIA, thus potentially facilitating frequency-based fault injection attacks. 
Consequently, for a x86-based \ac{GPU} \ac{TEE} prototype, we identify two attack surfaces (\cf \ding{193} and \ding{197}) regarding \ac{DVFS}-based attacks in \Cref{fig:intel_gputee_attacks}.

\noindent\textbf{On the \ac{CPU} side.} 
In many existing \ac{GPU} \acp{TEE} prototypes~\cite{volos2018graviton,jang2019heterogeneous,zhu2020enabling,lee2022tnpu,yudha2022lite,ivanov2023sage}, the part of the \ac{GPU} driver responsible for \ac{DVFS} parameter control lies outside their \ac{TCB}.
Consequently, an attacker can flexibly adjust \ac{GPU} voltage and frequency from the \ac{CPU} side to induce faults on the \ac{GPU} side.
Additionally, it warrants investigation whether previous attacks against \ac{SGX} could gain access to the relevant memory region to alter \ac{DVFS} parameters, even if the \ac{GPU} driver managing these parameters is protected by running in an \ac{SGX} enclave.

\noindent\textbf{On the \ac{GPU} side.} 
Due to the absence of protection on \ac{GPU} hardware, attackers with physical access to \ac{GPU} hardware could directly manipulate \ac{GPU} voltage, injecting faults into running \ac{GPU} kernels. 
For instance, ``VoltPillager''~\cite{chen2021voltpillager} can inject messages on the \ac{SVID} bus between the \ac{CPU} and the voltage regulator on the motherboard, enabling precise control over \ac{CPU} core voltage. 
Ultimately, this jeopardizes the confidentiality and integrity of Intel \ac{SGX} enclaves. 
Exploring whether attackers can inject malicious \ac{SVID} packets (or similar) into the respective bus on the \ac{GPU} card presents an intriguing avenue for further investigation.

\subsubsection{Physical Attacks on \ac{PCIe} Bus and \ac{GPU} Memory}
In many existing \ac{GPU} \acp{TEE} prototypes lacking vendor hardware support (\cf attack surfaces \ding{194} and \ding{196} in \Cref{fig:intel_gputee_attacks}), \acp{GPU} lack a trusted memory region, leaving data in \ac{GPU} memory exposed in plaintext. Consequently, direct physical access to \ac{GPU} memory can compromise sensitive data. Further, cold boot attacks~\cite{Halderman08coldboot} could be relevant, if an adversary can ensure (\eg by cooling) that \ac{TEE}-protected memory contents are retained and can be readout after a reset of the \ac{GPU}.

Additionally, as \ac{PCIe} interconnects are exposed, injecting malicious \ac{PCIe} packets via specialized hardware is feasible. 
Simply securing the routing path to the \ac{GPU} is not sufficient to protect \ac{GPU} control via \ac{MMIO} or \ac{DMA}~\cite{jang2019heterogeneous}. 
While most designs employ authenticated encryption to ensure data integrity during transfer, this alone is insufficient to protect data confidentiality and integrity.
For instance, certain designs lack a hardware root of trust necessary for secure authentication and establishment of trusted communication channels, involving secure storage for cryptographic keys and secure execution environments for cryptographic operations. 
Furthermore, ensuring secure boot and firmware integrity is crucial to guarantee the correctness of the secure channel establishment process.

\subsubsection{Architectural/Logical Attacks on \ac{GPU}}
Previous attacks on memory vulnerabilities (see \Cref{sec:al attack}) are also possible, as current \ac{GPU} \ac{TEE} designs often overlook memory clearing between different \ac{GPU} contexts (which is also relevant across resets in the case of cold boot attacks mentioned above). 
Therefore, a secure memory management strategy, including memory initialization and clearance, is critical to the integrity and security of the \ac{GPU} \ac{TEE} workflow.

\subsection{Potential Attacks against NVIDIA Hopper H100-based \ac{GPU} \ac{TEE}}

\begin{figure}
\centering
\includegraphics[width=0.6\linewidth]{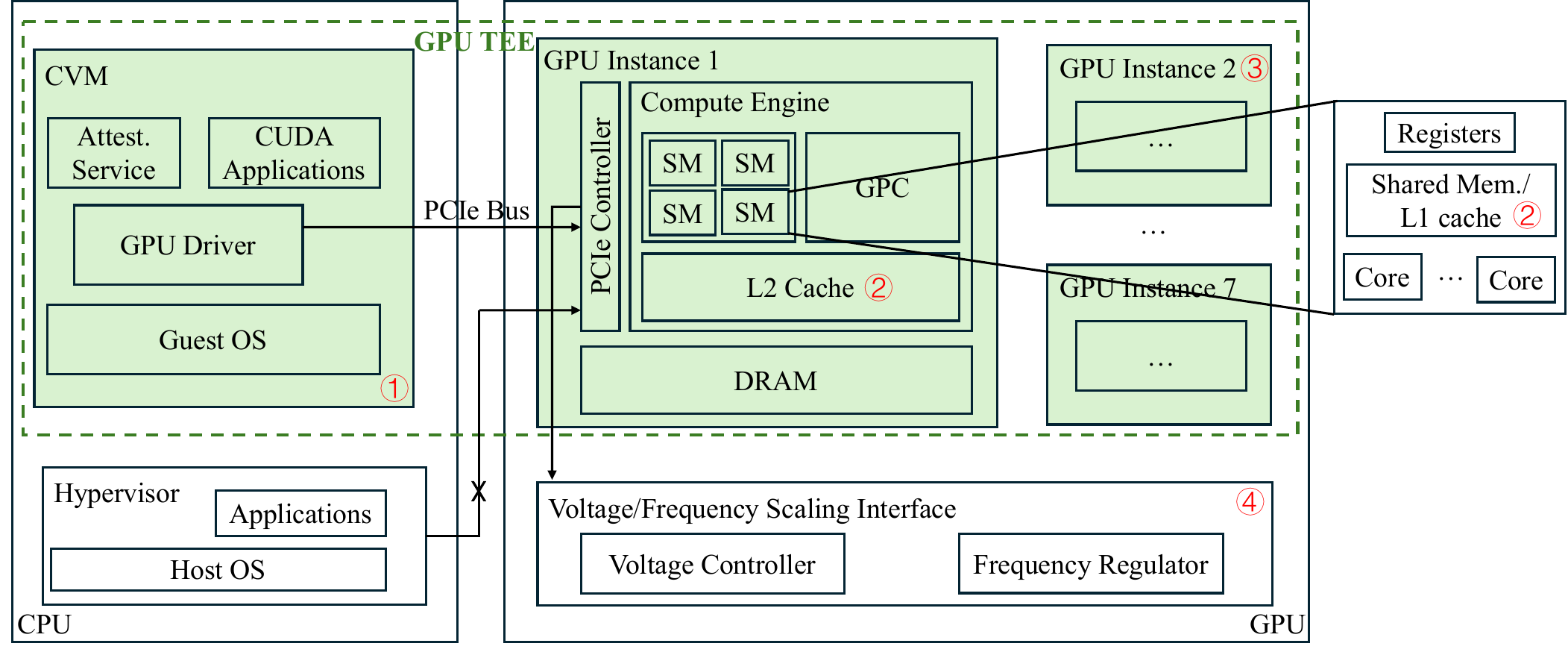}
\caption{Potential attacks against NVIDIA Hopper H100-based \ac{GPU} \ac{TEE}.}
\label{fig:nvidia_gputee_attacks}
\end{figure}

In addition to the state-of-the-art x86-based \ac{GPU} \ac{TEE} prototypes, NVIDIA’s Hopper H100 Tensor Core \ac{GPU}, based on the Hopper architecture, recently introduces advanced features for confidential computing (only available to select users at present). 
According to NVIDIA's whitepaper~\cite{NVIDIAh100}, the H100 relies on specific \ac{CPU} \acp{TEE}, such as Intel \ac{TDX}, AMD \ac{SEV}-SNP, or Arm \ac{CCA}, to enable confidential computing. 
Here, we provide a general system model of an NVIDIA Hopper H100-based \ac{GPU} \ac{TEE} based on the whitepaper in \Cref{fig:nvidia_gputee_attacks}.
Specifically, \emph{on the \ac{CPU} side}, this \ac{GPU} \ac{TEE} operates within the standard threat model of \ac{VM}-based \acp{TEE}~\cite{schluter2024wesee,schluter2024heckler,zhang2024cachewarp}. 
The untrusted hypervisor loads the \ac{VM} image into memory and manages the initial configurations. 
Remote attestation measures the \ac{VM}'s initial memory state before initiating the boot process. 
The software executing within the \ac{VM} (including the guest OS, user applications, and trusted modules for \acp{TEE}) is part of the \ac{TCB}. 
The specifications require the hypervisor to set up certain initial states (\eg the number of vCPUs, supported hardware features, and memory size). 
The hardware verifies these configurations and only enters the \ac{VM} if the setup is correct. 
Additionally, the hardware zeroes out certain values (\eg specific general-purpose registers) before exiting the \ac{VM} and manages context saving and restoring during \ac{VM} context switches.
\emph{On the \ac{GPU} side}, the H100 \ac{GPU} is expected to feature a memory encryption engine (most likely an \ac{MEE}~\cite{intelMEE} similar to \ac{SGX}) and other capabilities to ensure the confidentiality and integrity of data and code, even defending against basic physical attacks targeting the \ac{PCIe} bus and \ac{GPU} memory.
However, the whitepaper does not provide detailed technical information about these features.

\subsubsection{Existing Attacks against \ac{VM}-based \acp{TEE}}
Perhaps because \ac{VM}-based \acp{TEE} have only been proposed in recent years, there is currently limited research on attacks against them.
However, recent developments have shown that software-based fault injection attacks can be highly effective against AMD \ac{SEV}-\ac{SNP}. 
These attacks would allow an adversary to compromise the \ac{GPU} driver within the \ac{VM}, enabling precise control over sensitive tasks running on the \ac{GPU} (\cf attack surface \ding{192} in \Cref{fig:nvidia_gputee_attacks}).
For example, the ``WeSee'' attack~\cite{schluter2024wesee} targets AMD \ac{SEV}-\ac{SNP} used in NVIDIA H100-based \ac{GPU} \acp{TEE}, allowing an attacker to perform arbitrary memory writes by faking a $\mathtt{\#VC}$ for \ac{MMIO} reads. 
This capability would enabls the attacker to modify commands and data sent to the \ac{GPU}.
More critically, attacks like CacheWarp~\cite{zhang2024cachewarp}, which remain unmitigated in the latest \ac{SEV}-\ac{SNP} implementation, do not depend on the specifics of the guest \ac{VM}. 
Therefore, it is crucial to equip NVIDIA H100-based \ac{GPU} \acp{TEE} with protection against these attacks to preserve the confidentiality and integrity of \ac{GPU} data. 
It is also important to explore potential attacks against other \ac{VM}-based \acp{TEE}, such as Intel \ac{TDX}, as these are integral to NVIDIA H100-based \ac{GPU} \acp{TEE}.

\subsubsection{\ac{DVFS}-based Attacks}
Similar to \ac{GPU} \ac{TEE} prototypes, we identify two attack surfaces (\cf \ding{192} and \ding{195}) regarding \ac{DVFS}-based attacks in \Cref{fig:nvidia_gputee_attacks}.

\noindent\textbf{On the \ac{CPU} side.} 
Recall that existing attacks targeting \ac{VM}-based \acp{TEE} (\eg AMD \ac{SEV}-\ac{SNP}) allow attackers to escalate privileges to root and write any value of their choice to any location in the victim \ac{VM}.
Thus, attackers with these abilities might be able to manipulate the \ac{GPU}'s voltage and frequency using software instructions performed on the \ac{CPU} via the NVIDIA driver.

\noindent\textbf{On the \ac{GPU} side.} 
The H100 whitepaper~\cite{h100architecture} lacks detail about voltage and frequency management.
Assuming an attacker has physical access, they could potentially execute attacks like ``VoltPillager''~\cite{chen2021voltpillager} to inject malicious \ac{SVID} packets or equivalent into the relevant \ac{GPU} card's bus, thereby altering the \ac{GPU} voltage.

In summary, an attacker with the described capabilities could modify the voltage and frequency from both the \ac{CPU} and \ac{GPU} sides. 
However, it remains uncertain whether and how these changes affect tasks running on the NVIDIA H100 \ac{GPU}. 
Specifically, it is unclear if the induced faults significantly impact tasks, such as degrading \ac{ML} model accuracy or fully extracting private encryption keys. 
At the time of writing this survey, the complete confidential workflow of the NVIDIA H100-based \ac{GPU} \ac{TEE} is unavailable, as the corresponding NVIDIA libraries are still under development.

\subsubsection{Microarchitectural Side-Channel and Covert-Channel Attacks on \ac{GPU}}
The  H100 whitepaper~\cite{h100architecture} emphasizes the enhanced security features of the Hopper architecture for confidential computing, particularly in collaborative multi-party computing scenarios, namely \emph{single tenant with multiple \acp{GPU}} and \emph{multiple tenants with a single \ac{GPU}}.
However, vulnerabilities discussed in \Cref{sec:micro_sca_cca-attack} may potentially lead to security breaches on the NVIDIA H100 \ac{GPU}.

\noindent\textbf{Single tenant with multiple \acp{GPU}.} 
In the scenario of single tenant with multiple \acp{GPU}, a single \ac{VM} is concurrently assigned multiple physically independent \acp{GPU}, each connected via NVLink. 
Recall that \citeauthor{dutta2023spy}~\cite{dutta2023spy} demonstrate that an attacker on one \ac{GPU} could induce contention on the L2 cache of another \ac{GPU} via NVLink, enabling a covert channel and Prime+Probe attacks across \acp{GPU}. However, such attacks may face challenges in the NVIDIA H100 \ac{GPU} model due to two key factors (\cf attack surface \ding{193} in \Cref{fig:nvidia_gputee_attacks}). 
Firstly, the attacker's process needs access to a remote \ac{GPU}'s memory through NVLink, which may be restricted in the threat model of the NVIDIA H100 \ac{GPU} \ac{TEE}. Thus, the challenge lies in how to enable the spy process to run within one of the assigned \acp{GPU} and have access to another \ac{GPU}'s memory.
Secondly, the H100 \ac{GPU} features a 50\,MB L2 cache, $1.25{\times}$ larger than the A100 \ac{GPU}. Nevertheless, a larger L2 cache with more sets and ways can make Prime+Probe attacks more difficult, because the attacker needs to fill and monitor a larger number of cache lines.

\noindent\textbf{Multiple tenants with a single \ac{GPU}.} 
In the case of multiple tenants with a single \ac{GPU} (\cf attack surface \ding{194} in \Cref{fig:nvidia_gputee_attacks}), facilitated by the \ac{MIG} feature, an NVIDIA H100 \ac{GPU} can be partitioned into multiple \ac{GPU} instances, each isolated in terms of memory system paths, including the on-chip crossbar ports, L2 cache banks, memory controllers, and DRAM address busses.
However, as discussed in \Cref{sec:micro_sca_cca-attack}, \citeauthor{zhang2023t}~\cite{zhang2023t} demonstrate a design flaw of the \ac{MIG} feature where the last-level \ac{TLB} is not securely partitioned by all \ac{GPU} instances. 
Consequently, this flaw enables covert-channel attacks between different \ac{GPU} instances. 
Given that this mode supports multiple tenants, it is pertinent to investigate whether each \ac{GPU} instance created with \ac{MIG} truly has separate and isolated paths through other microarchitectural components such as L1 cache, warp scheduler, and command processor.

\subsection{Potential Attacks against Arm-based \ac{GPU} \ac{TEE} Prototype}

\begin{figure}
\centering
\includegraphics[width=0.7\linewidth]{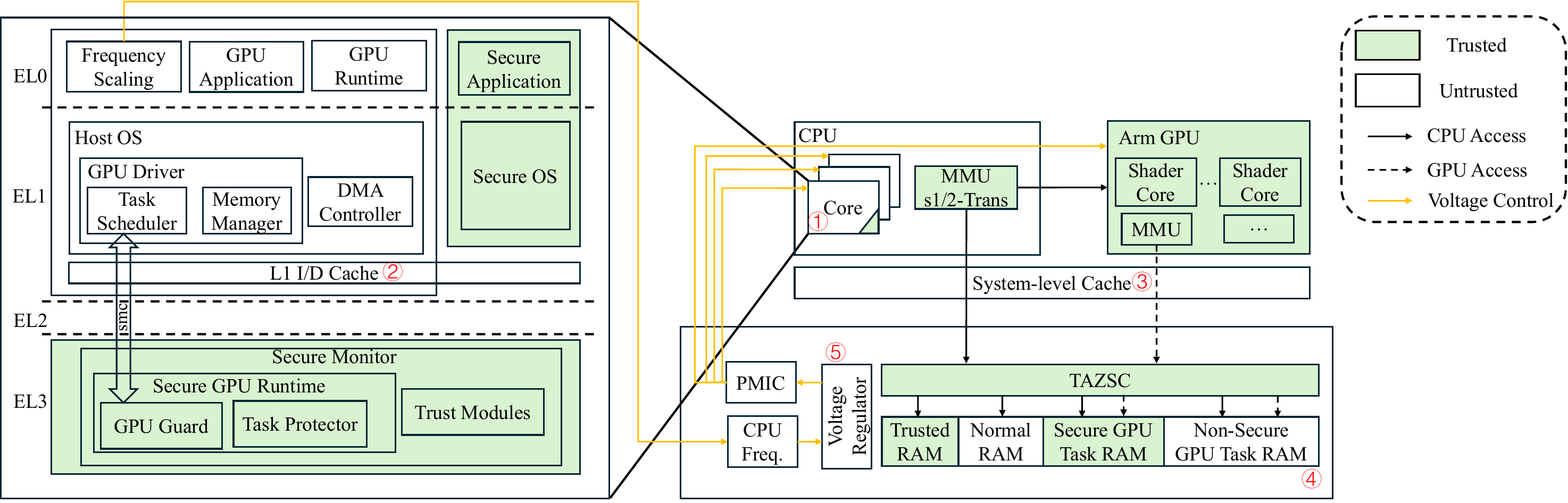}
\caption{Potential attacks against Arm-based \ac{GPU} \ac{TEE}.}
\label{fig:arm_gputee_attacks}
\end{figure}

Due to various architectural disparities in  \acp{TEE} for Arm \acp{GPU}, several studies have explored Arm-based \ac{GPU} \acp{TEE}. 
In \Cref{fig:arm_gputee_attacks}, we base it on the design proposed by StrongBox~\cite{deng2022strongbox}. 
StrongBox serves as a suitable foundation as it was recently introduced and does not rely on specific Arm endpoint features or necessitate hardware modifications to \ac{GPU} or \ac{CPU} chips.
Notably, while secure \ac{GPU} computing on Arm or Intel platforms may not align with Arm's \ac{CCA} realm-style architecture~\cite{wang2024cage}, we omit insights regarding this, as it closely resembles that of NVIDIA Hopper H100-based \ac{GPU} \ac{TEE}.
In \Cref{fig:arm_gputee_attacks}, the attacker gains control over the kernel and the entire \ac{GPU} software stack, including the \ac{GPU} driver and runtime.
This control enables them to manipulate sensitive data and code within \ac{GPU} applications. 
They can directly access unified memory used for \ac{GPU} tasks or manipulate peripherals to evade detection. Furthermore, by compromising the \ac{GPU} driver, the attacker can undermine the memory management of \ac{GPU} applications, potentially mapping sensitive data to unprotected regions. 
Similar to existing x86-based \ac{GPU} \acp{TEE}, trust is placed in the \ac{GPU} and the secure world. 
Based on this threat model, we discuss potential attacks in the following.

\subsubsection{Existing Attacks against Arm \ac{TZ}}
The attacks summarized in the supplemental material that target \ac{TZ} are also applicable to this Arm-based \ac{GPU} \ac{TEE} prototype. 
The impact of these side-channel attacks on \ac{TZ} indirectly affects Arm-based \ac{GPU} \ac{TEE} (\cf attack surfaces \ding{192} and \ding{193} in \Cref{fig:arm_gputee_attacks}). The potential consequences can be summarized as follows:
(i) An attacker could escalate privileges to root to submit numerous malicious tasks~\cite{beniamini2016trustzone,shen2015exploiting}. However, establishing covert channels between malicious and victim tasks is unlikely, as \ac{GPU} tasks are executed sequentially on Arm endpoint \acp{GPU}~\cite{deng2022strongbox}.
(ii) An attacker may monitor cache activity to extract \ac{AES} keys from \ac{TZ}~\cite{guanciale2016cache,zhang2016truspy,lipp2016armageddon}, thus compromising the confidentiality of data transferred between the \ac{CPU} and \ac{GPU}.

\subsubsection{Microarchitectural Side-Channel and Covert-Channel Attacks on Integrated \ac{GPU}}
As summarized in \Cref{tab:GPU attacks}, several studies have investigated side-channel and covert-channel attacks on integrated \acp{GPU}, with some of these potentially applicable to Arm-based \ac{GPU} \acp{TEE} prototype (\cf attack surface \ding{194} in \Cref{fig:arm_gputee_attacks}).

(i) Covert channels between \ac{CPU} and iGPU: \citeauthor{dutta2021leaky}~\cite{dutta2021leaky} establish bidirectional covert channels between a trojan process on the \ac{CPU} and a spy process on the Intel integrated \ac{GPU} (vice versa) via the \ac{LLC}. 
Considering this attack in an Arm-based \ac{GPU} \ac{TEE}, a possible scenario is that of using a trojan process launched on the normal world of the \ac{CPU}, communicating the bits to the \ac{GPU}. However, challenges arise due to undocumented \ac{LLC}/system-level cache behavior and the asymmetric view of cache utilization between the \ac{CPU} and \ac{GPU} in the Arm architecture.

(ii) The existing threat model of Arm-based \ac{GPU} \acp{TEE} may not effectively mitigate attacks exploiting data-dependent DRAM traffic and cache utilization induced by compression to reveal user pixels in the normal world~\cite{wang2024gpu}. Further testing is needed to assess the success rate of such attacks on Arm-based \ac{GPU} \acp{TEE}.

\subsubsection{Physical Attacks on \ac{GPU} Memory}
Most existing Arm-based \ac{GPU} \ac{TEE} prototypes primarily focus on delivering hardware-enforced isolation to protect sensitive code and data, rather than employing memory encryption mechanisms like Intel \ac{SGX}'s \ac{MEE}. 
Consequently, direct physical access to \ac{GPU} memory (again also possibly via cold boot attacks~\cite{Halderman08coldboot}) could potentially compromise sensitive data (\cf attack surface \ding{195} in \Cref{fig:arm_gputee_attacks}). 
For instance, in StrongBox~\cite{deng2022strongbox}, a region in the unified memory defined as Secure Task RAM serves as a fixed and non-secure memory region reserved for confidential \ac{GPU} applications. Its main purpose is to dynamically allocate memory and establish \ac{GPU} page table mappings and thus prevent unauthorized access control to this region.

\subsubsection{\ac{DVFS}-based Attacks}
In contrast to the \ac{DVFS} in NVIDIA \acp{GPU}, the voltage and frequency of an Arm \ac{GPU} (\eg Mali series) are typically controlled by the \ac{PMIC}. 
This component is responsible for managing power distribution to various system elements, including both \ac{CPU} and \ac{GPU}.
Thus, as depicted in \Cref{fig:arm_gputee_attacks} (\cf attack surface \ding{196}), attackers with privileges can potentially induce hardware faults by manipulating processor voltage and frequency settings through the \ac{PMIC}. 
Additionally, assuming an attacker gains physical access to the voltage regulator or clock control module within the \ac{SoC}, this prompts further investigation into whether hardware-based attacks targeting these components~\cite{chen2021voltpillager} are also applicable to Arm \acp{GPU}.


\subsection{Comparison of Attacks on Different GPU TEEs}
x86-based GPU TEE prototypes, NVIDIA Hopper H100, and Arm-based GPU TEEs adopt distinct trust models, hardware assumptions, and protection strategies.
x86-based designs (\eg Graviton, HIX, HETEE) typically build on Intel SGX, placing only essential GPU driver components inside the enclave to minimize the TCB. 
These prototypes often lack GPU-side memory encryption and leave out \ac{DVFS} and power management from protection, making them susceptible to software-based fault injection and physical attacks (\eg cold boot, SVID manipulation).
In contrast, NVIDIA Hopper H100 integrates vendor-supported confidential computing features, including memory encryption and compatibility with VM-based TEEs. 
While this offers system-wide isolation, it inherits known vulnerabilities of VM-based TEEs, such as software-based fault injection (\eg WeSee) and cache manipulation (\eg CacheWarp). 
Furthermore, the limited public documentation presents challenges for third-party validation and systematic defense design.
Arm-based GPU TEE designs, such as StrongBox, use TrustZone to provide software-based isolation without requiring hardware modifications. 
This approach improves portability and ease of deployment but sacrifices granularity. 
These systems typically lack hardware-enforced memory encryption and rely on unified memory and shared caches, which can expose sensitive data to physical and side-channel attacks. 
Operating in resource-constrained environments (\eg edge devices), Arm-based TEEs tend to have a smaller TCB but may not adequately defend against \ac{DVFS} abuse or microarchitectural leakage.

In summary, existing x86-based prototypes prioritize minimal hardware changes but depend on SGX’s integrity; 
NVIDIA H100 adopts holistic, hardware-enhanced isolation but inherits the complexity and weaknesses of VM-based TEEs; 
and Arm-based TEEs favor lightweight integration at the cost of fine-grained protection. 
From an attack perspective, x86 prototypes are vulnerable at the interface between CPU TEEs and unprotected GPUs, NVIDIA’s model is exposed through VM-level fault channels and microarchitectural sharing, and Arm designs are weakest at the hardware and cache/memory system level.  
These architectural differences reflect distinct trade-offs and highlight the need for architecture-aware defenses tailored to each GPU TEE design.

\section{Conclusion}
With the proliferation of heterogeneous systems integrating \acp{GPU}, \ac{GPU} \acp{TEE} are emerging as effective solutions for ensuring confidentiality and integrity.
In this survey, we first examined existing \ac{GPU} \acp{TEE} prototypes and NVIDIA's first commercial \ac{GPU} \ac{TEE}, providing key insights into their design implications and characteristics. 
We reviewed recent attacks extended across various types of \acp{GPU}, including integrated \acp{GPU}, single \acp{GPU}, and multiple \acp{GPU}, along with the efforts to mitigate these threats.
Additionally, we summarized popular attacks against common \ac{CPU} \acp{TEE} such as Intel SGX, Arm \ac{TZ}, and AMD \ac{SEV}.
Based on these insights, we explored potential attacks at both physical and software levels on x86-based and Arm-based \ac{GPU} \acp{TEE} prototypes, as well as the NVIDIA H100 \ac{GPU}, highlighting the need for careful analysis of the already-complex attack surface. 
With innovations in \ac{GPU} architecture and isolation mechanisms in \ac{GPU} \acp{TEE}, this attack surface will further expand, necessitating the evolution of secure designs and defence strategies as these systems are widely deployed across computing domains.

\section*{Acknowledgements}
This research is partially funded by the Engineering and Physical Sciences Research Council (EPSRC) under grants EP/X03738X/1, EP/V000454/1, and EP/R012598/1. The results feed into DsbDtech.


\bibliographystyle{ACM-Reference-Format}
\bibliography{ref}

\newpage
\appendix

\section{Security of \ac{CPU} \acp{TEE}}
\label{sec:CPU tee attacks}
In this online appendix, we introduce typical attacks against \ac{CPU} \acp{TEE}. As there already exist survey papers focusing on reviewing existing attacks on \ac{CPU} \acp{TEE}, and the aim of this survey is to explore potential attacks on \ac{GPU} \acp{TEE}, we simply categorize the most common attacks based on their technical details and existing surveys~\cite{fei2021security,nilsson2020survey,cerdeira2020sok,li2022understanding} here.
We provide a brief summary of these attacks regarding attack surfaces and countermeasures in \cref{tab:CPU tee attacks}.

\begin{table}[htbp]
\tiny
\centering
\caption{Summary of attacks on \ac{CPU} \acp{TEE}} \label{tab:CPU tee attacks}
    
\begin{tabular}{p{0.5cm}p{0.7cm}p{0.7cm}p{0.5cm}p{1.2cm}p{0.1cm}p{0.1cm}p{0.1cm}p{0.3cm}p{0.1cm}p{0.1cm}p{0.1cm}p{0.1cm}p{0.1cm}p{0.1cm}p{0.1cm}p{4cm}}

\toprule

\multicolumn{3}{l}{\textbf{CPU TEE}} & \multirow{3}{*}{\textbf{Attacks}} & \multirow{3}{*}{\textbf{Ref}} & \multicolumn{11}{l}{\textbf{Attack Surfaces}} & \multirow{3}{*}{\textbf{Countermeasures}} \\ \cline{1-3} \cline{6-16}
\multirow{2}{*}{\textbf{Arch.}} & \multicolumn{2}{l}{\textbf{Threat Model}} &  &  & \multirow{2}{*}{\textbf{PT}} & \multirow{2}{*}{\textbf{SE}} & \multirow{2}{*}{\textbf{CC}} & \multirow{2}{*}{\textbf{DRAM}} & \multirow{2}{*}{\textbf{BTB}} & \multirow{2}{*}{\textbf{PHT}} & \multirow{2}{*}{\textbf{RSB}} & \multirow{2}{*}{\textbf{IFC}} & \multirow{2}{*}{\textbf{HC}} & \multirow{2}{*}{\textbf{Bug}} & \multirow{2}{*}{\textbf{INT}} &  \\ \cline{2-3}
 & \textbf{Trusted} & \textbf{Untrusted} &  &  &  &  &  &  &  &  &  &  &  &  &  &  \\ \hline
\multirow{17}{*}{SGX} & \multirow{17}{*}{\makecell[l]{CPU\\ package,\\ EPC,\\ MEE}} & \multirow{17}{*}{OS/Kernel} & AT & \cite{xu2015controlled} & \CIRCLE & \Circle & \Circle & \Circle & \Circle & \Circle & \Circle & \Circle & \Circle & \Circle & \Circle & \multirow{3}{*}{Obfuscation, randomization \cite{brasser2019dr,ahmad2019obfuscuro}} \\
 &  &  &  & \cite{van2017telling,kim2019sgx} & \CIRCLE & \Circle & \Circle & \Circle & \Circle & \Circle & \Circle & \Circle & \Circle & \Circle & \Circle &  \\
 &  &  &  & \cite{gyselinck2018off} & \Circle & \CIRCLE & \Circle & \Circle & \Circle & \Circle & \Circle & \Circle & \Circle & \Circle & \Circle &  \\ \cline{4-17} 
 &  &  & \multirow{2}{*}{Cache} & \cite{brasser2017software,dall2018cachequote,gotzfried2017cache,moghimi2017cachezoom} & \Circle & \Circle & \CIRCLE & \Circle & \Circle & \Circle & \Circle & \Circle & \Circle & \Circle & \Circle & \multirow{2}{*}{Obfuscation, randomization, anomaly detection \cite{brasser2019dr,ahmad2019obfuscuro}} \\
 &  &  &  & \cite{schwarz2017malware} & \Circle & \Circle & \CIRCLE & \Circle & \Circle & \Circle & \Circle & \Circle & \Circle & \Circle & \Circle &  \\ \cline{4-17} 
 &  &  & \multirow{5}{*}{BP} & \cite{lee2017inferring} & \Circle & \Circle & \Circle & \Circle & \CIRCLE & \Circle & \Circle & \Circle & \Circle & \Circle & \Circle & \multirow{5}{*}{Obfuscation \cite{ahmad2019obfuscuro}} \\
 &  &  &  & \cite{chen2019sgxpectre} & \Circle & \Circle & \CIRCLE & \Circle & \CIRCLE & \Circle & \Circle & \Circle & \Circle & \Circle & \Circle &  \\
 &  &  &  & \cite{evtyushkin2018branchscope,huo2020bluethunder} & \Circle & \Circle & \Circle & \Circle & \Circle & \CIRCLE & \Circle & \Circle & \Circle & \Circle & \Circle &  \\
 &  &  &  & \cite{koruyeh2018spectre} & \Circle & \Circle & \CIRCLE & \Circle & \Circle & \Circle & \CIRCLE & \Circle & \Circle & \Circle & \Circle &  \\
 &  &  &  & \cite{van2018foreshadow} & \Circle & \Circle & \CIRCLE & \Circle & \Circle & \Circle & \Circle & \Circle & \Circle & \Circle & \Circle &  \\ \cline{4-17} 
 &  &  & \multirow{2}{*}{TE} & \cite{van2018foreshadow} & \CIRCLE & \Circle & \CIRCLE & \Circle & \Circle & \Circle & \Circle & \Circle & \Circle & \Circle & \Circle & \multirow{2}{*}{Microcode updates from Intel} \\
 &  &  &  & \cite{van2020sgaxe,ragab2021crosstalk} & \Circle & \Circle & \CIRCLE & \Circle & \Circle & \Circle & \Circle & \Circle & \Circle & \Circle & \Circle &  \\ \cline{4-17} 
 &  &  & \multirow{3}{*}{SwFIA} & \cite{murdock2020plundervolt,kenjar2020v0ltpwn,qiu2020voltjockey} & \Circle & \Circle & \Circle & \Circle & \Circle & \Circle & \Circle & \CIRCLE & \Circle & \Circle & \Circle & \multirow{3}{*}{\makecell[l]{Microcode updates from Intel, restricted access to\\ registers, ECC}} \\
 &  &  &  & \cite{lipp2021platypus,wang2022hertzbleed,chen22throttling} & \Circle & \Circle & \Circle & \Circle & \Circle & \Circle & \Circle & \CIRCLE & \Circle & \Circle & \Circle &  \\
 &  &  &  & \cite{jang2017sgx} & \Circle & \Circle & \Circle & \CIRCLE & \Circle & \Circle & \Circle & \Circle & \Circle & \Circle & \Circle &  \\ \cline{4-17} 
 &  &  & Hw & \cite{chen2021voltpillager,Chen23pmfault,lee20membuster} & \Circle & \Circle & \Circle & \Circle & \Circle & \Circle & \Circle & \Circle & \CIRCLE & \Circle & \Circle & Fuse-level lockouts, access pattern obfuscation, harden BMC/PMU firmware \\ \cline{4-17} 
 &  &  & Interface & \cite{van2019tale,alder2020faulty,cloosters2020teerex,khandaker2020coin,alder2024pandora} & \Circle & \Circle & \Circle & \Circle & \Circle & \Circle & \Circle & \CIRCLE & \Circle & \Circle & \Circle & State and register sanitization, minimize the ABI surface, automated runtime analysis \\ 
 
 \midrule
 
\multirow{4}{*}{TrustZone} & \multirow{4}{*}{\makecell[l]{SW,\\ Secure\\ monitor}} & \multirow{4}{*}{NW OS} & IB & \cite{beniamini2016trustzone,shen2015exploiting,Danielunbox} & \Circle & \Circle & \Circle & \Circle & \Circle & \Circle & \Circle & \Circle & \Circle & \CIRCLE & \Circle & Software-only verification \cite{ferraiuolo2017komodo,jin2022mipe} \\ \cline{4-17} 
 &  &  & Cache & \cite{zhang2016cachekit,guanciale2016cache,lipp2016armageddon,zhang2016truspy,cho2018prime+} & \Circle & \Circle & \CIRCLE & \Circle & \Circle & \Circle & \Circle & \Circle & \Circle & \Circle & \Circle & Cache isolation \cite{dessouky2021chunked} \\ \cline{4-17}
 &  &  & BP & \cite{ryan2019hardware} & \Circle & \Circle & \Circle & \Circle & \CIRCLE & \Circle & \Circle & \Circle & \Circle & \Circle & \Circle & Flush BTB \\ \cline{4-17} 
 &  &  & SwFIA & \cite{tang2017clkscrew,qiu2020voltjockey,Pierrerowhammer} & \Circle & \Circle & \Circle & \Circle & \Circle & \Circle & \Circle & \CIRCLE & \Circle & \Circle & \Circle & Restrict access to DVFS interfaces \\ 

\midrule
 
\multirow{5}{*}{SEV} & \multirow{5}{*}{\makecell[l]{PSP,\\ MEE}} & \multirow{5}{*}{\makecell[l]{OS/\\Hypervisor}} & \multirow{3}{*}{ME} & \cite{hetzelt2017security,buhren2017fault,du2017secure,morbitzer2018severed,wilke2020sevurity,li2021cipherleaks} & \Circle & \Circle & \Circle & \Circle & \Circle & \Circle & \Circle & \Circle & \Circle & \CIRCLE & \Circle & \multirow{3}{*}{Mitigated in AMD SEV-SNP} \\
 &  &  &  & \cite{morbitzer2018severed} & \CIRCLE & \Circle & \Circle & \Circle & \Circle & \Circle & \Circle & \Circle & \Circle & \Circle & \Circle &  \\
 &  &  &  & \cite{li2021crossline} & \Circle & \Circle & \CIRCLE & \Circle & \Circle & \Circle & \Circle & \Circle & \Circle & \Circle & \Circle &  \\ \cline{4-17} 
 &  &  & SwFIA & \cite{morbitzer2021severity, zhang2024cachewarp} & \Circle & \Circle & \Circle & \Circle & \Circle & \Circle & \Circle & \CIRCLE & \Circle & \Circle & \Circle & Microcode update from AMD \\ \cline{4-17} 
 &  &  & ID & \cite{schluter2024wesee,schluter2024heckler,wilke2024sev} & \Circle & \Circle & \Circle & \Circle & \Circle & \Circle & \Circle & \Circle & \Circle & \Circle & \CIRCLE & Isolate interrupt handlers, detect single-stepping \\ 

 \bottomrule
        
\end{tabular}

\begin{flushleft}
        \textbf{Attacks} contains Address Translation (AT), Cache, Branch Prediction (BP), Transient Execution (TE), Software-based Fault Injection (SwFIA), Interface, Hardware (Hw), Implementation Bug (IB), Memory Encryption (ME), Interrupt-driven (ID). 
        \textbf{Attack Surfaces} includes Page Table (PT), Segmentation (SE), CPU Cache (CC), Branch Target Buffer (BTB), Page History Table (PHT), Return Stack Buffer (RSB), Interface (IFC, containing \ac{DVFS} interfaces, enclave interfaces, \etc), Interrupt (INT, interrupt controller/interrupt handling logic).
\end{flushleft}

\end{table}

\subsection{Security of Intel \ac{SGX}}
\label{sec:sgx_attack}

\subsubsection{Controlled Channel and Address Translation-based Attacks}
Controlled channels attack, a concept introduced by \citeauthor{xu2015controlled}~\cite{xu2015controlled} in 2015, exploits the privileges of the \ac{OS} to monitor and manipulate page tables. This allows the attacker to uncover the memory access patterns of the enclave code, \cf also \cite{shinde2016preventing}. 
This attack is especially effective because, while Intel \ac{SGX} enclaves protect sensitive computations from a compromised \ac{OS}, the \ac{OS} still manages memory and can observe page faults, thereby inferring details about the enclave's operations. 
SGX-Step~\cite{van2017sgx} serves as a practical tool for, among others, executing such controlled channel attacks by configuring \ac{APIC} timers, issuing interrupts, and tracking page-table entries. SGX-Step enables deterministic single-stepping of enclave execution, allowing for precise monitoring of memory access patterns over time.

In fault-less page table attacks~\cite{kim2019sgx,van2017telling,wang2017leaky}, attackers exploit the page table itself rather than relying on page faults. This allows them to infer the target's memory access patterns by examining page table attributes and/or observing the caching behaviour of unprotected page table memory. 
However, these attacks are limited to page-level accuracy. 
Segmentation attacks~\cite{gyselinck2018off} allow attackers to deduce finer memory access patterns by manipulating the segmentation unit, but this only works on 32-bit enclaves, as segmentation is disabled in 64-bit enclaves~\cite{fei2021security}.

\subsubsection{Cache-based Attacks}
Typical cache timing-channel attacks (in the non-SGX scenario) comprise four primary variants, including \emph{Evict+Time}~\cite{tromer2010efficient}, \emph{Evict+Reload}~\cite{gruss2015cache}, \emph{Prime+Probe}~\cite{osvik2006cache}, and \emph{Flush+Reload}~\cite{yarom2014flush}, all relying on the time difference between cache hits and misses.
Recent studies~\cite{gotzfried2017cache,moghimi2017cachezoom,schwarz2017malware,brasser2017software,dall2018cachequote,moghimi2019memjam} show that \ac{SGX} enclaves are susceptible to the same cache attacks as any other software application.
In fact, they may be even more vulnerable due to the enhanced attackers within \ac{SGX}'s threat model.
Several papers~\cite{gotzfried2017cache,moghimi2017cachezoom,schwarz2017malware,brasser2017software} exploit L1 cache and the shared \ac{LLC} as the attack surface by employing Prime+Probe method.
\citeauthor{dall2018cachequote}~\cite{dall2018cachequote} take a slightly different approach by using Prime+Probe on Intel's provisioning enclave, allowing Intel themselves to compromise the \ac{EPID} unlinkability property.
Memjam~\cite{moghimi2019memjam} utilizes read-after-write false dependencies caused by the 4K aliasing of the L1 cache to conduct an Evict+Time-style attack.

\subsubsection{Branch Prediction Attacks}
Branch prediction attacks offer a more detailed level of utilization compared to address translation-based and cache-based attacks. 
They target the branch prediction unit, including the \ac{BTB}~\cite{lee2017inferring,chen2019sgxpectre}, the \ac{PHT}~\cite{evtyushkin2018branchscope,huo2020bluethunder}, and the \ac{RSB}~\cite{koruyeh2018spectre}.
Attackers exploit the \ac{BTB} to ascertain whether a particular branch instruction is taken by the victim. 
The \ac{PHT} attack seeks to provoke collisions between branches of victim and attacker processes and subsequently infer the direction of the victim process branch. 
\ac{RSB} attacks rely on contaminating the \ac{RSB} to incite mis-speculation, consequently resulting in the leakage.

\subsubsection{Transient Execution Attacks}
Transient execution attacks, which exploit out-of-order and speculative execution of instructions, have also been studied in the context of \ac{SGX}.
The aforementioned \ac{BTB} and \ac{RSB} attacks~\cite{chen2019sgxpectre,koruyeh2018spectre} can also fall into this class of attacks, stemming from Spectre~\cite{kocher2020spectre}.
Additionally, the Foreshadow attack~\cite{van2018foreshadow}, which targets \ac{SGX} by employing Meltdown-like techniques, breaches enclave execution confidentiality.
While \ac{MDS} attacks~\cite{canella2019fallout,van2019ridl,schwarz2019zombieload} also leverage speculative and out-of-order execution, they primarily exploit information leakage from various implementation-specific and undocumented intermediary buffers of the targeted microarchitecture. 
SGAxe~\cite{van2020sgaxe} and Crosstalk~\cite{ragab2021crosstalk} extend \ac{MDS}  to extract private keys from \ac{SGX} enclaves.

\subsubsection{Software-based Fault Injection and Side-Channel Attacks}
In fault injection attacks, the attacker disrupts the normal operation of a system by introducing malicious faults and then derives secret information from the faulty execution. 
Software-based fault attacks expand the threat model from a local attacker, who requires physical access to the target device, to a potentially remote attacker with only local code execution capabilities.
Several studies~\cite{murdock2020plundervolt,kenjar2020v0ltpwn,qiu2020voltjockey} exploit privileged \ac{DVFS} on x86 \acp{CPU} to reliably corrupt enclave computations, thereby compromising the confidentiality and integrity of enclaves. \citeauthor{Chen23pmfault}~\cite{Chen23pmfault} show that software fault attacks on \ac{SGX} can also be mounted by pivoting to management chips on Supermicro server systems, and can even permanently damage the host \acp{CPU}.
Apart from voltage scaling-based attacks, SGX-Bomb~\cite{jang2017sgx} leverages the Rowhammer bug to induce bit flips in enclave memory, resulting in failed data integrity checks and subsequent system lock-down.

In addition to these fault injection attacks,  \citeauthor{lipp2021platypus}~\cite{lipp2021platypus} exploits a side-channel attack by leveraging unprivileged access to the Intel \acp{RAPL} interface. This interface reveals values linked to power consumption, creating a power side channel exploitable across Intel server, desktop, and laptop \acp{CPU}. \citeauthor{wang2022hertzbleed}~\cite{wang2022hertzbleed} show that power leakage can be measured through indirect channels such as frequency scaling, and \citeauthor{chen22throttling}~\cite{chen22throttling} use the related feature of throttling to among others leak secrets from \ac{SGX} enclaves.

\subsubsection{Hardware-based Attacks}
Apart from software-based fault injection attacks, \citeauthor{chen2021voltpillager}~\cite{chen2021voltpillager,Chen23pmfault} explore hardware-level fault injection attacks on \ac{SGX} enclaves.
``VoltPillager'' is a low-cost tool that injects messages on the \ac{SVID} bus between the \ac{CPU} and the voltage regulator, breaching the confidentiality and integrity of Intel SGX enclaves~\cite{chen2021voltpillager}. \citeauthor{Chen23pmfault} later demonstrates that undervolting via the \ac{PMBus} can similarly break \ac{SGX} enclave integrity, bypassing Intel's countermeasures against previous software undervolting attacks~\cite{murdock2020plundervolt,kenjar2020v0ltpwn,qiu2020voltjockey}.
``MEMBUSTER''~\cite{lee20membuster} introduces an off-chip attack that breaks \ac{SGX} enclave confidentiality by snooping the memory bus (with high-spec, expensive lab equipment), extracting memory access patterns and increasing cache misses using techniques like critical page whitelisting and cache squeezing.

\subsubsection{Interface-based Attacks}
\ac{SGX} enclave security can be compromised if they fail to strictly adhere to secure interfaces between trusted and untrusted code. 
Despite well-designed open-source \acp{SDK}, \citeauthor{van2019tale}~\cite{van2019tale} identify several sanitization vulnerabilities in \ac{ABI} and \ac{API}, leading to memory safety and side-channel vulnerabilities.
\citeauthor{alder2020faulty}~\cite{alder2020faulty} explores the attack surface affecting floating-point computations in \ac{SGX} enclaves via \ac{ABI}, showing that control and state registers are not always properly sanitized.
Recently, there has been a focus on symbolic execution tools for \ac{SGX}~\cite{cloosters2020teerex,khandaker2020coin,alder2024pandora}. 
Pandora~\cite{alder2024pandora} enables runtime-agnostic symbolic execution of the exact attested enclave binary and validates the enclave shielding runtime. Pandora can autonomously discover 200 new and 69 known vulnerable code locations across 11 different SGX shielding runtimes.

\subsubsection{Countermeasures}

In this section, we review and categorize the most relevant countermeasures based on both prior surveys~\cite{nilsson2020survey,fei2021security} and potential solutions proposed in our summarized attack papers.

\emph{(i) System/Microcode-level countermeasures:}
A \ac{CPU}'s complex instructions are often managed by low-level software called microcode, rather than being entirely implemented in hardware. Thus, Intel can issue microcode updates, which are typically the most effective method.
For instance, in response to \ac{DVFS}-based attacks like Plundervolt~\cite{murdock2020plundervolt}, Intel released microcode updates that fully disable the software voltage scaling interface via MSR $\mathtt{0x150}$.

\emph{(ii) Compiler-based countermeasures:}
Beyond Intel's solutions, compiler-based approaches are also effective.
Deterministic multiplexing~\cite{shinde2016preventing} ensures that page-fault access patterns remain consistent regardless of input values by proactively accessing all data and code in a predetermined sequence.
``T-SGX''~\cite{shih2017t} use Intel's \ac{TSX} to redirect exceptions and interrupts to a specific page, terminating the program if a page fault is detected. 
\citeauthor{chen2017detecting}~\cite{chen2017detecting} propose shielded execution with reliable time measurement to check the execution time of target programs. 
``Cloak''~\cite{gruss2017strong} uses Hardware Transactional Memory (HTM) to protect against malicious cache observations.
Similarly, glitch detection~\cite{spensky2021glitching} can help mitigate \ac{DVFS}-based fault injection attacks.

\emph{(iii) Randomization:}
``SGX-Shield''~\cite{seo2017sgx} leverages \ac{ASLR} to create a secure in-enclave loader that secretly randomizes memory space layout.
\citeauthor{hosseinzadeh2018mitigating}~\cite{hosseinzadeh2018mitigating} propose runtime control flow randomization, implemented as compiler extensions on \texttt{llvm}. 
``DR.SGX''~\cite{brasser2019dr} disrupts memory observations by permuting data locations.

\emph{(iv) Application/Source code design:}
Early works like ``OBFUSCURO''~\cite{ahmad2019obfuscuro} enforce operations with \ac{ORAM} protection to prevent attackers from inferring access patterns. 
Several schemes~\cite{ohrimenko2016oblivious,poddar2020visor,wang2022enclavetree} utilize oblivious primitives, implemented in x86\_64 assembly, to design data-oblivious applications.

\subsection{Security of Arm \ac{TZ}}
\label{sec:tz_attack}

\subsubsection{Implementation Bugs}
\citeauthor{cerdeira2020sok}~\cite{cerdeira2020sok} examine bug reports obtained from public \ac{CVE} databases and vendor bulletin reports, categorising them into validation bugs, functional bugs, and extrinsic bugs. 
These bugs are pervasive and often exploited to escalate privileges, enabling attackers to fully hijack kernels in devices from Qualcomm~\cite{beniamini2016trustzone} and Huawei~\cite{shen2015exploiting} or compromise client applications like Samsung Pay~\cite{Danielunbox}.

\subsubsection{Cache-based Attacks}
Given that the cache is shared between \ac{NW} and \ac{SW} in \ac{TZ}-enabled processors, several studies capitalize on the contention arising from this cache coherency to extract sensitive information~\cite{zhang2016cachekit,guanciale2016cache,lipp2016armageddon,zhang2016truspy,cho2018prime+}.
\cite{lipp2016armageddon,zhang2016truspy,cho2018prime+} use the Prime+Probe method to monitor the cache activities within \ac{TZ}.

\subsubsection{Branch Prediction Attacks}
Similar to the attacks on Intel \ac{SGX} mentioned earlier, the branch predictor can also be exploited to target \ac{TZ}.
Since the \ac{BTB} is shared between \ac{NW} and \ac{SW}, techniques like Prime+Probe can be used to disclose secure information to \ac{NW}. 
For example, \citeauthor{ryan2019hardware}~\cite{ryan2019hardware} successfully recovered a 256-bit private key from Qualcomm’s hardware-backed keystore.

\subsubsection{Software-based Fault Injection Attacks}
Several studies have also expanded software-based fault injection attacks to \ac{TZ} or \ac{TZ}-based systems.
``CLKSCREW''~\cite{tang2017clkscrew} leverages manipulation on \ac{DVFS} to induce faulty computations to breach \ac{TZ} hardware-enforced boundaries, extract secret keys, and bypass code signing operations. 
Rowhammer-based attacks have also been adapted to undermine \ac{TZ} security. 
Given that overclocking typically involves running the processor at high frequencies, which can be relatively easy to detect and prevent (\eg through hardware frequency locking), ``VoltJockey''~\cite{qiu2020voltjockey} adopts a different strategy by manipulating voltages rather than frequencies to generate hardware faults on the victim cores. 
\citeauthor{Pierrerowhammer}~\cite{Pierrerowhammer} use bitflips to bypass security measures and access secure memory, enabling the extraction of \ac{RSA} keys and undermining the secure storage provided by \ac{TZ}.

\subsubsection{Countermeasures}

In this section, we review and categorize countermeasures based on both previous survey~\cite{cerdeira2020sok} and potential solutions from the summarized attack papers.
In addressing the architectural defences, some research efforts~\cite{sun2015trustice,brasser2019sanctuary} enhance isolation granularity between \ac{TEE} components and reduce the amount of code running in \ac{SW}, thereby limiting severe privilege escalation attacks within \ac{SW}. 
Countermeasures~\cite{jang2015secret,jang2018retrofitting} tackle the weaknesses of inadequate/weak authentication when accessing \ac{TEE} resources from \ac{NW}, and potentially insecure shared memory used for data exchange across the \ac{NW}-\ac{SW} boundary.
Unlike Intel \ac{SGX}, \ac{TZ} lacks built-in on-chip memory encryption, prompting some works~\cite{zhang2016case,yun2019ginseng} to design encryption strategies to bridge this gap. 
To further strengthen the verification of both \ac{TEE} and \ac{TA} binaries' integrity and identity, several studies~\cite{ferraiuolo2017komodo,lentz2018secloak} introduced additional primitives, such as enhanced remote attestation and sealed storage.
To address implementation bugs, ``RustZone''~\cite{evenchick2018rustzone} extends \ac{TZ} by implementing \acp{TA} using Rust, which inherently provides memory safety and thread safety. 

Cache side-channel attacks can be mitigated by carefully implementing cryptographic algorithms in software~\cite{guanciale2016cache} or using dedicated hardware~\cite{lipp2016armageddon}.
In terms of \ac{BTB} attacks, flushing the shared microarchitectural structures when transferring between \ac{NW} and \ac{SW} can prevent attacks from targeting keys through the \ac{BTB}. 
However, as observed in the Nexus 5X firmware by~\cite{ryan2019hardware}, the L1 cache is not flushed, making Prime+Probe-based attacks still feasible.

Countermeasures against \ac{DVFS}-based fault injection attacks are similar to those used for \ac{SGX}.
In comparison to other attacks on \ac{TZ}, Rowhammer attacks are relatively straightforward to mitigate. They rely on inducing bit flips in adjacent memory rows, making them feasible only if non-secure memory is situated next to a memory row within \ac{SW}. If secure memory resides on a separate memory storage device, Rowhammer attacks become impractical.

\subsection{Security of AMD \ac{SEV}}

\subsubsection{Memory Encryption Issues}
Since the release of AMD \ac{SEV}~\cite{amdsev}, numerous attacks have targeted its early memory security mechanisms~\cite{hetzelt2017security,buhren2017fault,du2017secure,morbitzer2018severed}.
\citeauthor{hetzelt2017security}~\cite{hetzelt2017security} utilize a return-oriented programming-like attack to exploit an unencrypted \ac{VM} control block, enabling arbitrary reading and writing of encrypted memory in the guest \ac{VM}. Such vulnerabilities have been addressed in \ac{SEV}-ES~\cite{kaplan2017protecting}.
\citeauthor{du2017secure}~\cite{du2017secure} leverage \ac{ECB} mode in memory encryption to conduct a chosen-plaintext attack via an HTTP server installed on the guest \ac{VM}. 
Sevurity~\cite{wilke2020sevurity} exploits vulnerabilities in the \ac{XEX} mode to insert arbitrary 2-byte instructions into encrypted memory, while \citeauthor{morbitzer2018severed}~\cite{morbitzer2018severed} breach confidentiality by manipulating nested page tables, altering the virtual memory of the guest \acp{VM}.
Most of these attacks arise from the lack of integrity protection, but \ac{SEV}-\ac{SNP}~\cite{sev2020strengthening} addressed them through the \acl{RMP}.
Conversely, ``CIPHERLEAKS''~\cite{li2021cipherleaks} allows a privileged adversary to infer guest \ac{VM} execution states or recover certain plaintext by monitoring changes in the ciphertext of the victim \ac{VM}.
The authors highlight that AMD has confirmed that \ac{SEV}-\ac{SNP} is also vulnerable to the CIPHERLEAKS attack~\cite{li2022understanding}.

Additionally, ``CROSSLINE'' and its variants by \citeauthor{li2021crossline}~\cite{li2021crossline,li2022understanding} extract the memory content of victim \acp{VM} by exploiting improper use of the \ac{ASID} for controlling \ac{VM} accesses to encrypted memory pages, cache lines, and \ac{TLB} entries. 
They also compromise the integrity and confidentiality of \ac{SEV} \acp{VM} by poisoning \ac{TLB} entries~\cite{li2021tlb}. It is noted that \ac{SEV}-\ac{SNP} is expected to address these issues, including \ac{TLB} misuse.

\subsubsection{Software-based Fault Injection Attacks}
\citeauthor{morbitzer2021severity}~\cite{morbitzer2021severity} exploit the system's lack of memory integrity protection to inject faults and execute arbitrary code within \ac{SEV} \acp{VM} through I/O channels, using the hypervisor to locate and trigger the execution of encrypted payloads. 
However, \ac{SEV}-\ac{SNP}~\cite{sev2020strengthening} can mitigate this issue through \ac{RMP}.
\citeauthor{buhren2021one}~\cite{buhren2021one} present a voltage fault injection attack enabling the execution of custom payloads on AMD secure processors, subsequently decrypting a \ac{VM}’s memory by deploying custom \ac{SEV} firmware.
``CacheWarp''~\cite{zhang2024cachewarp} presents a new software-based fault attack targeting AMD \ac{SEV}-ES and \ac{SEV}-\ac{SNP}, exploiting the possibility to revert modified cache lines of guest \acp{VM} to their previous (stale) state at the architectural level. 
This attack only requires interrupting the \ac{VM} at a point chosen by the attacker to invalidate modified cache lines without writing them back to memory.

\subsubsection{Interrupt-driven Attacks}
Similar to SGX-Step framework~\cite{van2017sgx}, SEV-Step~\cite{wilke2024sev} introduces reliable single-stepping against \ac{SEV} \acp{VM}.
In addition to this functionality, SEV-Step facilitates access to common attack techniques, such as page fault tracking and cache attacks targeting \ac{SEV}.
More recently, \citeauthor{schluter2024wesee}~\cite{schluter2024wesee} introduce a novel attack where the hypervisor injects well-crafted, malicious $\mathtt{\#VC}$ (a new exception by AMD facilitating communication between the \ac{VM} and untrusted hypervisor) into a victim \ac{VM}’s \ac{CPU}, enabling the attacker to induce arbitrary behaviour in the \ac{VM}.
In another related attack, ``HECKLER''~\cite{schluter2024heckler} shows that the hypervisor can inject malicious non-timer interrupts to compromise \acp{VM}. They leverage interrupt handlers with global effects, allowing manipulation of a \ac{VM}'s register states to alter data and control flow. 
In general, \ac{SEV}-\ac{SNP}  remains vulnerable to these attacks.

\subsubsection{Countermeasures}
Most memory-related attacks have been addressed in recent AMD product updates. Here, we focus on countermeasures against fault injection attacks.
For malicious voltage drops or glitches, similar to countermeasures used for \ac{SGX} and \ac{TZ}, hardware-level voltage monitoring circuits and software-based detection mechanisms can be implemented. 
To counter fault attacks that enable fine-grained memory write suppression through specific hypervisor instructions, restricting the use of those instructions at the hardware level or implementing compiler-level solutions can ensure correct memory operations. 
For attacks that inject malicious interrupts, one can block or detect the injections by monitoring external interrupts or disabling the interrupt handler.

\end{document}